\documentclass{aa}

\usepackage{graphicx}
\usepackage{txfonts}
\usepackage{upgreek}
\usepackage[dvipsnames]{xcolor}
\usepackage[squaren]{SIunits}
\usepackage{hyperref}
\usepackage{amsfonts}
\usepackage{amsmath}
\usepackage{booktabs}
\usepackage{orcidlink}

\usepackage{subcaption}
\let\savedegree\degree
\let\degree\relax
\let\savecdot\cdot
\let\cdot\relax
\let\savesecond\second
\let\second\relax
\let\savefourth\fourth
\let\fourth\relax
\usepackage{mathabx}
\let\degree\savedegree
\let\cdot\savecdot
\let\second\savesecond
\let\fourth\savefourth

\usepackage{amssymb}
\usepackage{tikz}
\hypersetup{
    colorlinks=true,
    citecolor=blue,
    linkcolor=blue,
    filecolor=blue,
    urlcolor=blue
}

\newcommand{\tensorA}{\boldsymbol{\mathcal{A}}}
\newcommand{\tensorO}{\boldsymbol{\mathcal{O}}}
\newcommand{\tensorPhi}{\boldsymbol{\varPhi}}
\newcommand{\tensorD}{\boldsymbol{\mathcal{D}}}
\newcommand{\tensorS}{\boldsymbol{\mathcal{S}}}
\newcommand{\tensorP}{\boldsymbol{\mathcal{P}}}
\newcommand{\tensorE}{\boldsymbol{\mathcal{E}}}
\newcommand{\tensorSest}{\boldsymbol{\hat{\mathcal{S}}}}
\newcommand{\tensorPest}{\boldsymbol{\hat{\mathcal{P}}}}

\newcommand{\Halpha}{\ensuremath{\mathsf{H}_{\upalpha}}\xspace}
\newcommand{\Hbeta}{\ensuremath{\mathsf{H}_{\upbeta}}\xspace}
\newcommand{\PaBeta}{\ensuremath{\mathsf{Pa}_{\upbeta}}\xspace}
\newcommand{\BrGamma}{\ensuremath{\mathsf{Br}_{\upgamma}}\xspace}

\begin{document}

    \title{Stellar halo subtraction alternative for accreting companions' characterization with integral field spectroscopy}
    
    \subtitle{Analytical and on-sky demonstration on the PDS70, HTLup, and YSES1 systems}
    
    \author{
        Rémi Julo\inst{1}\fnmsep\inst{2} \orcidlink{0009-0009-7275-944X},
        Mickaël Bonnefoy\inst{1} \orcidlink{0000-0001-5579-5339},
        Florent Chatelain\inst{2} \orcidlink{0000-0002-0949-8103},
        Olivier Flasseur\inst{3} \orcidlink{0000-0003-4885-3275},
        Olivier Michel\inst{2} \orcidlink{0000-0002-0890-0383}, \\
        Sebastián Jorquera\inst{4}\fnmsep\inst{5}\fnmsep\inst{6} \orcidlink{0000-0001-6822-7664},
        Philippe Delorme\inst{1} \orcidlink{0000-0002-2279-410X}, and
        Gaël Chauvin\inst{7}\fnmsep\inst{4}\fnmsep\inst{1} \orcidlink{0000-0003-4022-8598}
    }

    \institute{
        UGA, Institut de Planétologie et d'Astrophysique de Grenoble, CNRS, Saint-Martin-d'Hères, 38400, France
        \and
        UGA, Grenoble Images Parole Signal Automatique, Grenoble INP, CNRS, Saint-Martin-d'Hères, 38400, France
        \and
        UCBL, Centre de Recherche Astrophysique de Lyon UMR 5574, ENS de Lyon, CNRS, Villeurbanne, F-69622, France
        \and
        Université Côte d’Azur, Nice, 06304, France
        \and
        Universidad de Chile, Las Condes, 7550000, Chile
        \and
        Millennium Nucleus for Planet Formation, Valparaíso, 2340000, Chile
        \and
        Max-Planck-Institut für Astronomie, Königstuhl 17, Heidelberg, 69117, Germany
    }

    \abstract
    {
    Medium-resolution integral field spectrographs (IFS), such as the Multi-Unit Spectroscopic Explorer (MUSE) instrument at the Very Large Telescope (VLT), are equipped to detect the emission lines (e.g., $\Halpha, \Hbeta$) of faint accreting companions when associated with dedicated stellar halo subtraction methods. We recently proposed a new approach based on polynomial modulations of a stellar spectrum estimate across the field of view, with orthogonal polynomials and lines masking. This new technique is designed to better preserve both continuum and emission lines of accreting companions.
    }
    {
    We seek to highlight and quantify analytically and on real data the benefits of this new approach over the one classically used, particularly with regard to distortions of the extracted spectra. We also examine both operating regimes.
    }
    {
    We carried out analytical calculations based on simple toy models of spectra to identify and quantify the main theoretical problems of the state-of-the-art technique, the proposed corrections of our new method, and the remaining limitations of the latter. Simulations of the most extreme situations identified were used to highlight these problems and corrections. Archival VLT/MUSE data of the young PDS70 and HTLup systems were used to vet the detection and characterization capabilities using on-sky observations. New images of the YSES1 planetary system were used to further illustrate the gains.
    }
    {
    We show that the state-of-the-art stellar halo subtraction method, based on low-pass filtering, can lead to the self-subtraction of the emission lines and modify the neighboring continua, depending on the line contrast to neighboring continuum contrast ratios. We show that the proposed technique corrects these characterization problems, while maintaining the same detection capabilities. The two protoplanets PDS70 b and c were detected with 5$\sigma$ significance. The $\Halpha$ line estimate of the HTLup B stellar companion was improved by $\sim$30\% for the integrated flux and by $\sim$8\% for the 10\%-width. As for YSES1 b, we found it uniquely displays a combination of \Halpha, \Hbeta, CaII H\&K triplet, and HeI lines in emission that can be attributed to accretion and/or chromospheric activity. We derived an accretion rate at \Halpha of $\sim1.45 \times 10^{-9}$ $\mathrm{M_{Jup}/year}$ with our new method, versus $\sim1.11 \times 10^{-9}$ $\mathrm{M_{Jup}/year}$ with the reference method, namely, $\sim$30\% less. These new results are compatible with values derived for other companions in this mass range. We note that YSES1 c was not detected in our observations.
    }
    {
    The proposed subtraction method better preserves the spectral information, notably the emission line fluxes and profiles, while achieving similar detection performance. Based on a linear and parametric approach, it can be extended and/or combined with additional faint signal search algorithms.
    }

    \keywords{
        Techniques: imaging spectroscopy --
        Planets and satellites: detection --
        Line: profiles --
        Accretion, accretion disks
    }

    \titlerunning{Stellar halo subtraction alternative for integral field spectroscopy}
    \authorrunning{R. Julo, M. Bonnefoy, F. Chatelain, O. Flasseur, O. Michel}
    
    \maketitle

\section{Introduction}\label{intro}

A growing number of exoplanets are being detected around young stars (age $<$ 50 Myr), either through transit monitoring \citep[e.g.,][]{Mann_2016, Plavchan_2020, Bouma_2020, Barber_2020} or radial velocity measurements \citep[e.g.,][]{Lagrange_2019}. Historically, however, the first young exoplanets were identified via high-contrast imaging \citep[e.g.,][]{Chauvin_2005, Marois_2008, Lagrange_2010, DeRosa_2023}. At such a young age, the continuum of a Jovian exoplanet becomes bright enough to be detected directly thanks to adaptive optics (AO), enabling instruments to reach contrasts up to $10^{-6}$ \citep[e.g., VLT/SPHERE, Gemini/GPI, Subaru/SCExAO;][]{Beuzit_2019, Macintosh_2014, Guyon_2010}. Young planets are of particular interest to test the most recent predictions of formation models since they retain information about their orbital radius and composition at formation. Among that population, planets undergoing active accretion from surrounding gas provide direct constraints on formation timescales and the physics of accretion in the planetary-mass range, which is believed to set their early evolutionary pathway \citep[the so-called hot- and cold-starts;][]{Marley_2007}, spins \citep{Batygin_2018}, internal structure \citep{Cumming_2018}, planet-disk migration \citep{Pierens_2016}, and the composition and formation of exomoons \citep{Heller_2018}.

A handful of accreting planetary-mass companions have been discovered \citep{Betti_2023} so far. The emission lines (\Halpha at $6563\text{\AA}$, \PaBeta at $12820\text{\AA}$, \BrGamma at $21660\text{\AA}$, HeI lines, etc.) are inventoried, and the characterization of line fluxes, profiles, and variability levels has begun, allowing the community to peer into the physics of accretion \citep[e.g.,][]{Betti_2022, Demars_2023, Ringqvist_2023, Aoyama_2024}. Among that sample, PDS70 stands out as a young system with two firmly established protoplanets, at $\sim$170 mas and $\sim$220 mas projected separation \citep{Wang_2021}. They are both nested within the cavity of a primordial disk surrounding the host star \citep{Keppler_2018, Muller_2018, Wagner_2018, Haffert_2019}. The detection of the second planet in the system was made possible by the Multi-Unit Spectroscopic Explorer \citep[MUSE;][]{Bacon_2010} integral field spectrograph (IFS) at the Very Large Telescope (VLT), which provides hyperspectral imaging at optical wavelengths coupled to AO \citep{Arsenault_2008, Stroele_2012}. The medium resolution of MUSE ($\mathrm{R}_{\lambda}$ of 1770 at $4750\text{\AA}$ to 3590 at $9350\text{\AA}$) allows us to pinpoint the narrow \Halpha line emission of accreting companions \citep[e.g.,][]{Haffert_2019, Eriksson_2020}; at the specific wavelength of the \Halpha emission line, the contrast with the star becomes modest \cite[$10^{-2}$ to $10^{-3}$;][]{Mordasini_2017}. IFS such as MUSE also provide the data diversity needed for an efficient removal of the stellar flux and stellar light scattered by circumstellar disks that can overlap the signal of the companions. Emission line fluxes and profiles of accreting companions such as PDS70 b or c can be extracted, after halo subtraction, from the MUSE data cubes. These studies have triggered new theoretical developments and provided constraints on the accretion rates, in-line extinction by surrounding material, and the physics of accretion \citep[e.g.,][]{Aoyama_2019, Thanathibodee_2019, Szulagyi_2020, Hashimoto_2020, Marleau_2022}.

Campaigns to detect analogues of PDS70 b and c have thus far targeted \Halpha, \PaBeta, and \BrGamma emissions in young systems with circumstellar disks. Most of them have relied on differential imaging in narrow-band filters, whose transmission is centered on the line and the surrounding continuum \citep{Cugno_2019, Uyama_2020, Zurlo_2020, Huelamo_2022, Follette_2023, Chaushev_2024}. Reported candidates \citep[LkCa 15 b, AB Aur b;][]{Sallum_2015, Currie_2022} await confirmation. AO-fed IFS at medium spectral resolution have been used to target a few systems thus far, looking for \PaBeta emissions \citep[OSIRIS at Keck;][]{Uyama_2017, Uyama_2021, Biddle_2024} as well as \Halpha lines \citep[MUSE at VLT;][]{Xie_2020}. Only the stellar companion around HTLup has been reported as a positive detection with MUSE, at a separation similar to the protoplanets around PDS70 \citep[$\sim$160 mas;][]{Jorquera_2024}. No new protoplanet has been reported from archival data or ongoing observations. Variability of the lines or in-line extinction from circumstellar and/or circumplanetary material are proposed to explain this poor yield \citep{Szulagyi_2020, Marleau_2022, Demars_2023, Alarcon_2024}.

\cite{Hoeijmakers_2018} presented a methodology named high-resolution spectral differential imaging (HRSDI) to remove the stellar light from VLT/SINFONI AO-fed IFS observations ($R\sim 4000$; $19600$ - $24500\text{\AA}$). The method was adapted by \cite{Haffert_2019} and \cite{Xie_2020} to work on MUSE data. However, although it is currently the reference method for planet detection and characterization from IFS data, it can distort the companion emission spectrum. Moreover, being non-linear and non-parametric, it prevents regularization and further extensions. Alternatively, spline-based models for complex halos/speckles thus have also emerged \citep{Agrawal_2023, Ruffio_2023}.

In that context, in \cite{Julo_2024}, we laid the foundations of a new halo subtraction methodology aimed at paving the way for some of the developments mentioned above. In that work, we demonstrated the feasibility of our approach, but we did not provide a more extended demonstration of its capabilities on-sky.

We build upon the study of \cite{Julo_2024}, and present a more thorough investigation of the biases introduced by the state-of-the-art method using an analytical approach in Sect. \ref{sgf}, while providing a demonstration of our method using archival and new on-sky data in Sect. \ref{lpm}. Then, we examine the remaining limitations and give our conclusions, remarks, future prospects in Sect. \ref{conclu}. A preliminary definition of the proposed model was given in \cite{Julo_2024}. The present paper also provides a more accurate and complete description of both methodologies in Appendix \ref{meths} for better understanding and reproducibility.

\section{Quantification of the biases}\label{sgf}

The HRSDI method can be decomposed into two steps:
\begin{itemize}
    \item First, a “reference stellar spectrum” is built by averaging a selection of spaxels of the field of view. For each spaxel of the full field of view, the ratio of the data to the reference spectrum is smoothed by a low-pass Savitzky-Golay filter (to remove the high-frequency planetary components of the observations). Each of the spaxels of the field is subtracted from the product of the reference spectrum and this filtered ratio. The division and filtering problems are highlighted in the following section using a custom toy model.
    \item Second, principal component analysis (PCA) is used on the cube of residuals to remove static patterns in the field of view. It assumes (improperly) the relevance of a low-rank model of the companion in the post-subtraction residuals of the data, and could introduce self-subtraction of its signal. 
\end{itemize}
Because of this uncontrolled self-subtraction, we prefer to avoid this second step. Therefore, we chose to focus our work on the first step and name it the Savitzky-Golay filtering (SGF) method.

\subsection{Analytical study with toy models}\label{sgf_toys}

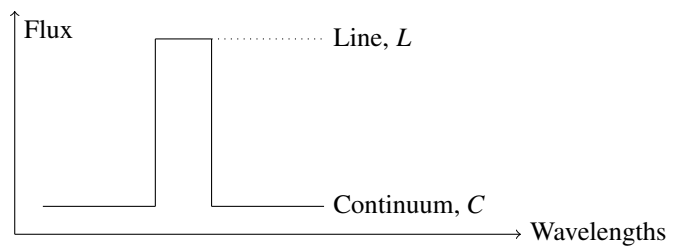
\begin{figure}[t!]
\begin{tikzpicture}
\draw (0*0.74,0*0.74) -- (2*0.74,0*0.74);
\draw (2*0.74,0*0.74) -- (2*0.74,3*0.74);
\draw (2*0.74,3*0.74) -- (3*0.74,3*0.74);
\draw (3*0.74,3*0.74) -- (3*0.74,0*0.74);
\draw (3*0.74,0*0.74) -- (5*0.74,0*0.74) node [right] {Continuum, $C$};
\draw[dotted] (3*0.74,3*0.74) -- (5*0.74,3*0.74) node [right] {Line, $L$};
\draw[->] (-0.5*0.74,-0.5*0.74) -- (8.5*0.74,-0.5*0.74) node [right] {Wavelengths};
\draw[->] (-0.5*0.74,-0.5*0.74) -- (-0.5*0.74,3.5*0.74) node [right, anchor=north west, align=left] {Flux};
\end{tikzpicture}
\caption{Toy model with rectangular line and flat neighboring continuum.}\label{fig:toy_model}
\end{figure}

The toy model is introduced in Fig. \ref{fig:toy_model}. $C_S$ and $L_S$ (respectively, $C_P$ and $L_P$) denote the continuum and line fluxes of the stellar (respectively, planetary) component of a data spaxel. The latter thus exhibits $C=C_S+C_P$ and $L=L_S+L_P$. Furthermore, $C_S \gg C_P$ and $L_S \gg L_P$ are assumed. These hypothesis allow us to infer the quantities $C_S$ and $L_S$ to a certain factor through the stellar reference spectrum. The major issue related to this step is then to evaluate this (position-dependent) factor, referred to as $\alpha$ (assumed locally spectrally constant in the toy model).

\begin{figure*}[t]
\centering

\begin{subfigure}[b]{0.45\linewidth}
\captionsetup{labelformat=parens,labelsep=space}
\renewcommand\thesubfigure{a}
\begin{tikzpicture}
\draw (0*0.74,0*0.74) -- (2*0.74,0*0.74);
\draw (2*0.74,0*0.74) -- (2*0.74,3*0.74);
\draw (2*0.74,3*0.74) -- (3*0.74,3*0.74);
\draw (3*0.74,3*0.74) -- (3*0.74,0*0.74);
\draw (3*0.74,0*0.74) -- (5*0.74,0*0.74) node [right] {$h_\mathrm{min}=\alpha\left(1 + \frac{C_P}{C_S}\right)$};
\draw[dotted] (3*0.74,3*0.74) -- (5*0.74,3*0.74) node [right] {$h_\mathrm{max}=\alpha\left(1 + \frac{L_P}{L_S}\right)$};
\draw[dotted] (2*0.74,0*0.74) -- (2*0.74,-0.5*0.74);
\draw[dotted] (3*0.74,0*0.74) -- (3*0.74,-0.5*0.74);
\draw[<->] (4*0.74,0*0.74) -- (4*0.74,3*0.74);
\draw[<->] (2*0.74,-0.5*0.74) -- (3*0.74,-0.5*0.74);
\node[right] at (4*0.74,1.5*0.74) {$\bar{\b{H}} = \alpha\left(\frac{L_P}{L_S} - \frac{C_P}{C_S}\right)$};
\node[below] at (2.5*0.74,-0.5*0.74) {$\mathrm{R}\bar{\b{W}}$};
\draw[->] (-0.6*0.74,-1.4*0.74) -- (9.6*0.74,-1.4*0.74) node [right, anchor=north east] {Wavelengths};
\draw[->] (-0.6*0.74,-1.4*0.74) -- (-0.6*0.74,3.6*0.74) node [right, anchor=north west, align=left] {Spaxel to\\spectrum\\ratio};
\end{tikzpicture}
\caption{}\label{fig:toy_division}
\end{subfigure}
\hspace{\baselineskip}
\begin{subfigure}[b]{0.5\linewidth}
\centering
\captionsetup{labelformat=parens,labelsep=space}
\renewcommand\thesubfigure{c}
\begin{tikzpicture}
\draw (0*0.74,0*0.74) -- (1*0.74,0*0.74);
\draw (1*0.74,0*0.74) -- (1*0.74,1*0.74);
\draw (1*0.74,1*0.74) -- (2*0.74,1*0.74);
\draw (2*0.74,1*0.74) -- (2*0.74,4*0.74);
\draw (2*0.74,4*0.74) -- (3*0.74,4*0.74);
\draw (3*0.74,4*0.74) -- (3*0.74,1*0.74);
\draw (3*0.74,1*0.74) -- (4*0.74,1*0.74);
\draw (4*0.74,1*0.74) -- (4*0.74,0*0.74);
\draw (4*0.74,0*0.74) -- (5*0.74,0*0.74) node [right] {$\hat{C}_S = C_S + C_P$};
\draw[dotted] (4*0.74,1*0.74) -- (5*0.74,1*0.74) node [right] {$\tilde{C}_S = C_S + C_P + C_S~\mathrm{R}\left(\frac{L_P}{L_S} - \frac{C_P}{C_S}\right)$};
\draw[dotted] (3*0.74,4*0.74) -- (5*0.74,4*0.74) node [right] {$\hat{L}_S = L_S\left(1+\frac{C_P}{C_S} + \mathrm{R}\left(\frac{L_P}{L_S} - \frac{C_P}{C_S}\right)\right)$};
\draw[->] (-0.3*0.74,-0.5*0.74) -- (11.8*0.74,-0.5*0.74) node [right, anchor=north east] {Wavelengths};
\draw[->] (-0.3*0.74,-0.5*0.74) -- (-0.3*0.74,4.5*0.74) node [right, anchor=north west, align=left] {Flux};
\end{tikzpicture}
\caption{}\label{fig:toy_stellar_est}
\end{subfigure}

\vspace{\baselineskip}

\begin{subfigure}[b]{0.45\linewidth}
\captionsetup{labelformat=parens,labelsep=space}
\renewcommand\thesubfigure{b}
\begin{tikzpicture}
\draw (0*0.74,0*0.74) -- (1*0.74,0*0.74);
\draw (1*0.74,0*0.74) -- (1*0.74,1*0.74);
\draw (1*0.74,1*0.74) -- (4*0.74,1*0.74);
\draw (4*0.74,1*0.74) -- (4*0.74,0*0.74);
\draw (4*0.74,0*0.74) -- (5*0.74,0*0.74) node [right] {$h_\mathrm{min}$};
\draw[dotted] (4*0.74,1*0.74) -- (5*0.74,1*0.74) node [right] {
  $h_\mathrm{min} + \mathrm{R}
    \makebox[0pt][l]{
      \smash{$\underbrace{\phantom{\alpha\frac{L_P}{L_S} -- \frac{C_P}{C_S}}}_{\bar{\b{H}}}$}}
    \alpha\left(\frac{L_P}{L_S} - \frac{C_P}{C_S}\right)$
};
\draw[dotted] (0.5*0.74,1*0.74) -- (1*0.74,1*0.74);
\draw[dotted] (1*0.74,0*0.74) -- (1*0.74,-0.5*0.74);
\draw[dotted] (4*0.74,0*0.74) -- (4*0.74,-0.5*0.74);
\draw[<->] (0.5*0.74,0*0.74) -- (0.5*0.74,1*0.74);
\draw[<->] (1*0.74,-0.5*0.74) -- (4*0.74,-0.5*0.74);
\node[left] at (0.5*0.74,0.5*0.74) {$\mathrm{R}\bar{\b{H}}$};
\node[below] at (2.5*0.74,-0.5*0.74) {$\bar{\b{W}}$};
\draw[->] (-0.6*0.74,-1.4*0.74) -- (9.6*0.74,-1.4*0.74) node [right, anchor=north east] {Wavelengths};
\draw[->] (-0.6*0.74,-1.4*0.74) -- (-0.6*0.74,3.6*0.74) node [right, anchor=north west, align=left] {Spaxel to\\spectrum\\ratio};
\end{tikzpicture}
\caption{}\label{fig:toy_filtering}
\end{subfigure}
\hspace{\baselineskip}
\begin{subfigure}[b]{0.5\linewidth}
\centering
\captionsetup{labelformat=parens,labelsep=space}
\renewcommand\thesubfigure{d}
\begin{tikzpicture}
\draw (0*0.74,0*0.74) -- (1*0.74,0*0.74);
\draw (1*0.74,0*0.74) -- (1*0.74,-1*0.74);
\draw (1*0.74,-1*0.74) -- (2*0.74,-1*0.74);
\draw (2*0.74,-1*0.74) -- (2*0.74,2*0.74);
\draw (2*0.74,2*0.74) -- (3*0.74,2*0.74);
\draw (3*0.74,2*0.74) -- (3*0.74,-1*0.74);
\draw (3*0.74,-1*0.74) -- (4*0.74,-1*0.74);
\draw (4*0.74,-1*0.74) -- (4*0.74,0*0.74);
\draw (4*0.74,0*0.74) -- (5*0.74,0*0.74) node [right] {$\hat{C}_P=0$};
\draw[dotted] (4*0.74,-1*0.74) -- (5*0.74,-1*0.74) node [right] {$\tilde{C}_P = -C_P~\mathrm{R}\left(\frac{C_S}{C_P}\frac{L_P}{L_S}-1\right)$};
\draw[dotted] (3*0.74,2*0.74) -- (5*0.74,2*0.74) node [right] {$\hat{L}_P=L_P\left(1-\mathrm{R}\right)\left(1-\frac{L_S}{L_P}\frac{C_P}{C_S}\right)$};
\draw[->] (-0.3*0.74,-1.7*0.74) -- (11.8*0.74,-1.7*0.74) node [right, anchor=north east] {Wavelengths};
\draw[->] (-0.3*0.74,-1.7*0.74) -- (-0.3*0.74,3.3*0.74) node [right, anchor=north west, align=left] {Flux};
\end{tikzpicture}
\caption{}\label{fig:toy_planetary_est}
\end{subfigure}

\caption{Signal processing steps leading to post-subtraction planetary spaxel estimates. (a) Data spaxel ratio with regard to the reference spectrum. (b) Low-pass filtering with Savitzky-Golay filter. (c) Stellar spaxel estimation with overfitting. (d) Planetary spaxel estimation with self-subtraction.}

\end{figure*}
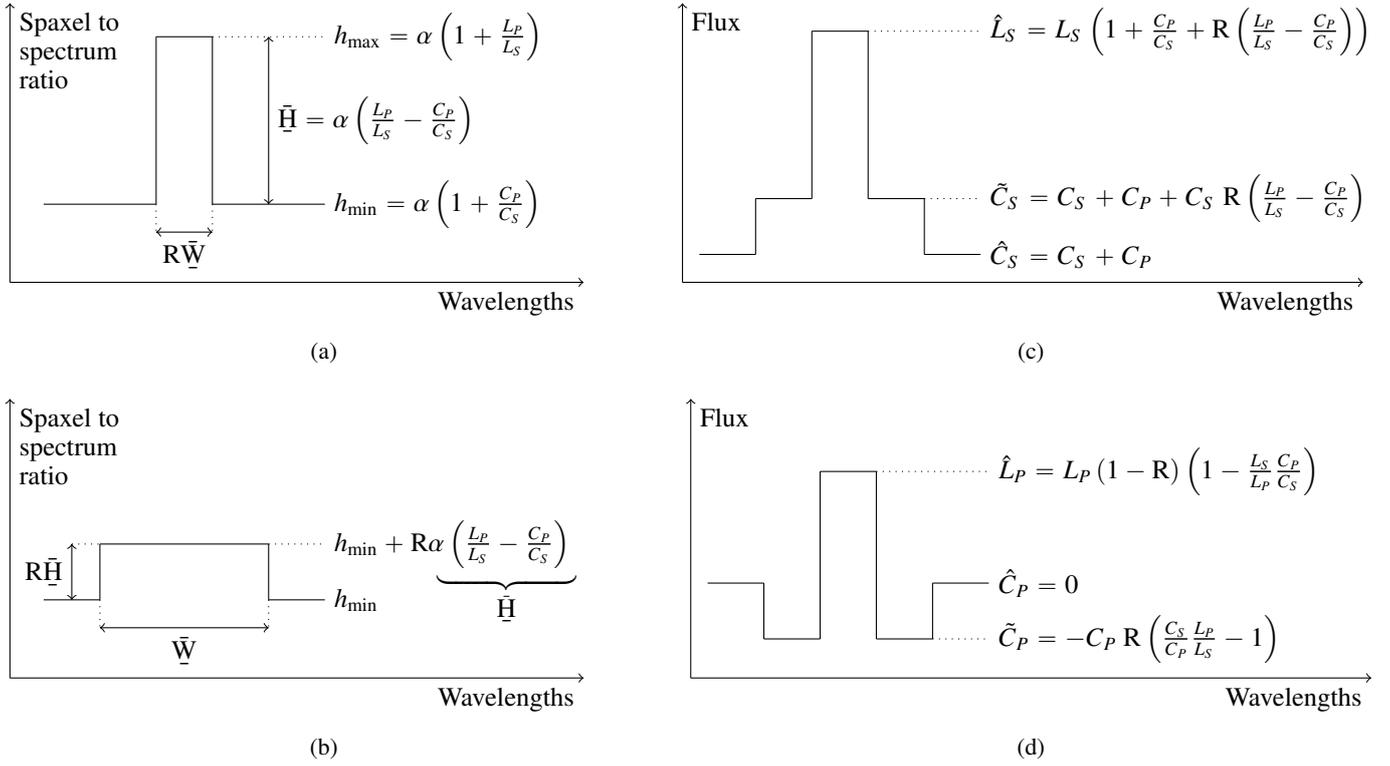

To achieve this, the spaxel is divided entrywise by the stellar reference spectrum, yielding the signal shown in Fig. \ref{fig:toy_division}.

By neglecting the planetary continuum at flux level $C_P$ (which is less informative than the planetary lines), $h_\mathrm{min}$ (which is by construction the data to stellar continuum ratio) is indeed equal to the factor $\alpha$. The task then reduces to filtering the peak induced by the higher contrast planetary line at flux level $L_P$.

Here, this is done by low-pass (Savitzky-Golay) filtering, which results in the signal in Fig. \ref{fig:toy_filtering}. This filtering step leaves a residue that depends on physical and processing configurations. The width of this residual is $\bar{\b{W}}$ and its height is $\mathrm{R}\bar{\b{H}}$, with $\bar{\b{W}}$ the filtering window width, $\mathrm{R}$ the ratio between the line width and the filtering window width, and $\bar{\b{H}}$ the peak height of the spaxel relative to the stellar reference spectrum. Technical details of the calculations are given in Appendix \ref{filter_effect}.

The entrywise product of this deformation with the stellar reference spectrum gives the estimation of the stellar component of the data spaxel shown in Fig. \ref{fig:toy_stellar_est}. Subtracting this estimation from the initial data spaxel gives the estimation of the planetary component of the data shown in Fig. \ref{fig:toy_planetary_est}. Yet, the propagation of the artifact described in the previous steps is clearly highlighted and quantified with this toy model.

Next, we explore two extreme (but still realistic) scenarios, to better understand the limitations of this approach:

\begin{itemize}

\item If $L_P/L_S \gg C_P/C_S$, then $\bar{\b{H}} \gg \alpha$ and consequently $\mathrm{R}\bar{\b{H}} \gg \mathrm{R}\alpha$. This leads to the overfitting of the flux of the stellar spaxel defined in Fig. \ref{fig:toy_stellar_est}, such that $\tilde{C}_S -\hat{C}_S\gg \mathrm{R}C_S$. This results, in turn, in an oversubtraction in the flux of the planetary spaxel estimate defined in Fig. \ref{fig:toy_planetary_est}, such that $\tilde{C}_P-\hat{C}_P \ll -\mathrm{R}C_P$. This is referred to as “self-subtraction” and potentially limits the relevance of physical interpretations. Importantly, the oversubtraction is not uniform across the spectrum: this differs between line wavelengths and neighboring continuum ones (i.e., $L_S-\hat{L}_S\neq\tilde{C}_S$). Consequently, this process induces an irrecoverable loss of information (since $L_S\neq\hat{L}_S+\tilde{C}_S$).

\item If $L_P/L_S \approx C_P/C_S$, then $\bar{\b{H}} \approx 0$ and consequently $\mathrm{R}\bar{\b{H}} \approx 0$. As a result, the artifacts are negligible in the final estimates (i.e., $\tilde{C}_P-\hat{C}_P \approx 0$), but the planetary signal is minimally recovered (i.e., $\hat{L}_P \ll L_P$). In fact, the full process relies on the peak of the stellar spaxel to spectrum ratio: if it is too high, this leads to self-subtraction; if it is too low, this hinders the recovery of the planetary signal. This reflects the intrinsic difficulty of separating signals with similar contrast ratios.

\end{itemize}

These extreme phenomena are illustrated below (Sect. \ref{sgf_simus}) through a few simulations. These effects compound the intrinsic numerical instabilities introduced by the division step depicted in Fig. \ref{fig:toy_division}, especially at low-flux (noise-dominated) wavelengths.

\subsection{Illustration from simulations}\label{sgf_simus}

While the reduction in line height is intrinsic to the problem of separating signals that are too close together, the self-subtraction around the line is only specific to the SGF method. Following the calculations carried out on our simple toy model, its width can simply be approximated to the width $\bar{\b{W}}$ of the Savitzky-Golay filter window. Its depth with respect to the 0 flux level, which is the expected estimation for the continuum, can be expressed as:
\begin{equation}
\frac{\tilde{C}_P}{\hat{L}_P} = -\left(\frac{\mathrm{R}}{1-\mathrm{R}}\right)\frac{C_S}{L_S}\,,
\label{SGF_depth}
\end{equation}
where $\hat{L}_P$ and $\tilde{C}_P$, introduced in Fig. \ref{fig:toy_planetary_est}, represent, respectively, the estimation of the planetary line height and the estimation of the planetary neighboring continuum depth with regard to the $0$ value. The interest of Eq. \eqref{SGF_depth} is to quantify the self-subtraction effect using only the post-subtraction quantities ($\hat{L}_P$ and $\tilde{C}_P$) and the stellar continuum-to-line ratio, $C_S/L_S$ (since the planetary continuum-to-line ratio, $C_P/L_P$, simplifies in the calculation).

\onecolumn

\begin{figure}[h!]
\centering
\includegraphics[width=0.99\linewidth]{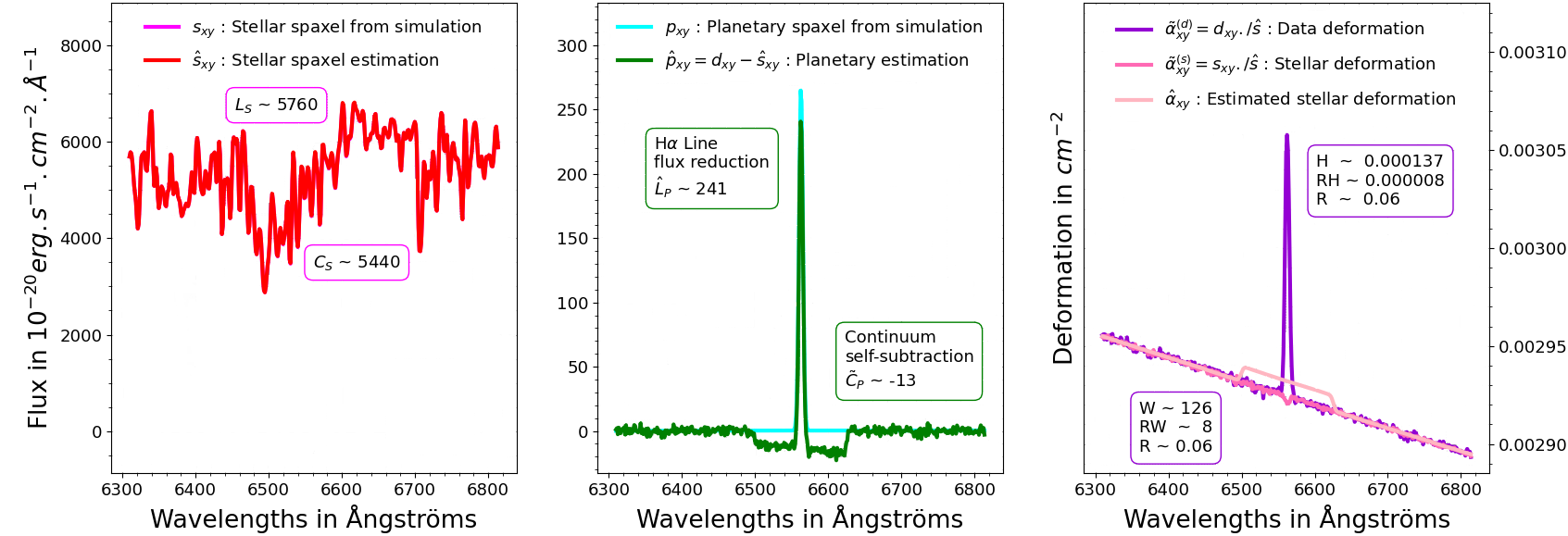}
\caption{Subtraction results at the position of a planet with the SGF method. Left: Overlapping simulation (ground truth) and estimation of the stellar spaxel. There is no emission line in this simulation. Middle: The green spaxel is the estimation of the planetary spaxel, while the cyan one is the associated ground truth. The self-subtraction problem is clearly visible in this situation. Right: The reason for the self-subtraction is evidenced. This is the deformation estimation, in light pink, overfitting the planetary line component of the data to stellar spectrum estimation ratio, in purple. Yet, this overfitting is relatively greater than the planetary and stellar continuum-to-line ratios are different. The real deformation is in dark pink.
\label{fig:sgf_with_out_line}}
\end{figure}

\begin{figure}[h!]
\centering
\includegraphics[width=0.99\linewidth]{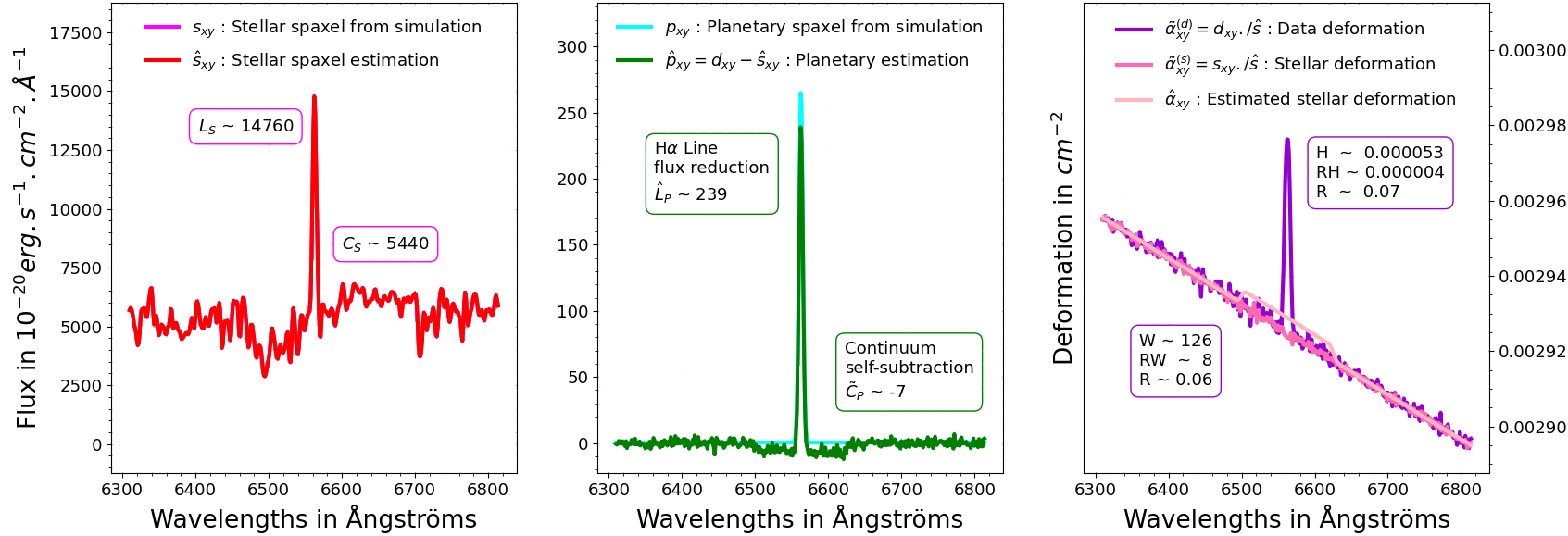}
\caption{Same as Fig. \ref{fig:sgf_with_out_line}, but with a small stellar line (total flux of $5\times10^{-13}$ $\text{erg}.\text{s}^{-1}\text{cm}^{-2}$\AA$^{-1}$ in the 6560.3-6565.3\AA~band).}
\label{fig:sgf_with_small_line}
\end{figure}

\begin{figure}[h!]
\centering
\includegraphics[width=0.99\linewidth]{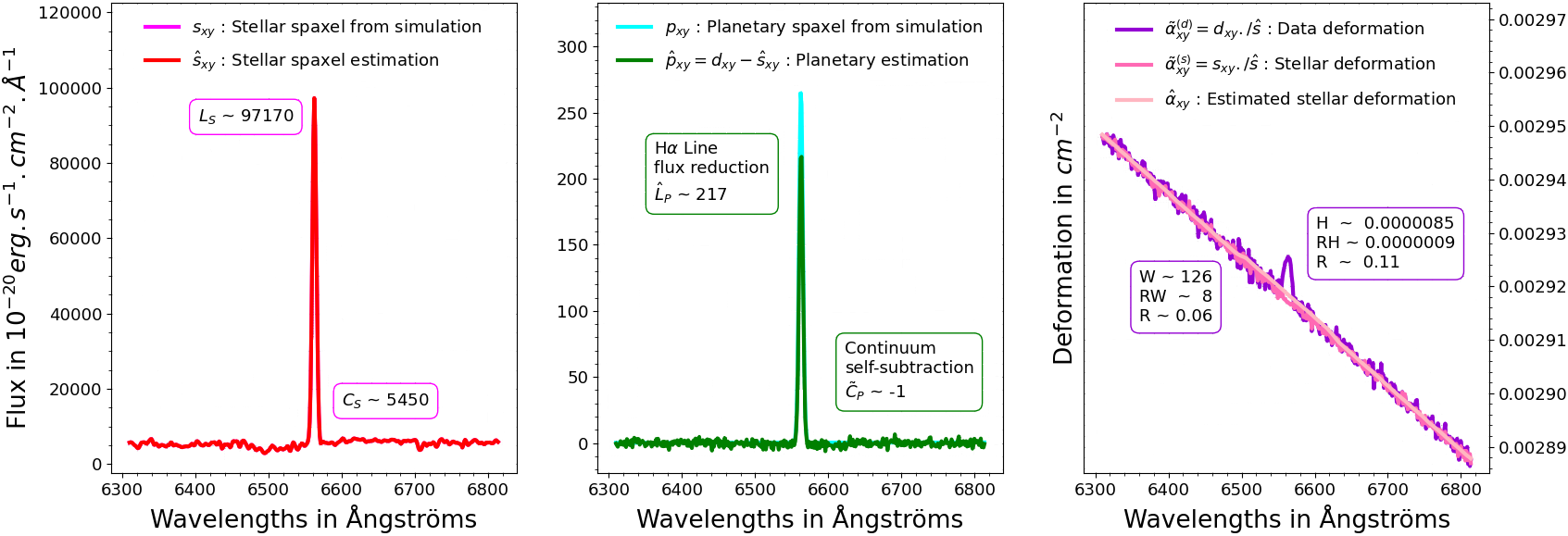}
\caption{Same as Fig. \ref{fig:sgf_with_small_line}, but with a brighter stellar line (total flux of $5\times10^{-12}$ $\text{erg}.\text{s}^{-1}\text{cm}^{-2}$\AA$^{-1}$ in the 6560.3-6565.3\AA~band).}\label{fig:sgf_with_big_line}
\end{figure}

\twocolumn

This provides a generalized framework for estimating the post-subtraction residuals behavior using a limited amount of free parameters. We vetted it using more realistic simulations, as illustrated in Figs. \ref{fig:sgf_with_out_line}, \ref{fig:sgf_with_small_line}, and \ref{fig:sgf_with_big_line}. They allowed us to confirm that the greater the stellar continuum-to-line ratio, $C_S/L_S$, the clearer the self-subtraction in the post-subtraction residuals. However, it is important to note that $\tilde{C}_P/\hat{L}_P$ reflects the self-subtraction depth, $\tilde{C}_P$, in relation to the line size estimate, $\hat{L}_P$, not only to the self-subtraction depth, $\tilde{C}_P$. As a consequence, in addition to being more pronounced when the line of the stellar spectrum, $L_S$, is small compared to its continuum, $C_S$, this effect is also more pronounced when the line of the planetary spectrum, $L_P$ (likely approximately equal to $\hat{L}_P$), is high (relative to the noise level). Furthermore, it is also important to note that this indicator only expresses the depth of the self-subtraction in relation to the line size estimate, $\hat{L}_P$, not to the true line size, $L_P$. Thus, although it may be counter-intuitive, the deeper the self-subtraction around the line, $\tilde{C}_P$, and the higher the line size estimate, $\hat{L}_P$. Indeed, the higher the stellar line, the more the stellar spaxel resembles the planetary spaxel to a certain factor and the more intrinsically difficult the separation problem becomes, hence the lower the planetary line estimate. As a matter of fact, the self-subtraction of the neighboring continuum and the post-subtraction flux loss of the line are distinct phenomena evolving in opposition to two intrinsically linked causes.

\section{Application on VLT/MUSE data}\label{lpm}

To limit the self-subtraction effect caused by the filtering step, and the numerical errors caused by the division step, we recently proposed in \cite{Julo_2024}\footnote{\samepage\sloppy\url{https://github.com/RemiJulo/Stellar_Subtractor}} an alternative halo subtraction technique based on a linear parametric model of the chromatic deformation of the halo. The estimates of the stellar spaxels are polynomial modulations of an estimate of the stellar spectrum minimizing the distance with the data in the least squares sense. The numerical errors caused by matrix inversion are mitigated using orthogonal Legendre polynomials, while the direct use of the analytical solution enables \Halpha line masking to avoid the self-subtraction caused by planetary flux overfitting in the estimate of the stellar component. Details are presented in Appendix \ref{meths} for better understanding and reproducibility. We name this method the Legendre polynomial modulation (LPM) method.

In this section, we illustrate multiple operating scenarios of this new method with real data, notably in order to highlight its performance with regard to detection and characterization. It is also the opportunity to illustrate the new understanding of the subtraction process that this method brings, notably through three ways for selecting the polynomial degree of modulation.

\subsection{Description of on-sky data}\label{data}

\subsubsection{Archival data of the PDS70 and HTLup systems}\label{data_old}

We re-analyzed the archival observations of PDS70 and HTLup, obtained, respectively, on June 20, 2018 (program 60.A-9100) and March 25, 2021 (program 106.21EN), with the narrow field mode of MUSE (MUSE-NFM) located at the VLT/UT4 \citep{Bacon_2010}. Details of the observation sequences can be found in \cite{Haffert_2019} and in \cite{Jorquera_2024}. Raw data were calibrated using the ESO MUSE pipeline, version 2.8 \citep{Weilbacher_2020}. It produced 6 cubes of PDS70 observations and 11 cubes of HTLup observations calibrated in wavelengths (the data used are summed up in the Appendix \ref{obs}). All data cubes have, approximately, a squared field of view of $7.5\times7.5"$ with square spaxels of $25\times25$ mas, covering the $4750$–$9350\text{\AA}$ wavelength interval with spectral channels of $1.25\text{\AA}$ in width. The host stars were approximately centered at target acquisition.

The custom routines described in \cite{Jorquera_2024} have been used to measure the star position in each cube slice since it is known to drift with wavelength due to the partial correction of atmospheric refraction by the instrument and its atmospheric dispersion compensator. The star was later shifted to a common position at the center of the field of view.

\subsubsection{New investigation of the YSES1 system}\label{data_new}

YSES1 (TYC 8998-760-1) is a young ($16.7\pm1.4$ Myr) solar-mass star with a K3 spectral type, around which $14\pm3$ $\mathrm{M_{Jup}}$ (YSES1 b) and $\mathrm{6\pm1 M_{Jup}}$ (YSES1 c) companions have been identified at projected separations of 160AU (1.71”) and 320AU (3.37”), respectively \citep{Bohn_2020_MNRAS, Bohn_2020_AJL}. 
Spectroscopy at K band revealed that YSES1 b displays a faint \BrGamma emission line (2.16 $\mu$m) indicative of active accretion \citep{Zhang_2021}. JWST observations of YSES1 revealed excess infrared emission produced by a disk around the b planet, strengthening accretion as the origin of the emission line, but could not identify one around the c planet \citep{Hoch_2025}.

We observed the YSES1 system with the VLT/MUSE-NFM on April 27, 2023 (program 109.22YA). We recorded 9$\times$300 s exposures, with the field of view being rotated by 45° between exposures to filter out static features related to the slicer at the data reduction step. We followed the same reduction procedure as above to produce 9 new data cubes calibrated in wavelengths. The star is centered and corrected from the barycentric velocity.

\subsection{Hyperparameters and parameters selection}\label{params}

\begin{table}[t!]
    \centering
    \renewcommand{\arraystretch}{1.46}
    \begin{small}
    \begin{tabular}{ccc}
        \toprule
        & SGF method & LPM method \\
        \midrule
        Hyperparameters & ($\mathrm{d}$;~$\bar{\b{W}}$) = (1;~101) & $\partial$ = 4 \\
        \midrule
        Spatial masks & $\mathrm{F}_{xy}$ < 0.01 $\mathrm{F_{max}}$ & $\mathrm{F}_{xy}$ < 0.01 $\mathrm{F_{max}}$ \\
        & $\mathrm{F}_{xy}$ > 0.1 $\mathrm{F_{max}}$ & $\mathrm{F}_{xy}$ > 0.1 $\mathrm{F_{max}}$ \\
        \midrule
        Spectral masks & $\lambda_i$ $\lesssim$ 6051 $\text{\AA}$ & $\lambda_i$ $\lesssim$ 6051 $\text{\AA}$ \\
        & $\lambda_i$ $\gtrsim$ 7075 $\text{\AA}$ & $\lambda_i$ $\gtrsim$ 7075 $\text{\AA}$ \\
        ($\Halpha$ mask) & -- & 6562 $\text{\AA}$ $\lesssim$ $\lambda_i$ $\lesssim$ 6564 $\text{\AA}$ \\
        \bottomrule
    \end{tabular}
    \end{small}
    \caption{Parameters sum up of both the halo subtraction methods.}
    \label{table:param}
\end{table}

A summary of the two sets of parameters of the two stellar halo subtraction methods is presented in Table \ref{table:param}. Firstly, the stellar spectrum estimate was made using a set of spaxels excluding both the least bright ones (as noisier) and the most bright ones (as most likely distorted by the MUSE pipeline). The spaxels were selected by comparing their total flux, $\mathrm{F}_{xy}$ (determined by spectral integration), with the maximum of the total fluxes, $\mathrm{F_{max}}$. Secondly, we chose to select only a given wavelength interval in order to reduce the risk of polynomials being disturbed by abnormal spectrum distortions (particularly because of abusive interpolations by the MUSE pipeline when pixels are considered to be outliers). Arbitrarily, the lower bound of the kept selected spectral window was set to the upper bound of the AO band, and its upper bound was chosen so that the $\Halpha$ line was centered. Moreover, the LPM method being parametric, the $\Halpha$ line could be masked too (note that the same could be done by rewriting the Savitzky-Golay in a parametric form). Thirdly, SGF parameters were set as indicated in \cite{Haffert_2019}, namely, with a first-order Savitzky-Golay filter, with a degree $\mathrm{d}=1$ and a window width $\bar{\b{W}}=101$ spectral channels ($\sim126.25\text{\AA}$). As for the LPM method, the hyperparameter, $\partial$ (i.e., the polynomial degree of the modulation), was set as per the results of Sect. \ref{params_analytics}, \ref{params_psf}, and \ref{params_coeffs}, i.e., following analytical results (with applications to simulations mimicking real data as much as possible) and on-sky data results (thanks to the interpretability results given by the orthogonality of the Legendre polynomials, with decomposition of real data). In any case, this leads to the consensus: $\partial=4$.

After subtracting the stellar halo of the different expositions, the noise was reduced by median of the residual cubes.

\subsubsection{Tuning from analytical expressions and simulations}\label{params_analytics}

\begin{figure}[t!]
\includegraphics[width=\columnwidth]{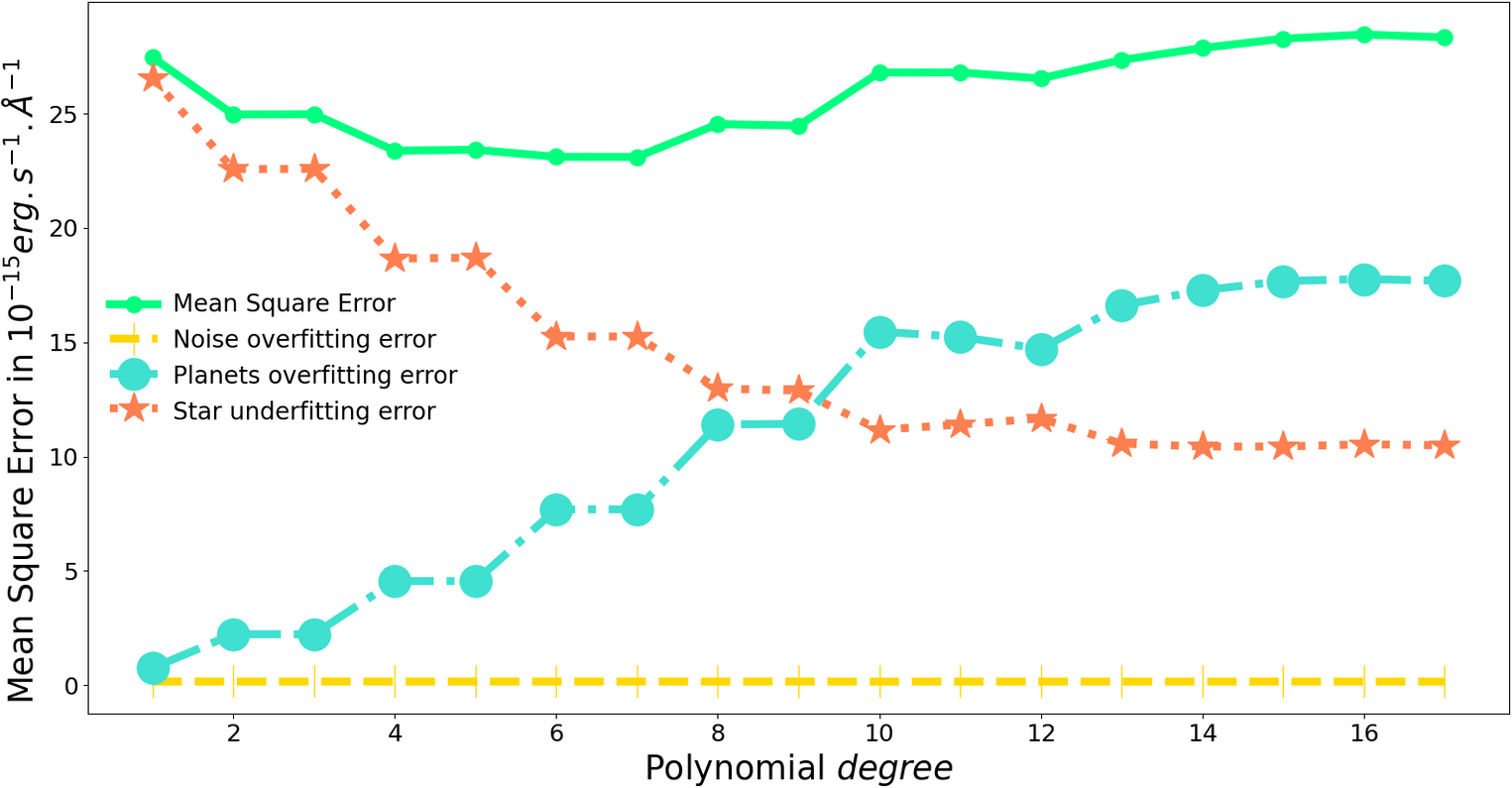}
\caption{MSE decomposition at planet position at \Halpha line wavelength, following Eq. \eqref{MSE_SPE}. The MSE (solid green line) is the sum of the noise overfitting error (dashed yellow line), planets overfitting error (dash-dotted turquoise line), and star underfitting error (dotted orangy line).\label{fig:MSE}}
\end{figure}

The polynomial degree, $\partial,$ is the key hyperparameter that must be correctly adjusted for good stellar contribution estimation (see Appendix \ref{meths_lpm}). An underestimation of the stellar contribution is caused by overly small values of $\partial$ as the model is too simple and underfits the deformation of the stellar spectrum estimation giving the stellar spaxel. Conversely, too large values of $\partial$ lead to the overestimation of the planetary contribution (and thus lead to oversubtraction). Setting the hyperparameter, $\partial$, consequently amounts to solving a critical bias-variance trade-off. The strategy adopted here to set the hyperparameter, $\partial$, consists in minimizing the mean square error (MSE). As detailed in Appendix \ref{resids_mse}, this latter is analytically decomposable into three components.

In a simulated framework, each component is actually known and the MSE can be evaluated from the simulated data and the analytical expression of the post-subtraction residuals. It can even be shown (see Eq. \ref{MSE_SPE}) that data modeled as the sum of three components (star, planets, and noise) lead to an MSE that is decomposable into three (associated) components. These errors are, respectively, named “star underfitting error,” “planets overfitting error,” and “noise overfitting error.” With the simulation described in Appendix \ref{simus_process} and the masks chosen as in the Table \ref{table:param} (to be consistent with real data subtractions), MSE as a function of the degree, $\partial$, can be plotted as shown in Fig. \ref{fig:MSE}. Firstly, from a theoretical point of view, such a decomposition confirms the expected behavior of the MSE with regard to $\partial$, particularly with the star underfitting error decreasing with $\partial$ and the planets overfitting error increasing with $\partial$. Secondly, from a more practical point of view, the curve also indicates that the minimization of the MSE is achieved for polynomial degrees, $\partial$, between $4$ and $7$ (in slightly equivalent ways), under these conditions. Although the simulation is a simplified version of real data, it is expected that it is sufficient to transfer these results to real data processing. Then, the simplest of these valid models can be selected: $\partial=4$.

We note, however, that the optimal value of $\partial$ depends on both the data and the masks used. Firstly, the stellar spaxel to stellar spectrum deformation depends on many parameters, from the observing conditions to the target used for tip-tilt sensing and even the use of the pipeline. Secondly, the set of spaxels used for the estimation of the stellar spectrum as well as the number of spectral channels used for the stellar spaxel estimation change the complexity of the model to be used, and, equivalently, the degree, $\partial$, of the polynomial to be used. Consequently, in order to optimize subtraction performances, all of the analysis described in this section should be done for each dataset and questioned in light of its final results.

\subsubsection{Tuning from on-sky data and PSF decomposition}\label{params_psf}

\begin{figure*}[t!]
\includegraphics[width=\linewidth]{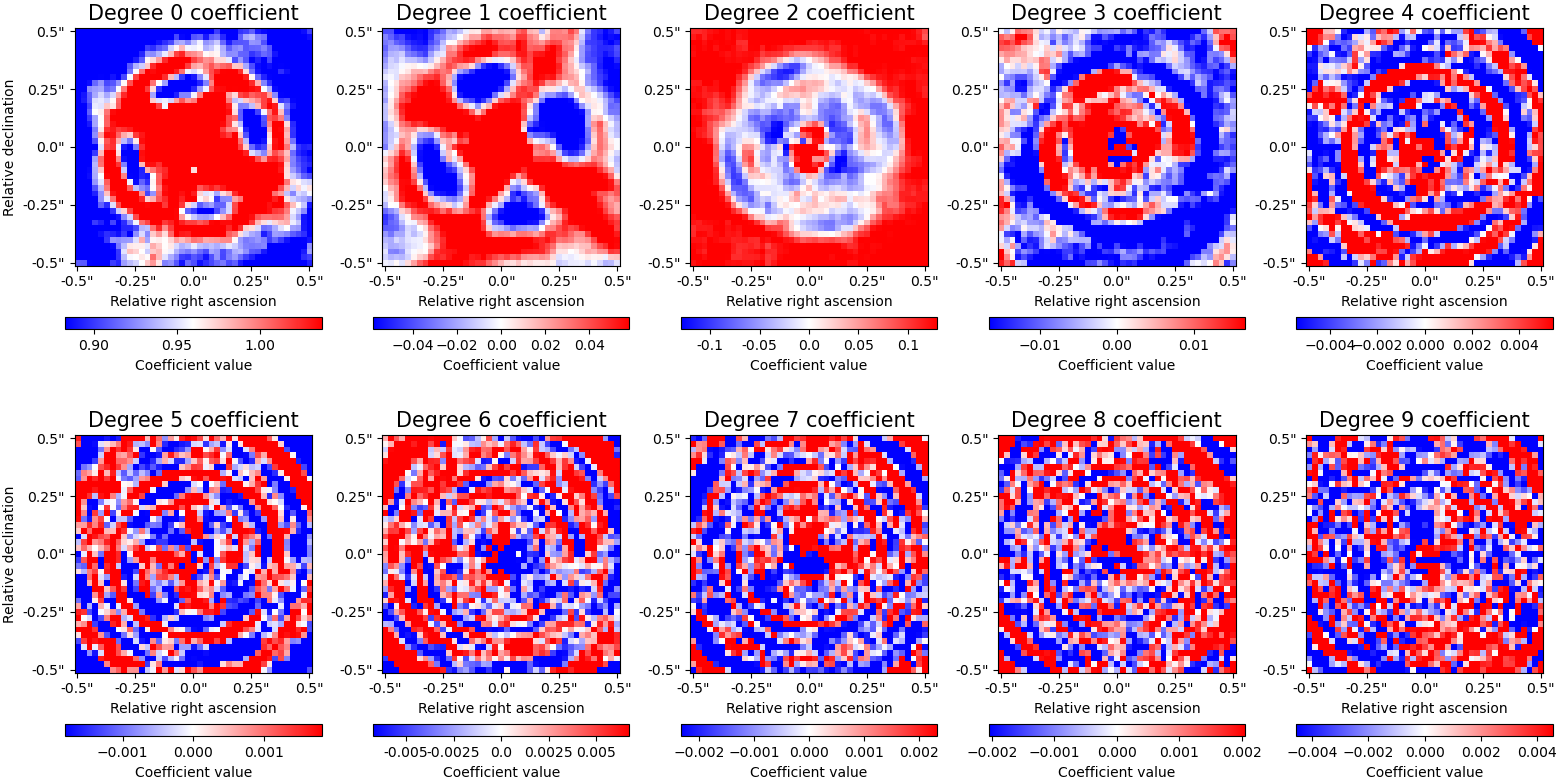}
\caption{Maps of the polynomial coefficients of estimation of the stellar spaxels of real data of observation of the PDS70 system. The color scale is saturated at the 25\% and 75\% quartiles and centered on 0 by extension in the negative or in the positive (except for the degree of 0 coefficient as corresponding to the total flux). For each spaxel, before estimating the coefficients, each modulation by the base polynomials is L2 normalized (after subtraction of the average, except for the degree of 0, responsible for the total flux). Various spatial components of the PSF are identifiable thanks to the orthogonality of the polynomial basis used. Notably, the degree of 0 and 1 coefficient maps reveal the diffraction spikes in addition to the AO radius, the degree of 2 and 3 coefficient maps reveal other related patterns as well as the heart of the Airy disk, and the degree of 4 and more coefficient maps reveal the secondary rings of the Airy disk (with their outward displacement as the degree of the polynomial increases).\label{fig:PDS70_PSF}}
\end{figure*}

In addition to mitigating numerical errors, the orthogonality of the Legendre polynomials improves the interpretability of the polynomial coefficients estimated to model the stellar nuisance. By doing so, each Legendre polynomial coefficient expresses only one frequency component (in the spectral direction) of the spatial point spread function (PSF). To have a complete view of this decomposition, the median cube of 4 observations of the PDS70 system (see Appendix \ref{obs} for details) was processed with the LPM method with Legendre polynomials of degree $\partial=9$ (to extend the decomposition), and over the full range of wavelengths (except those around the $\Halpha$ line). Fig. \ref{fig:PDS70_PSF} shows the maps of the obtained coefficients, with some color-scale saturations for illustration purposes. The adaptive optics radius, the diffraction spikes, the Airy disk (core and secondary rings), as well as other patterns are decomposed in this way. Knowing that adding a higher degree polynomial to the Legendre basis does not change the estimated values taken by the coefficients of the lower-degree polynomials (thanks to their orthogonality), this gives, this time, a physical argument for the choice of the polynomial degree, $\partial$. This confirms the relevance of the choice: $\partial=4$. Beyond that, the modulation is just adjusting ever fainter secondary rings of the Airy disk (with low coefficient values). Moreover, we note that the circularity of most patterns justifies the treatments on successive nested rings presented in the following. Ultimately, this decomposition would help to regularize (both spatially and spectrally) the stellar estimation problem, e.g., by constraining the values of these coefficients.

\subsubsection{Tuning from estimated polynomial coefficients}\label{params_coeffs}

\begin{figure*}[t!]
\includegraphics[width=\linewidth]{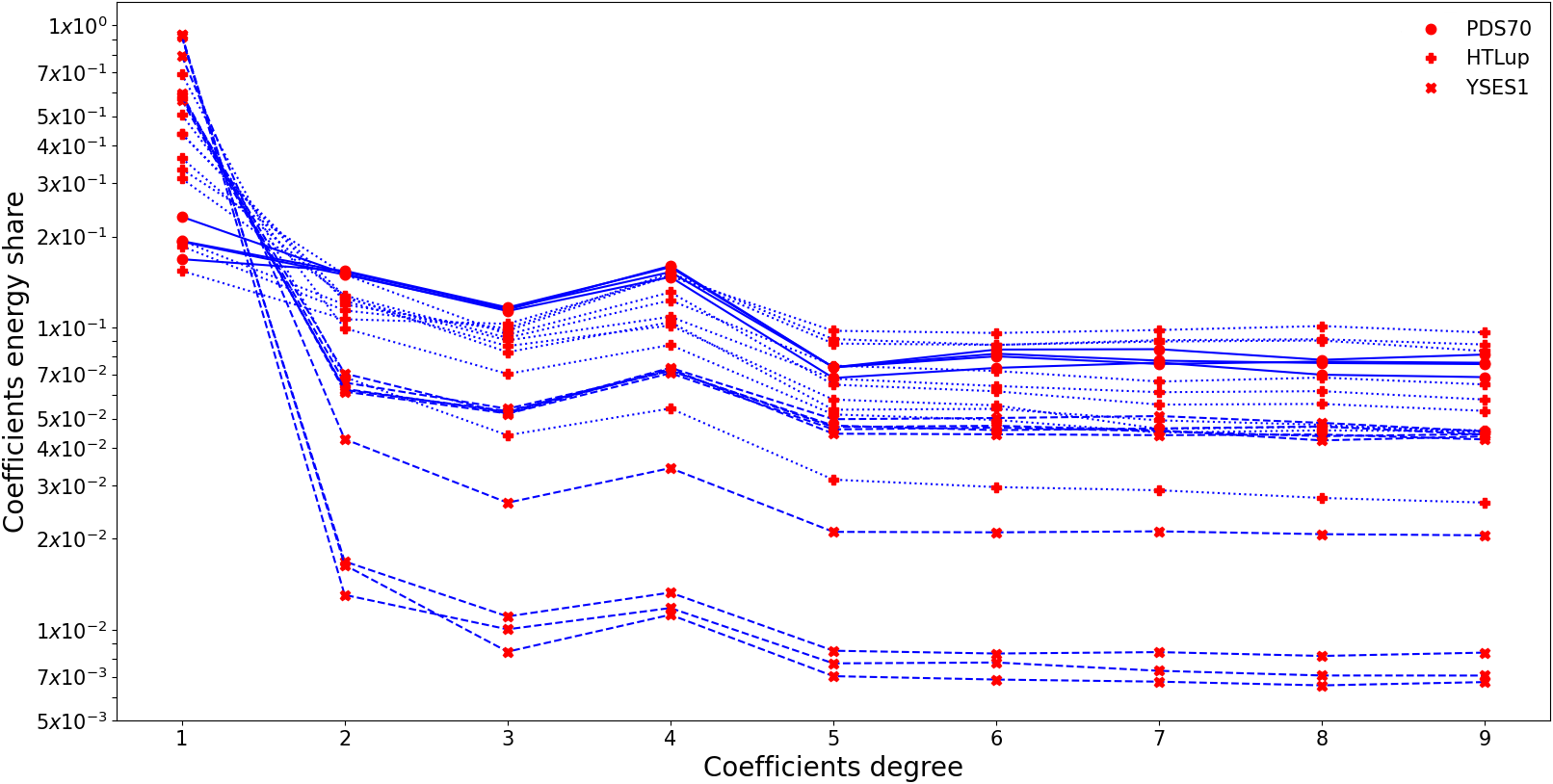}
\caption{Energy share curves of the estimated polynomial coefficients for each of the degrees (except that of degree of 0 as accounting for the total flux and not for the variations) of the basis of decomposition. Before estimating the coefficients (except that of degree of 0), the modulations by the base polynomials are, firstly, subtracted from their average, secondly, L2 normalized. For each of the degrees, the energy share is calculated as the median over the field of view of the squared coefficients, divided by the sum of all the median squared coefficients (except that of degree of 0). This was done for PDS70, HTLup, and YSES1 data cubes, each time for individual exposure. This shows that the degree of 4 coefficients are the last useful ones. Above, the energy of the high-degree coefficients converges towards a minimum where it is mainly fitting noise. It can be seen here that this minimum is reached for the degree of 5 coefficient for each of the 23 observations, indicating that this choice may well generalize.\vspace*{-0.69cm}\label{fig:LPM_coefficients}}
\end{figure*}

To go a step further and quantify the results of Sect. \ref{params_psf}, we calculated, degree by degree, the energy of the coefficients (i.e., the sum over the field of their squared values) to understand their relative importance. Once again, the polynomials of the chosen basis being orthogonal, these energies are related to the given data cube and not to the degree of modulation used. Fig. \ref{fig:LPM_coefficients} shows these calculations for several systems. The optimal polynomial degree being above all linked to the physical behavior of the PSF, all trends are similar from one observation to another. Again, among these different choices, the best one is: $\partial=4$. Beyond that, the coefficients take on the same orders of magnitude as those for very large degrees, namely, those that only fit the noise.

\onecolumn

\twocolumn

\subsection{Impact on detection: the planetary system PDS70}\label{pds70bc}

\subsubsection{Matched filtering}\label{pds70bc_match}

The protoplanets PDS70 b and c were detected with the two methods at \Halpha wavelengths. Spatio-spectral matched filtering was proposed to reduce the (symmetric) noise, in order to push the detection limits even further, using the spectral and spatial a priori that were unused by the spaxel-by-spaxel subtraction.

To model the \Halpha line, a Gaussian with a 3 spectral channels ($\sim3.75\text{\AA}$) full width at half maximum (FWHM) is chosen for the spectral model. The wavelength of the latter being known, the first step is therefore the dot product of each of the spaxels of the median residual cube with this spectral model centered on the wavelength of the \Halpha line ($\sim6562.8\text{\AA}$). To model the associated spatial PSF, the spatial model is derived from the dot product of the spaxels of the pre-subtraction cube with the spectral model; indeed, under the small-field hypothesis, the star spreads are the same as those of the planets (with planets and noise being minor compared with the star). Both models are normalized to 1.

\begin{figure}[t!]
\includegraphics[width=\columnwidth]{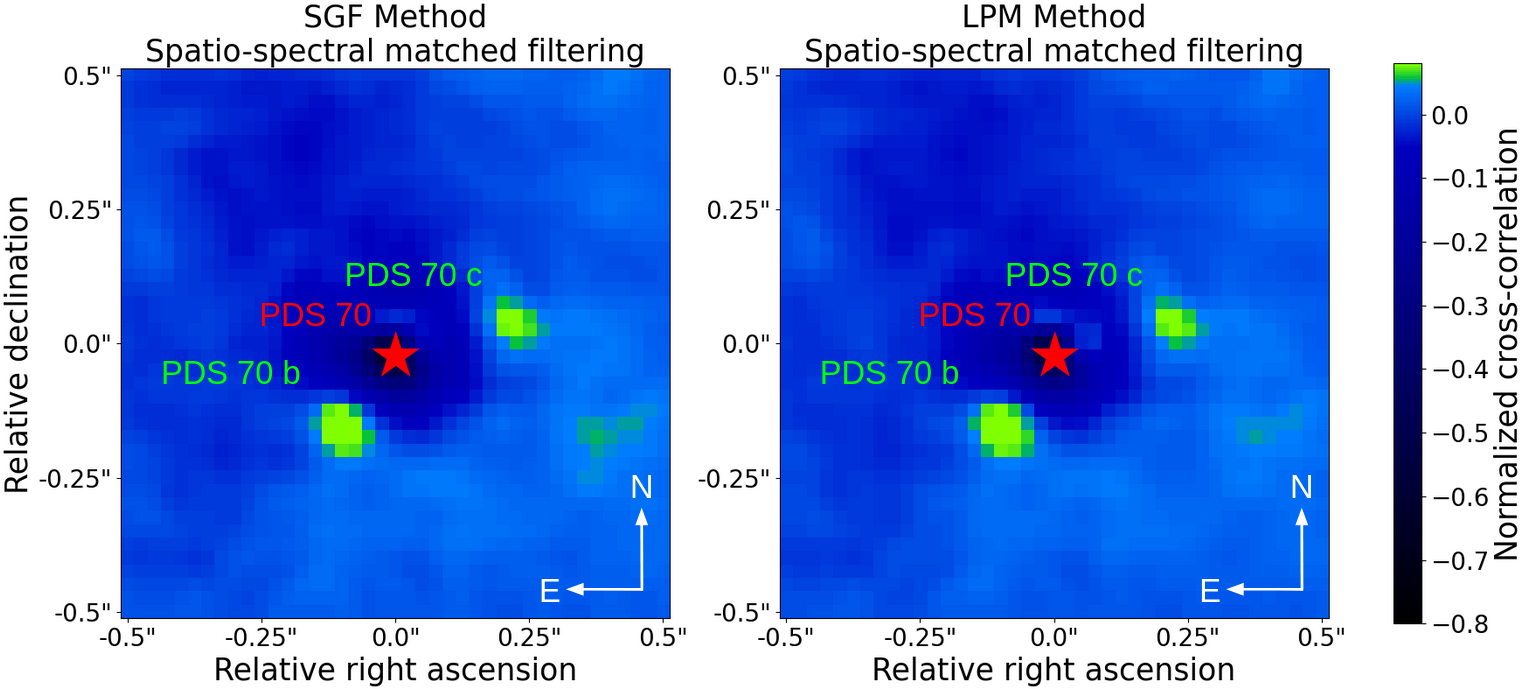}
\caption{Spatio-spectral match filtering of the median PDS70 data cube, after stellar halo subtraction by the SGF (left) and LPM (right) methods. A Gaussian is chosen for the spectral model to search for \Halpha line signal. The corresponding pre-subtraction monochromatic images are chosen for the spatial model to search for spatial PSF patterns.\label{fig:PDS70bc_detection}}
\end{figure}

Fig. \ref{fig:PDS70bc_detection} illustrates the similar detection capabilities of the two considered methods. These are confirmed in the general case in Sect. \ref{pds70bc_roc} thanks to ROC curves built from these same maps.

\subsubsection{Contrast curves}\label{pds70bc_scc}

Following \cite{Xie_2020}, we constructed the contrast curves from fake planet injections at different contrasts and separations. In practice, fake planets were injected at contrasts varying from $1 \times10^{-5}$ to $1 \times10^{-2}$ in steps from $1 \times10^{-5}$ to $5\times10^{-4}$. A first series of fake planet injections was done at separations varying from 0.025” to 0.5” in steps of 0.025” and at four position angles 0°, 90°, 180°, and 270°. A second series of injections was done from $\sim$0.035” to $\sim$0.7” in steps of $\sim$0.035” at four other position angles 45°, 135°, 225°, and 315°. To standardize the sampling step in separation, the field of view was paved in concentric rings of pixels around the center of the images. The \cite{Andres_1994} circle tracing algorithm was used to do so. Thus, each injection could be associated with a ring, each of them being associated with a separation (by averaging the distances to the center of the pixels making up the rings). This led to as many separations as there are rings in the field of view. To test and compare both halo subtraction methods, several cubes of observation of the PDS70 system were considered (see Appendix \ref{obs} for more information on the cube selection) as true noise and nuisance realizations. At chosen positions and flux values, cubes containing only the fake planets were simulated according to the description given in Sect. \ref{simus_process}, and added to the real data cubes. Care was taken to inject the fake planets at positions where only noise is present in the real data cubes (i.e., far from the two protoplanet positions). Then, both stellar halo subtraction methods were applied to each of the four injected PDS70 cubes, and the residuals were merged into one median cube (for each of the injections).

Following \cite{Xie_2020}, the three spectral channels around the \Halpha line (i.e., channels 1450, 1451, and 1452) were averaged to get narrow-band images around these wavelengths; we recovered around half of the planetary fluxes as the FWHM of the spectral PSF is approximately three spectral channels. From these images, the planetary fluxes were estimated from the integration of a $3\times3$ pixels square centered on the positions of injection. The mean, $\mu$, and standard deviation, $\sigma$, of the noise were estimated by integrating over the same circles, at identical separations but different angular positions than the fake planets. In practice, this comes to using pixels of the same \cite{Andres_1994} rings, but at different angular positions. The noise characteristics (i.e., $\mu$ and $\sigma$) being noticed to be varying with the wavelengths (the noise is notably more important at the \Halpha line wavelengths), they were estimated from the images at the \Halpha line wavelengths only (and not from other residuals images at other wavelengths).

To sum up, for each separation and for each flux, we have, for each of the eight angular positions, an integrated planetary flux value, and a certain number of integrated noise fluxes from which an average value, $\hat{\mu}$, and a standard deviation value, $\hat{\sigma}$, can be computed. For each of these situations, the detection of an injected planet is validated if its integrated flux is greater than $\hat{\mu}$+5$\hat{\sigma}$, and invalidated otherwise. The contrast curves were built for each separation, by decreasing the contrast until less than half of the planets of the corresponding situation were detected (under the assumption of a symmetrical noise distribution). The contrast curve represents the minimum contrast for which more than 50\% of the planets were detected. To estimate an error, the same curves were plotted for 25\% and 75\% detection thresholds.

\begin{figure}[t!]
\includegraphics[width=\columnwidth]{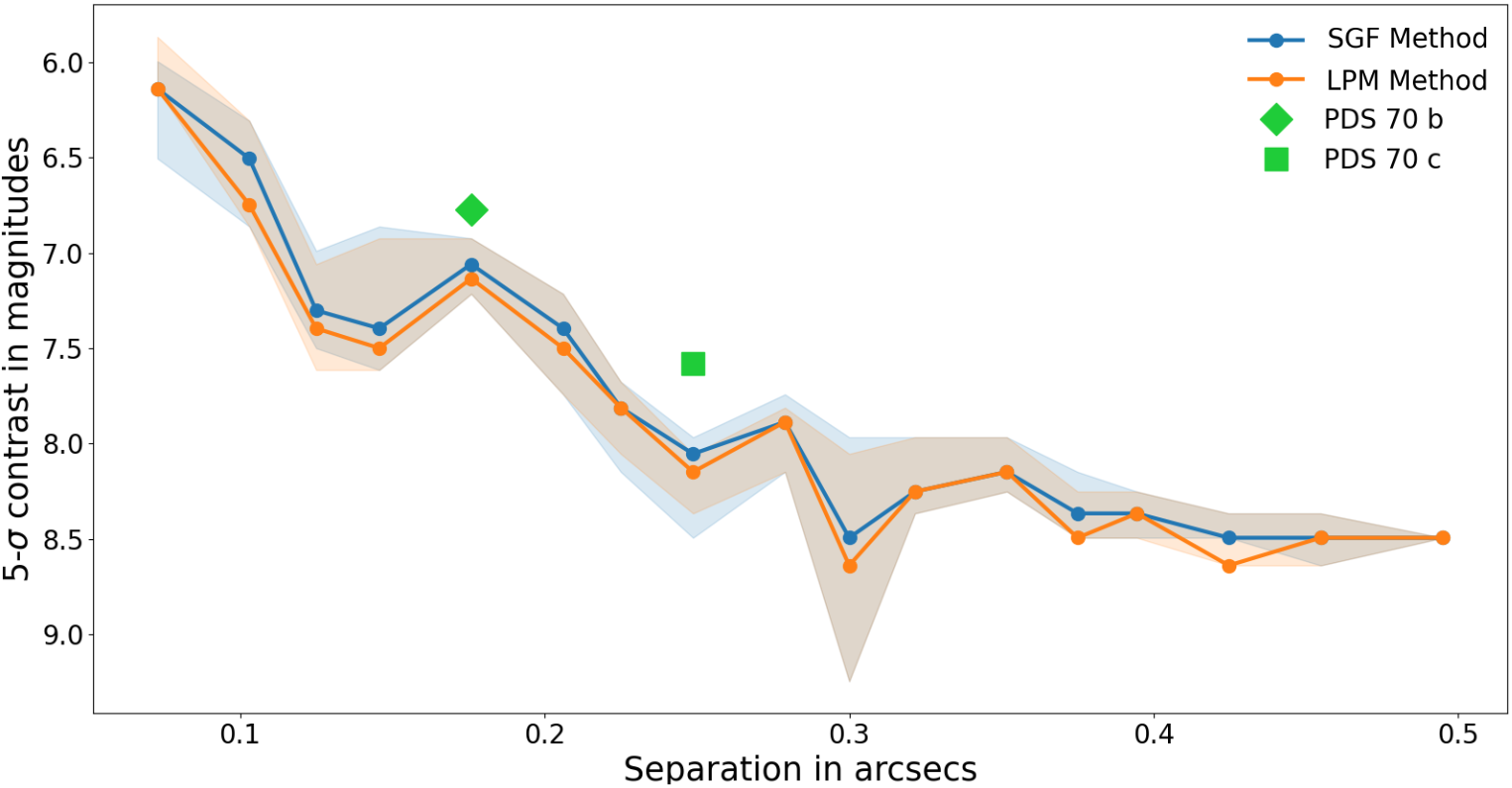}
\caption{5-$\sigma$ contrast curve of both the SGF and LPM methods with 50\% detection in solid line as well as 25\% and 75\% delimiting colored areas. Planets PDS70 b and c are also placed at their contrast and separation. Both are detected by both methods.\label{fig:SCC}}
\end{figure}

The results are presented in Fig. \ref{fig:SCC}. Given that the SGF method already retained almost the entire \Halpha line, and that the LPM method is based on the same principles as the former, it was expected that the detection limits would not be significantly improved. Yet, despite the lack of noise diversity preventing any sound conclusions, it seems that the detection performance is slightly improved. As a matter of fact, the new method improves the integrity of the planetary spectra estimates (see Sect. \ref{htlupb}), without degrading the detection performance.

\begin{figure*}[t!]
\includegraphics[width=\linewidth]{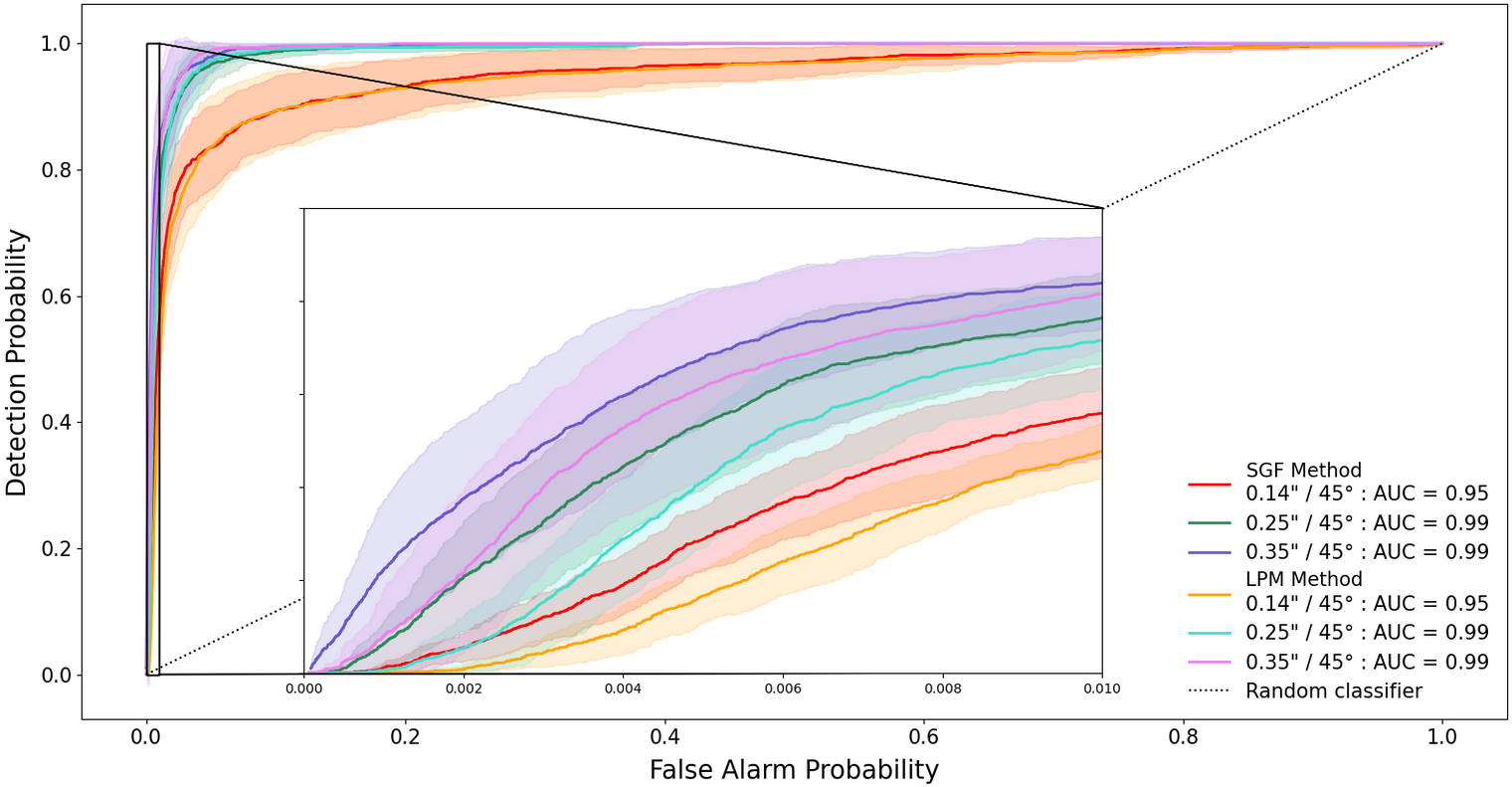}
\caption{Bundle of ROC curves, for both halo subtraction methods, and for various fake planet injections separations; a zoom for FAP between $0$ and $10^{-2}$ is included in the insert. The random classifier, which validates or invalidates detections with the same probability of 0.5 is shown as a dotted diagonal line for comparison. AUC stands for area under the curve; by definition, the AUC of the random classifier is equal to 0.5. Contrast curves are located on ROC curves at DP = 0.5 (as well as 0.25 and 0.75 to construct error margins for example) and FAP = $2.87\times 10^{-7}$ (corresponding to a 5-$\sigma$ confidence interval, as in Fig. \ref{fig:SCC}). Data is lacking to produce an accurate ROC curve, especially around FAP = $2.87\times 10^{-7}$, but this still shows that, within the 1-$\sigma$ margin of error, the two stellar halo subtraction methods have relatively similar overall robustness as well.\vspace*{1.0cm}\label{fig:ROC}}
\end{figure*}

\begin{figure*}[t!]
\includegraphics[width=\linewidth]{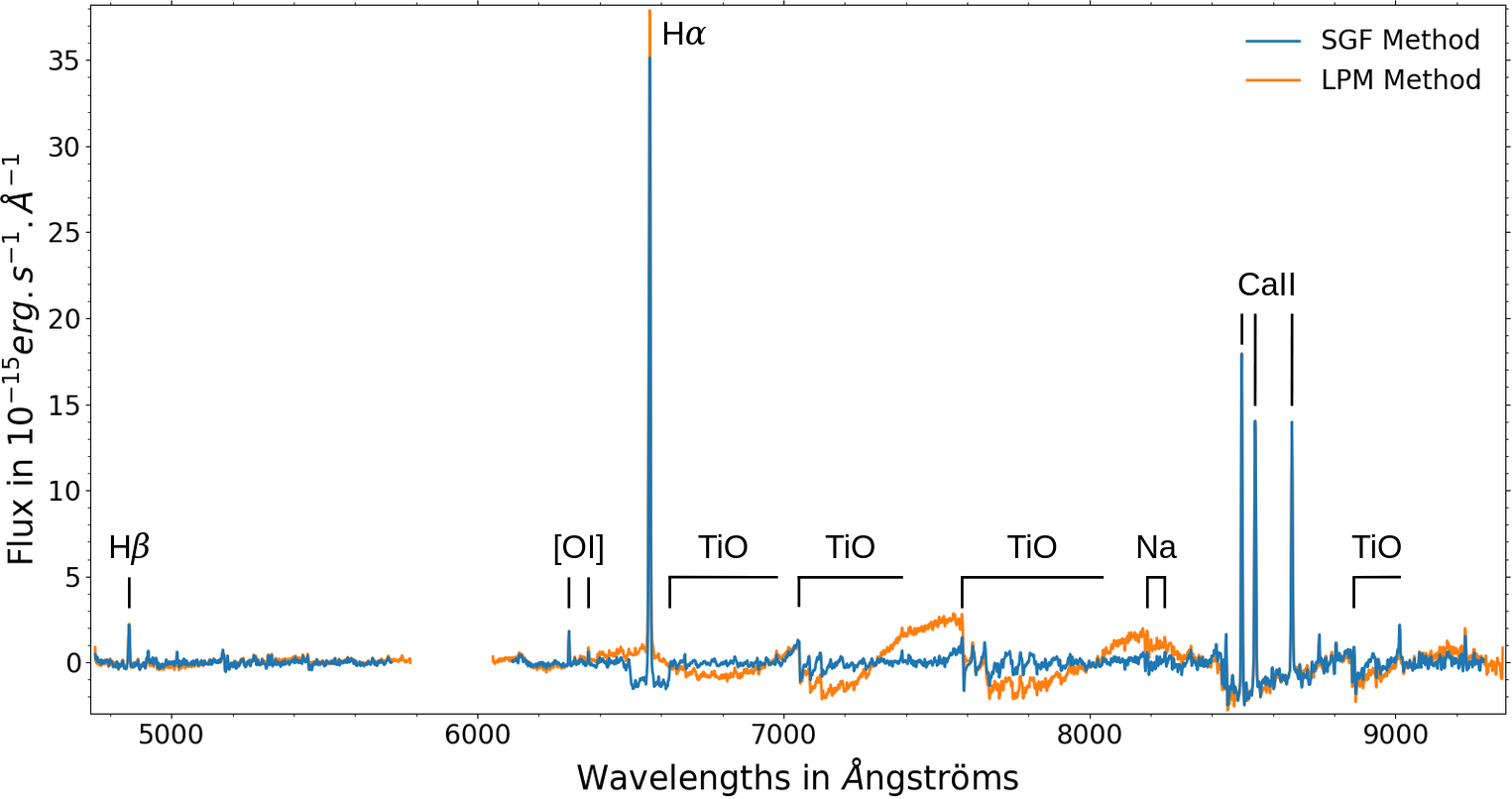}
\caption{HTLup B spectrum estimates by the SGF (in blue) and LPM (in orange) methods. The $\Halpha$, $\Hbeta$, and [OI] lines as well as the CaII infrared triplet are detected by the SGF method. Absorptions of TiO and Na are revealed by the LPM method in addition, thanks to its greater preservation of the continuum. On the contrary, it reveals the absence of any VO absorptions, suggesting that the star should be a late-M type. Further improvements of the LPM method are planned to improve the continuum recovery and its quantification (see Sect. \ref{conclu_discs}) so that it would be indeed usable for physical interpretations. A more complex study is beyond the scope of this paper, which is primarily focused on emission lines.\vspace*{0.7cm}\label{fig:HTLupB_characterization}}
\end{figure*}

\subsubsection{ROC curves}\label{pds70bc_roc}

Alternatively, a common approach to assess the robustness of detection of a binary classifier is to build its receiver operating characteristic (ROC) curve. Thresholding of a detection image (whether it comes from the averaging of monochromatic images of interest or from matched filtering with spatio-spectral models) being a binary classification, it can be evaluated in this way.

The ROC curve of a binary classifier is meant to represent its detection probability (DP) and its false alarm probability (FAP), when the detection threshold varies. Thus, an ideal detector is characterized by both a probability of false alarm equal to 0 and a probability of detection equal to 1. On the contrary, a classifier that leads to DP = FAP is not better than the random classifier accepting or rejecting the detection with the same probability of 0.5. A typical way to assess the performance of such a classifier is to compute the area under the curve (AUC) expressing DP as a function of FAP \citep{Kay_1997}. Following what is said just above, the AUC values range from 0.5 (for the random classifier) to 1 (for the perfect classifier). Below, any classifier can be replaced by its inverse classifier for strictly better detection performance.

Another advantage of the ROC curve is that its construction can be experimental, without requiring any assumption on the noise distribution. The FAP can be estimated from real data that may contain buggy pixels, ghosts, or speckles \citep{Refregier_2007}. Yet, the drawback is that it illustrates only one scenario, for example at a given separation and contrast in our situation. Several bundles of curves for different scenarios (e.g., different separations) can be plotted to overcome this problem.

Six of these curves are shown in Fig. \ref{fig:ROC} for two methods and three separations. We used again fake planet injections for that. A general overview of the ROC curves construction method is summarized in Fig. \ref{fig:ROC_process}. The same planetary models as for the contrast curves were used (see Sect. \ref{pds70bc_scc}). An injection contrast of $1.7\times10^{-3}$ was chosen to illustrate a PDS70 b-like scenario, namely, at the detection limit. We chose injection positions at three separations: $\sim0.14$”, $\sim0.25$”, and $\sim0.35$”. This ensured that the injected planets were all within the correction radius of the adaptive optics without being too close to the star core.

Classification was performed by thresholding the matched filter map of the post-subtraction residuals. The same models as those described in Sect. \ref{pds70bc_match} were used for matched filtering; thus, the thresholding was performed on values of the output of the normalized matched filter. The FAP and DP were evaluated by applying thresholding chosen to adequately sample the output dynamic of the matched filter \citep{Helstrom_1994}. This latter thus varied between -1 and 0.5 in steps of 0.0001 for the construction of these curves. The number of noise realizations was $4\times310$ by injecting the planets into the four cubes at our disposal, and in the 310 spectral channels 1090 to 1399 and 1503 to 1811 by steps of two spectral channels, namely at wavelengths $6112.8\text{\AA}$ to $6497.8\text{\AA}$ and $6627.8\text{\AA}$ to $7012.8\text{\AA}$ by steps of $2.5\text{\AA}$ (knowing that the \Halpha line is centered at spectral channel 1451 i.e., at a wavelength of $6562.8\text{\AA}$); the two true planets PDS70 b and c did not influence the detection of the fake planets for either of the two methods in this way. The contrasts were set according to the stellar flux values at the corresponding wavelengths.

For each of the threshold values, the detection probabilities were estimated by the ratio of the number of detected planets (i.e., with fluxes at their positions above the threshold value) to the number of detectable planets (i.e., $4\times310$). Similarly, for each of the threshold values, the false alarm probabilities were estimated by the ratio of the number of false alarms (i.e., where the fluxes are above the threshold, and outside planetary zones) to the total number of potential false alarms (i.e., where the field has been restricted, and outside planetary zones). The planetary zones are positions around the fake planets that are not counted as false alarms because of their fluxes mainly coming from the planets (whose fluxes are spread around by the PSF). For these planetary zones, we chose the $3\times3$ pixels square around the planet positions, which corresponds approximately to the FWHM of the spatial PSF at these injection fluxes.

Finally, while the discussion in Sect. \ref{pds70bc_scc} demonstrates the similarity of the detection capabilities of both methods based on a first detection metric, this section is aimed at showing that the LPM method provides the same detection capabilities as the SGF method (to the given margin of error), this time with respect to robustness (i.e., precision-return trade-off). A slight lack of noise diversity would be deplored for correctly estimating the curve at the regime of interest, i.e., low false alarm probabilities (e.g., $\textrm{FAP}~<~2.87\times 10^{-7}$ and $\textrm{DP} = 0.5$ for the construction of a contrast curve). However, it can be assumed that the behavior of the curves with respect to each other remains the same whatever the false alarm regime chosen, especially given the importance of the margin of error (plotted at 1-$\sigma$, here, with $\sigma$ the standard deviation along the four post-subtraction residuals).

\begin{figure*}[t!]
\begin{center}
\begin{tabular}{cc}
\includegraphics[width=\columnwidth]{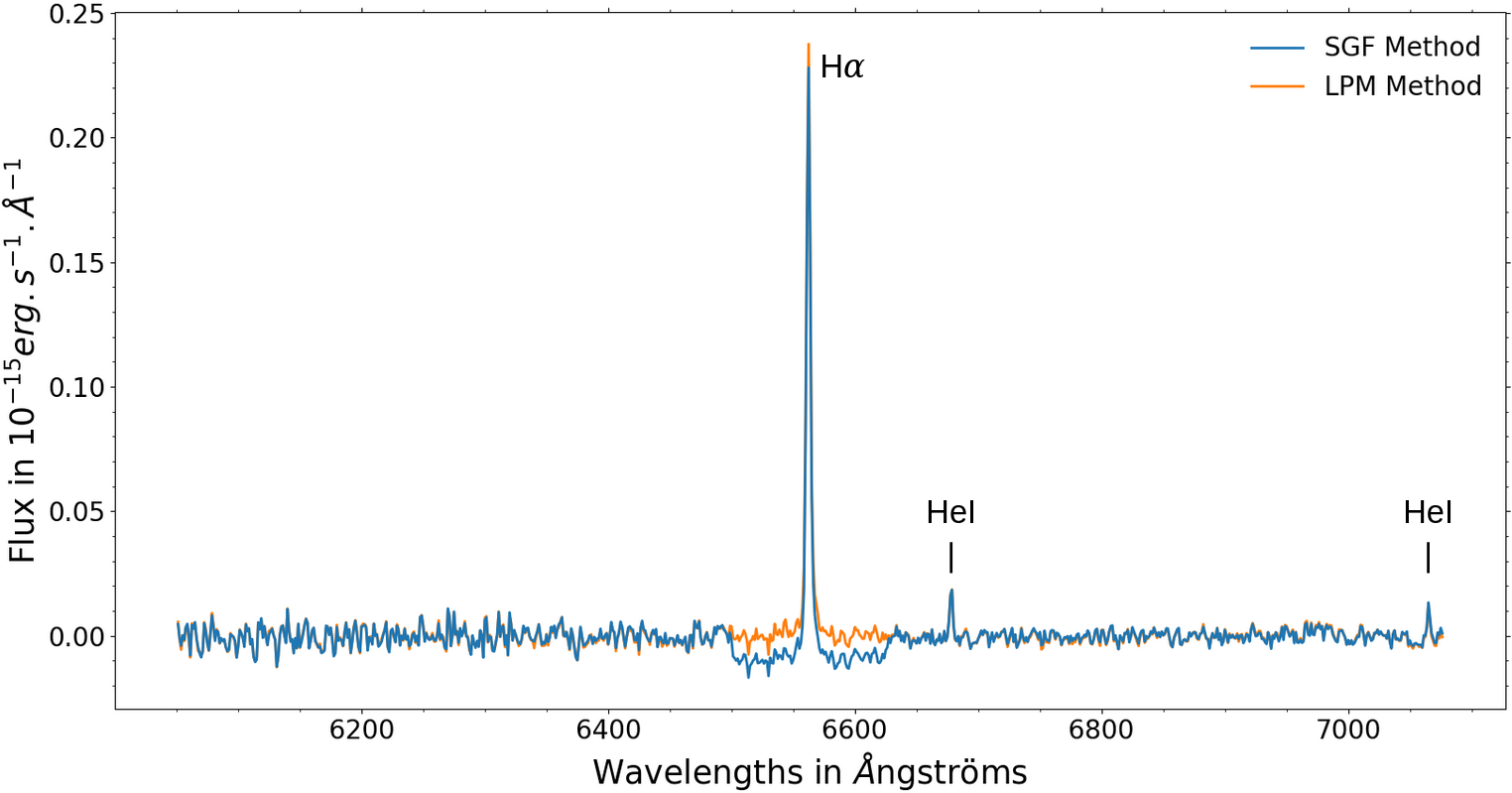} & \includegraphics[width=\columnwidth]{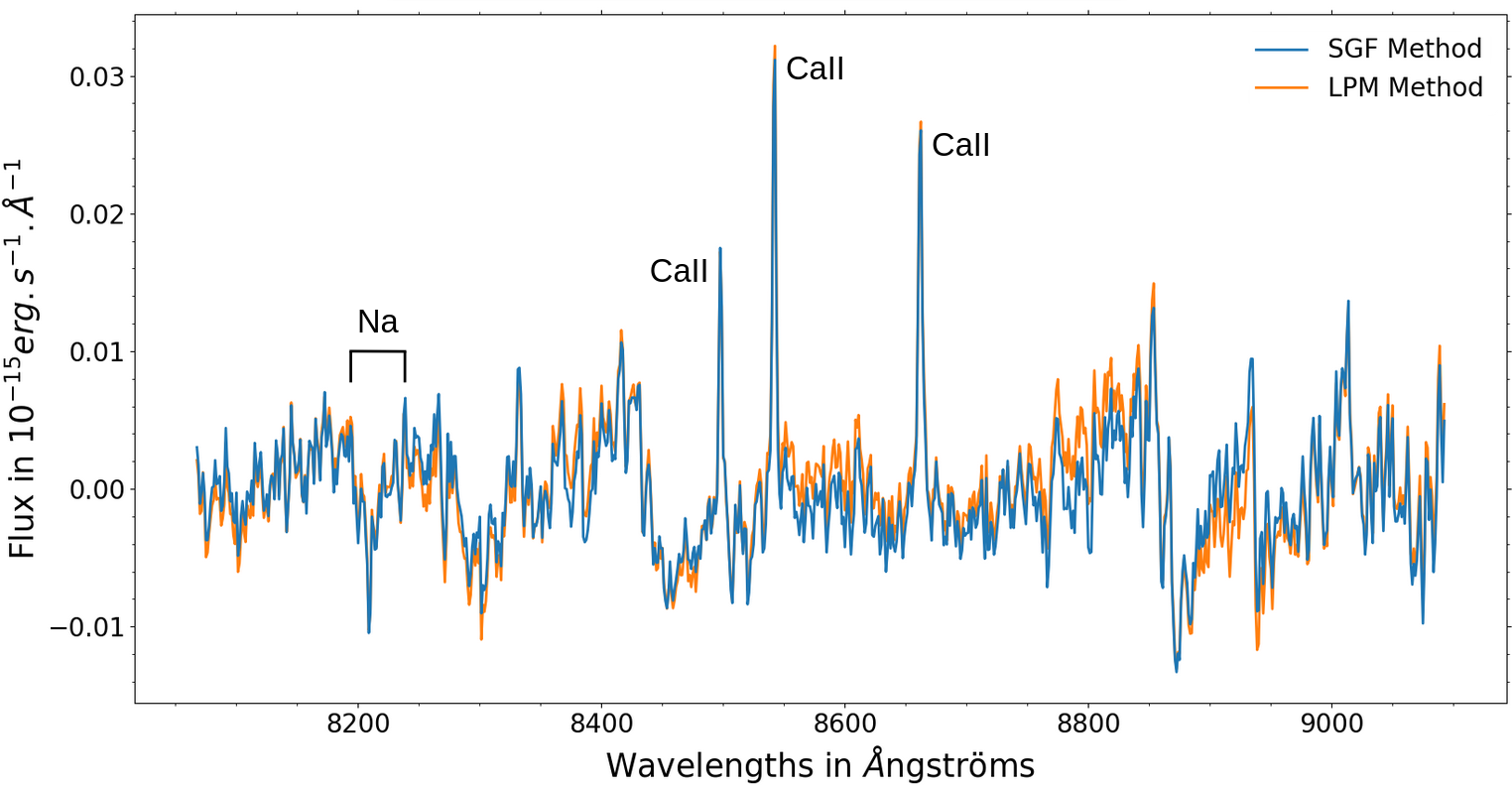} \\
\end{tabular}
\end{center}
\caption{YSES1 b spectrum estimates, by the SGF (in blue) and LPM (in orange) methods. Left:\ Zoom in on 821 channels around the \Halpha line. The self-subtraction caused by the SGF method as well as its correction by the proposed LPM method are visible. Two HeI lines are also revealed. Right: Zoom in on 821 channels around the CaII infrared triplet. The self-subtraction around the lines is less visible here, because both the lines are weaker and the continuum is (likely) stronger at these wavelengths (yet, in turn, the heights of the lines are theoretically further underestimated); the mask used by the LPM method during this estimation process was adapted to cover the full triplet. A noticeable Na absorption is also revealed.\label{fig:YSES1b}}
\end{figure*}

\subsection{Impact on characterization: the binary system HTLup}\label{htlupb}

We used HTLup observations to explore the potential of both methods in the case of low contrasts, tight pairs ($\sim$170 mas), and relatively close spectral features between the binary components (the primary is a K3V star while the companion is an M dwarf; see more details below). In this section, we confirm the impact of the signal distortion caused by the SGF method in such an extreme scenario, and how the LPM method alleviates this issue.

After subtracting the stellar halo of each of the selected cubes (see Appendix \ref{obs}) and with each of the two methods, an aperture of $3\times3$ pixels (with the spatial PSF having a $\sim3$ pixels FWHM) was used to integrate the companion flux at every wavelengths. To estimate the losses outside of this aperture, a compensation ratio was calculated from the stellar flux before subtraction, as the ratio of the integration over the entire field to that over only the $3\times3$ pixel aperture. At each wavelength, the total flux was thus finally estimated by integration over the $3\times3$ aperture, each time multiplied by a different compensation ratio, function of the wavelength to account for the spectral dependence of the spatial PSF (leading to the spectral dependence of the losses too).

Finally, the median of the spectra (shown in Fig. \ref{fig:HTLupB_characterization}) obtained with both methods reveals, as expected, two different extracted spectra. Indeed, the high contrast of HTLup B with respect to the noise level leads to critical overfitting with the SGF method, and subsequently to noticeable self-subtraction \citep[also reported in][]{Jorquera_2024}. All emission lines revealed with the SGF method by \cite{Jorquera_2024} were also recovered by the LPM method, namely the \Halpha and \Hbeta lines, as well as the CaII infrared triplet. However, unlike the SGF method, the LPM method also partially reconstructed fragments of the continuum fluctuations. Although it is not fully quantifiable in a trustworthy way, it allows us to recognize patterns identifiable as TiO and Na absorptions, while unveiling the absence of others such as VO absorptions; this suggests that HTLup B might be a late-M type star. What is more quantifiable, nonetheless, is the effect of the LPM method on the \Halpha line compared to the SGF method. Markedly, measurements gave a difference in the integrated flux of $\sim$30\%, and differences in velocity at the 10\% and 50\% of the maximum flux of the line of $\sim$8\%; importantly, these measures can be used for deriving accretion rates (the estimation given by the routines built for \cite{Jorquera_2024} was improved from $\mathrm{2.4\times10^{-9}~M_{Sun}/year}$ to $\mathrm{3.2\times10^{-9}~M_{Sun}/year}$ in this situation) or discriminating between chromospheric activity and accretion as line emission mechanisms.

To qualify these results, we note that when self-subtraction is noticeable, corrective treatments of the estimates are possible; especially to artificially correct negative fluxes around the lines. For example, in \cite{Jorquera_2024}, the negative values of the estimated spectrum of HTLup B around the $\Halpha$ line being obviously non-physical, an estimation of the loss was added to the recovered spectrum prior to any analysis with line models (which cannot deal with negative fluxes). This was done through fake planets injections at the same contrast and separation. Then, the estimated loss was the difference between the line height of the injected planet and the one of the recovered fake planet after stellar halo subtraction. This procedure yielded lines relatively similar to those estimated by the LPM method, leading to less significant physical consequences in practice.

We note that when the lines are too faint compared to the noise level, as for protoplanets PDS70 b and c, only very few changes can be seen between the SGF and the LPM methods. Additionally, in their case, the continuum seems to be too faint to be detected by the LPM method.

\subsection{Looking for line emission in the YSES1 planetary system}\label{yses1b}

\begin{figure*}[t!]
\includegraphics[width=\linewidth]{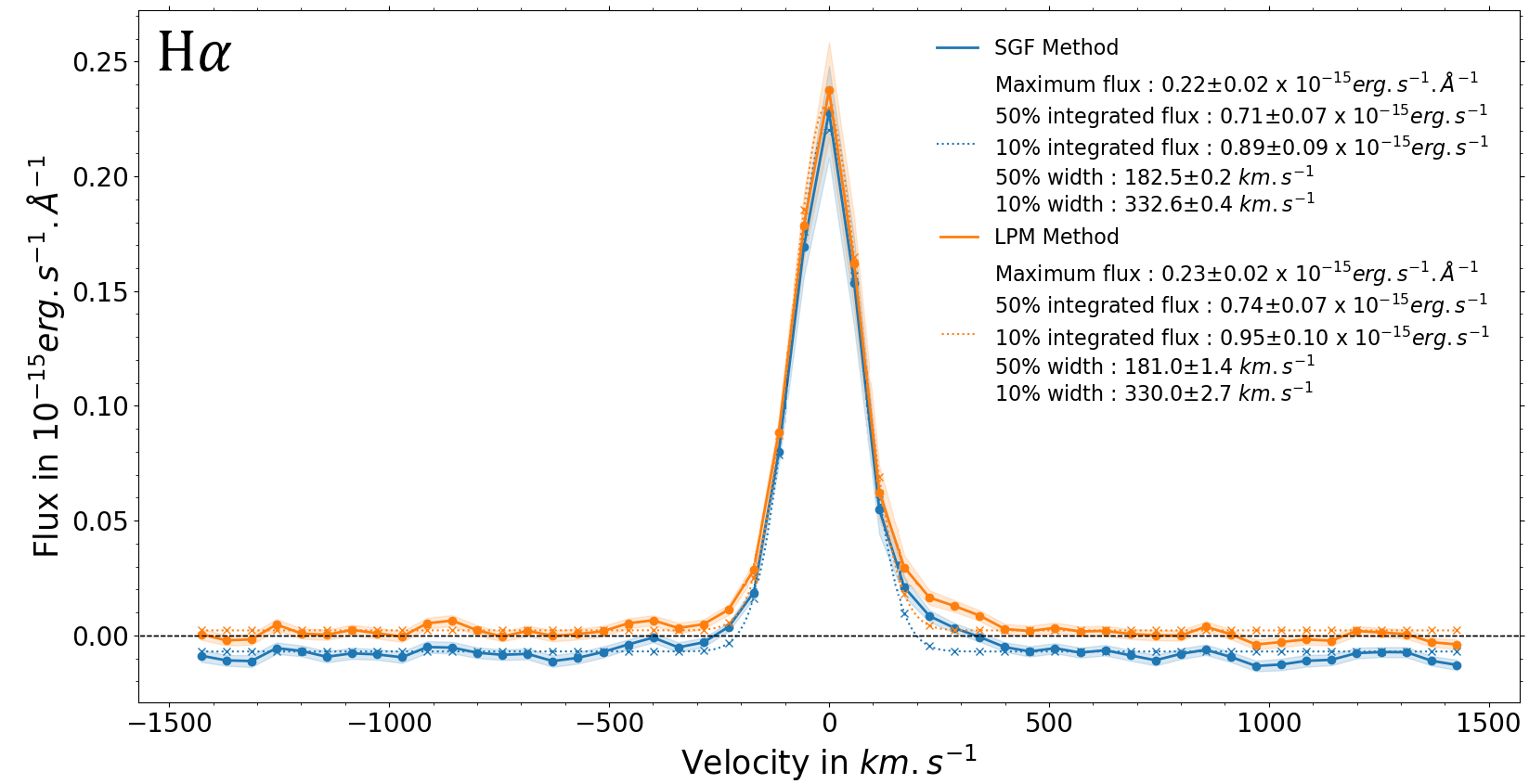}
\caption{Zoom in on 51 channels around the \Halpha line of the YSES1 b spectrum estimates, both with the SGF (in blue) and LPM (in orange) methods. Colored areas delimit the 1-$\sigma$ confidence intervals with regard to noise. Dotted Gaussians are fitted to the lines to estimate their main parameters (presented in the legend, with confidence intervals derived from Gaussian fits to the lower and upper bounds of those of the spectrum estimates). The horizontal dots indicate the zero-flux level, i.e., the theoretical flux lower bound. Unlike the LPM method (whose estimate baseline is positive), the SGF method yet cause abnormal negative values (i.e., which cannot be explained by additive electronic noise), as exposed by the Gaussian fits; that is mainly these negative values, unanticipated by the models, that disrupt the usual routines subsequently applied to the estimates. Moreover, even when the Gaussian parameters are calculated relative to their baselines (as is done here) in order to account for this effect as much as possible, it still leads to biased parameter estimations; in fact, self-subtraction not only shifts the lines but also vertically distorts them (as shown in Sect. \ref{sgf}).\label{fig:YSES1b_Halpha}}
\end{figure*}

\begin{figure*}[t!]
\includegraphics[width=\linewidth]{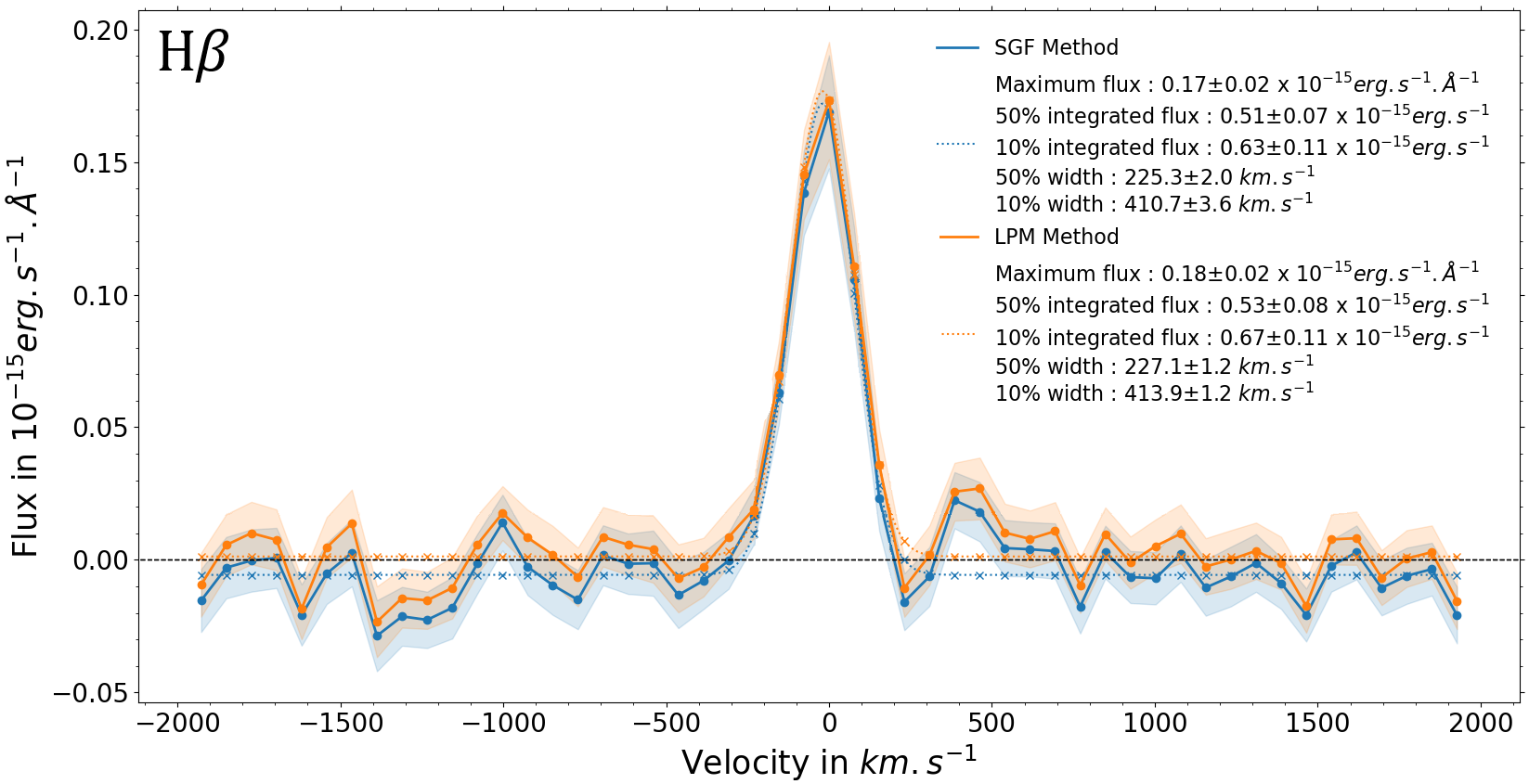}
\caption{Same as Fig. \ref{fig:YSES1b_Halpha}, but for the \Hbeta line (the mask used by the LPM method during the estimation process being centered on the \Hbeta line).\label{fig:YSES1b_Hbeta}}
\end{figure*}

YSES1 observations offer a chance to further validate the new method on the sky. YSES1 b and c are early-L and late-L type objects \citep{Bohn_2020_AJL}, with spectral features that diverge from those of their host star YSES1, and with larger separations. We processed every individual data cube of YSES1 with both halo subtraction methods and median-combined the residuals. The YSES1 c companion was not detected at \Halpha~wavelengths (Jorquera et al., in prep), whose accretion model predicts to be the brightest emission \citep{Aoyama_2020}. We focus on the comparison of the YSES1 b spectrum estimates, and in particular on the emission lines (presented in Fig. \ref{fig:YSES1b}, \ref{fig:YSES1b_Halpha}, and \ref{fig:YSES1b_Hbeta}). There, the colored areas are delimiting the 1-$\sigma$ confidence intervals with regard to noise. They were estimated as the root mean square deviation from the mean of the integrated flux on $3\times3$ apertures at different angular positions at the same separation. They were also corrected by the ratio of total flux to $3\times3$ aperture flux.

We detected \Halpha and \Hbeta lines on YSES1 b (Fig. \ref{fig:YSES1b_Halpha} and \ref{fig:YSES1b_Hbeta}), as well as HeI lines (at 6678$\AA$ and 7065$\AA$) and the CaII H\&K triplet (at 8498$\AA$, 8542$\AA$, and 8662$\AA$). YSES1 b is thus the only second object (following Delorme 1 AB b) in this mass range known to display HeI emission lines \citep{Eriksson_2020} and the only object in this mass range to display this full combination of lines. All these lines are usually observed on accreting and/or active stars, but the presence of a circumplanetary disk around the companion advocates for accretion as an origin of the lines.

Figures \ref{fig:YSES1b}, \ref{fig:YSES1b_Halpha}, and \ref{fig:YSES1b_Hbeta} reveal the self-subtraction caused by the SGF method, as well as its correction by the LPM method. Estimations of the main parameters of the lines were made from Gaussian fitting to compare the results (recorded in the legends). Notably, the \Halpha and \Hbeta lines exhibit maximum flux differences of $\sim$5\% and $\sim$6\%, as well as 10\% integrated flux differences of $\sim$7\% and $\sim$6\%. Confidence intervals were determined from Gaussian fitting on the lower and upper bounds of the ones of the spectra. Using the routines built for \cite{Jorquera_2024}, we derived, for the SGF and LPM methods, accretion rates of $\mathrm{1.11\times10^{-9\pm0.19}M_{Jup}/year}$ and $\mathrm{1.45\times10^{-9\pm0.19}M_{Jup}/year}$ from the \Halpha line flux (the errors are systematic errors derived from model fits by the routines). Thus, we found the SGF method underestimates the accretion rates by $\sim$30\%. The accretion rate is comparable to those of objects of similar masses \citep{Betti_2023}. A follow-up study, focused on the characterization of the accretion mechanisms on YSES1 b using these observations and complementary data, is in progress (Jorquera et al. in prep.).

To go even further, we injected fake planets into all YSES1 cubes (following Appendix \ref{simus_process}) and performed both stellar halo subtraction methods. The results are presented in Appendix \ref{fake}. Qualitatively and quantitatively, with MSE calculation for the line and for the continuum, they confirm all the self-subtraction issues and corrections highlighted in this paper.

\newpage

~

\newpage

~

\section{Concluding remarks}\label{conclu}

\subsection{Remaining limitations of the LPM method}\label{conclu_lpm}

An interesting take on the LPM method is that estimating the stellar component via an orthogonal projection of the data on the vector space of the modulation matrix, while estimating the planetary component via the subtraction of this stellar estimate, is equivalent to the orthogonal decomposition of the data. This means that the subtraction quality depends on the orthogonality of the stellar and planetary spectra vectors.

The proposed LPM method is thus optimal for extracting companions with spectra orthogonal to the one of their host star, and, on the contrary, blind to every companions with spectra collinear with the one of their host star. Assuming positive values for the spectra, the angle between two spectra is close to $\pm~\pi/2$ when, for given total fluxes, their inner product is close to $0$. Thus, assuming a stellar spectrum dominated by the continuum, our proposed method is particularly adapted for detecting and characterizing companions with sparse spectra (e.g., either with accretion or chromospheric activity producing emission lines). On the contrary, the approach is still limited when the stellar spaxel and the planetary spaxel have identical flux ratios over the entire wavelength (or, as shown in Fig. \ref{fig:sgf_with_big_line}, more realistically, when they exhibit a similarity up to a certain factor).

We quantified this behavior via the calculations presented in Appendix \ref{qual}, looking for the best possible planetary estimate when the stellar component is perfectly estimated. This gives both a relative angle and an upper bound on the norms ratio.

For all these reasons, more information will be needed to push the method further (e.g., by spatial information injection). These two metrics will allow better comparisons of the future performance with the relevant theoretical approaches.

\subsection{Discussions}\label{conclu_discs}

The primary aim of this article is to lay the foundations for a new halo subtraction method, adapted to medium- to high-resolution integral field spectrographs, and which can be complemented by other advanced signal processing techniques, to improve both the detection and the characterization of accreting companions.

Whether in the case of simulations or real data observations, the LPM and SGF methods have similar detection capabilities. This result is confirmed by the contrast curves and ROC curves (on the one hand to express the recall when the precision is fixed, and on the other hand to explore the precision-recall trade-off).

Regarding characterization, however, a comparative analysis of the \Halpha line estimates, from the two halo subtraction methods, indicates that the accretion rate was underestimated by $\sim$30\% ($\mathrm{2.4\times10^{-9}~M_{Sun}/year}$ instead of $\mathrm{3.2\times10^{-9}~M_{Sun}/year}$), in the case of the low-mass stellar accreting companion HTLup B. In contrast, no significant change could be noticed regarding the fainter PDS 70 b and c; consequently, tour conclusions on the interpretations of the lines remain unchanged \citep{Aoyama_2019, Thanathibodee_2019, Hashimoto_2020}. Yet, for the intermediate YSES1 b case, changes were revealed by the LPM method as well, indicating that the accretion rate was underestimated by $\sim$30\% when using the SGF method ($\mathrm{1.11\times10^{-9}~M_{Jup}/year}$ instead of $\mathrm{1.45\times10^{-9}~M_{Jup}/year}$).

In fact, self-subtraction not only disrupts subsequent routines with unexpected negative flux values, but also vertically distorts the lines so that, whatever is attempted, it becomes impossible to recover their actual parameters: information is truly lost.

Importantly, the improvements offered by the LPM method are applicable to other techniques derived from the SGF method, such as molecular mapping (see Sect. \ref{intro}). The lack of filtering at the stellar halo subtraction step guarantees that information on the molecular lines of cool companions is preserved; in turn, this could secure the estimate of their molecular abundances and rotation velocities \citep[e.g.,][]{Petrus_2021}. 

And finally, even more importantly, our proposed approach moves from a non-linear and non-parametric model of the data to a linear and parametric one. This offers to use more advanced machine learning approaches and signal processing techniques. Future refinements of the method could include weighting of the estimates by precision matrices based on covariance matrices to improve robustness to noise and defaults. Better local estimates of the stellar spectrum would now also be usable to deal with the various instrumental defaults. Ultimately, spatial and spectral regularization (in a greedy approach) would allow us to recover planetary continua and absorptions. This would be a new source of information to that provided by the lines, opening up brand new research directions, both in signal processing, from inverse problems to faint signal search algorithms, and in astrophysics, from molecular mapping to planet formation processes.

\begin{acknowledgements}
    We acknowledge support from the French National Research Agency (ANR) through project grant ANR-20-CE31-0012 and the Programmes Nationaux de Planétologie et de Physique Stellaire (PNP and PNPS). We thank Maud Langlois and Eric Thiébaut for fruitful discussions about mathematical modeling of the problem and its solution. We thank Chen Xie, Jun Hashimoto, and Sebastiaan Haffert for fruitful discussions about the data analysis and for sending their data cubes for comparison. S.J. acknowledges support from the National Agency for Research and Development (ANID), Scholarship Program, Doctorado Becas Nacionales/2020 - 21212356. We thank the anonymous referee for her/his careful reading of the manuscript as well as her/his insightful comments and suggestions.
\end{acknowledgements}

\bibliographystyle{aa}
\bibliography{refs}

\begin{appendix}

\section{Stellar halo subtraction methods}\label{meths}

\subsection{Model of the observations}\label{meths_mod}

Let $\tensorA\in\mathbb{R}^{\ell \times m \times n}$ stands for the ideal theoretical hyperspectral data cube, with $\ell$ spectral channels, $m$ pixels in the horizontal direction, and $n$ pixels in the vertical direction. For $i\in[\ell]$\footnote{The set of the $\ell$ first integers is denoted as $[\ell]=\{1,\ldots,\ell\}$}, the monochromatic image at wavelength $\lambda_i$ is defined by the matrix $\boldsymbol{A}_i\in\mathbb{R}^{m \times n}$ of the elements of $\tensorA$ along the second and third axes. For $(x,y)\in[m]\times[n]$, the spaxel at position $(x,y)$ is defined by the vector $\boldsymbol{a}_{xy}\in\mathbb{R}^{\ell}$ of the elements of $\tensorA$ along the first axis. Following these notations, the ideal data cube containing $u$ unresolved astronomical bodies at positions $(x_j,y_j)$ for $j\in[u]$ is:
\begin{equation}
\label{A_aD}
\tensorA = \sum_{j=1}^{u}\boldsymbol{\check{a}}^{(j)} \otimes\boldsymbol{\Delta}^{(j)}\,,
\end{equation}
where $\otimes$ stands for the outer product (e.g., here, between a vector and a matrix representing a spaxel and a monochromatic image). The matrix $\boldsymbol{\Delta}^{(j)} \in \mathbb{R}^{m\times n}$ satisfies $\boldsymbol{\Delta}^{(j)} = \delta(x-x_j)~\delta(y-y_j)$ and equals $1$ at position $(x_j,y_j)$ and $0$ otherwise. It indicates the position of the $j^{\text{~th}}$ object for which the vector $\boldsymbol{\check{a}}^{(j)}\in\mathbb{R}^{\ell}$ is the ideal spectrum. The use of $\boldsymbol{\Delta}^{(j)}$ comes from the assumption that the observed objects (typically a host star and several accreting companions in the context of this article) are unresolved by the optical system and thus appear as point sources (which seems reasonable in the case of images from the MUSE instrument, each of its pixels covering an area of $0.025\times0.025$” and thus, typically at a distance of 100 pc, receiving the flux of $2.5\times2.5$AU area in the sky -- note also that the chances of an object straddling two pixels are consequently very low under these conditions).

For an actual imaging system (assumed linear), we introduce the observed data cube noted $\tensorO$:
\begin{equation}
\label{O_AOmega}
\tensorO = \tensorA\ast\boldsymbol{\Omega}\,,
\end{equation}
where $\ast$ stands for the Fredholm operator. The 6-dimensional tensor $\boldsymbol{\Omega}\in\mathbb{R}^{(\ell \times m \times n) \times (\ell \times m \times n)}$ is the tensor of the 3-dimensional point spread function (PSF) of the optical system (IFS, telescope, AO, atmosphere). Its effect is to spread the flux $a_{i'x'y'}$ of each pixel of position $(x',y')$ and wavelength $\lambda_{i'}$ of $\tensorA$ on each pixel of position $(x,y)$ and wavelength $\lambda_{i}$ of $\tensorO$:
\begin{equation}
\label{O_sumAOmega}
\tensorO = \sum_{i'=1}^\ell\sum_{x'=1}^m\sum_{y'=1}^n a_{i'x'y'}~\boldsymbol{\Omega}_{i'x'y':::}\,,
\end{equation}
where $\boldsymbol{\Omega}_{i'x'y':::}$ is the 3-dimensional response of the total system at position $(x',y')$ and wavelength $\lambda_{i'}$. Conversely, each element $o_{ixy}$ of $\tensorO$ is the sum of each of the flux contributions $a_{i'x'y'}$ of $\tensorA$, more or less spread out from their initial spatio-spectral position.

For convenience, this 3D point spread function, $\boldsymbol{\Omega}_{i'x'y':::}$, is commonly spatio-spectrally separated into a 1D spectral PSF (known as the LSF for line spread function) and a 2D spatial PSF (known as the FSF for field spread function), both wavelength dependent. The spatial dependence is neglected using the small-field hypothesis, whose use is relevant here by reducing the study to the AO correction radius \citep[note however that the PSF could spatially vary in some integral field spectrograph images due to imperfect calibration of the cubes, instrument flexure or even scattered light, as shown for example in][]{Xie_2020}. This results in the PSF expression at wavelength $\lambda_{i'}$:
\begin{equation}
\label{PSF_LSFFSF}
\text{PSF}_{i'}(i,x,y) = \text{LSF}_{i'}(i)\times\text{FSF}_{i'}(x,y)\,,
\end{equation}
We note that for each pixel, this means neglecting all the flux scattered towards or from pixels both at neighboring positions and at neighboring wavelengths (i.e. along the cube's diagonals).

The presence of this PSF induces a spread on the spectral axis at each position $(x,y)$ and a spread in the image plane at each wavelength $\lambda_{i}$. The integration of the PSF over the whole field and over the spectral direction is set to 1 to satisfy the flux conservation constraint. The separation assumption of Eq. \eqref{PSF_LSFFSF} implies that this flux constraint must also be satisfied separately for both the LSF and the FSF.

Consequently, in practice, only spectrally spread versions of the spectra $\boldsymbol{a}^{(j)}=\boldsymbol{\check{a}}^{(j)} \ast \boldsymbol{\psi}$ are observed, the matrix $\boldsymbol{\psi}\in\mathbb{R}^{\ell\times\ell}$ containing LSF information for each spectral channel of index $i$ and thus expressing the wavelength dependency of the spectral spread with the Fredholm operator. As the LSF is little known and the spectra are (very) noisy, it is usual to settle for these imperfect spectrally spread spectra, $\boldsymbol{a}^{(j)}$, rather than attempt their spectral deconvolution (the models being convolved with an LSF approximation instead).

Similarly, these are $u$ spreads $\boldsymbol{\phi}_{i}^{(j)} = \boldsymbol{\phi}_{i} \ast \boldsymbol{\Delta}^{(j)}$ on each of the monochromatic images at wavelength $\lambda_i$ that are observed at positions $(x_j,y_j)$, the matrix $\boldsymbol{\phi}_{i}\in\mathbb{R}^{m \times n}$ expressing the spectral dependency of the spatial spread with the Fredholm operator by containing FSF information for each spectral channel of index $i$.

The separation assumption of Eq. \eqref{PSF_LSFFSF} allows us to express the observed image at wavelength $\lambda_i$ as:
\begin{equation}
\label{O_aphi}
\boldsymbol{O}_i = \sum_{j=1}^{u} a_i^{(j)} \boldsymbol{\phi}_{i}^{(j)},\quad\forall i\in[\ell]\,,
\end{equation}
where the scalar $a_i^{(j)}\in\mathbb{R}$ is the flux observed from the $j^{\text{~th}}$ object at wavelength $\lambda_i$ and is given by the $i^{\text{~th}}$ entry of the vector $\boldsymbol{a}^{(j)}$. Because of the PSF normalization, $a_i^{(1)}$ is the total integrated flux of the image $\boldsymbol{O}_i$ when a single unresolved object is observed.

\begin{figure}[t!]
\includegraphics[width=\columnwidth]{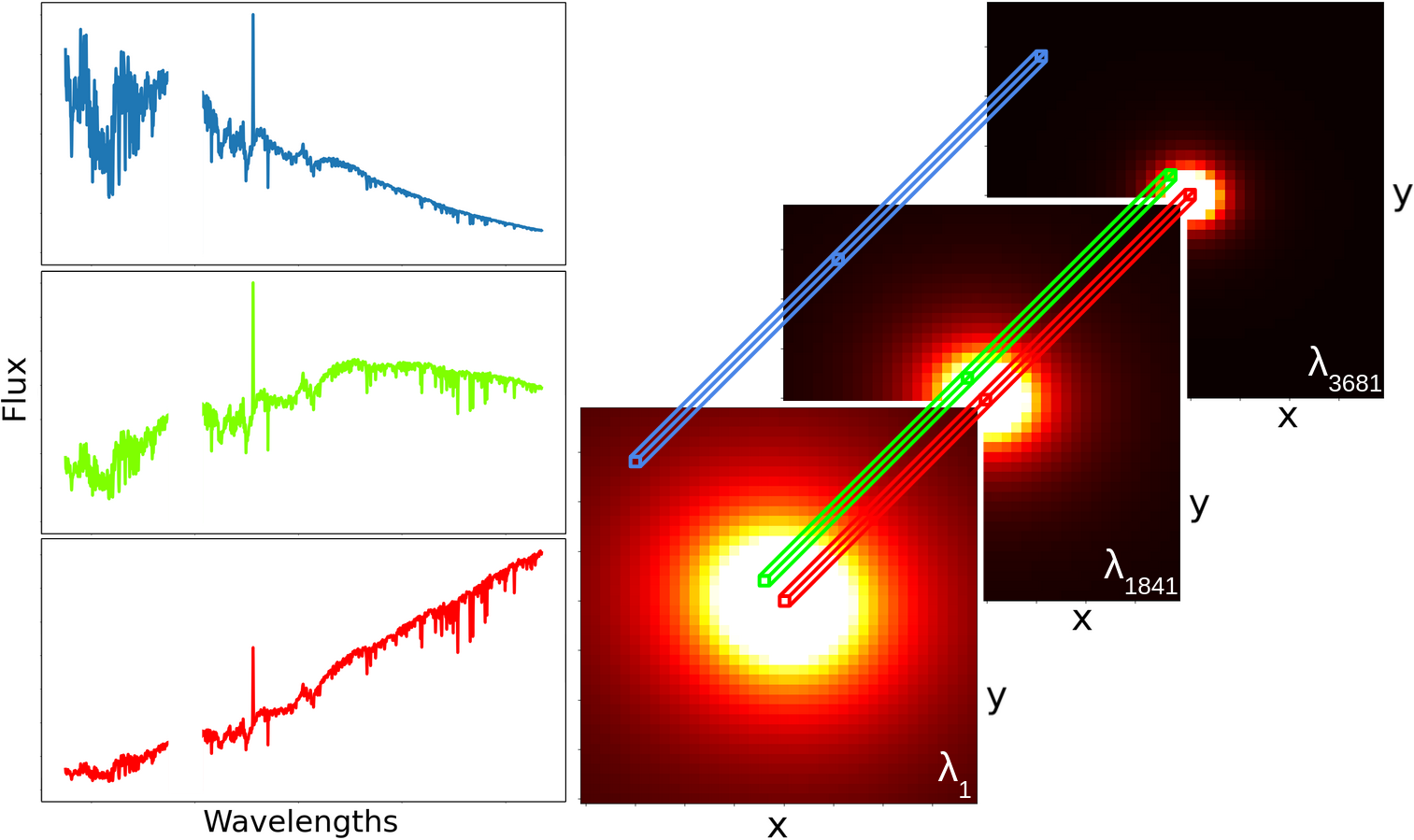}
\caption{Spatio-spectral simulation of a star observation. On the right, three monochromatic images are shown at wavelengths $\lambda_1$, $\lambda_{1841}$, $\lambda_{3681}$ to highlight the spatial spread variation with wavelength. On the left, three spaxels are shown at three different separations to highlight the spectrum deformation caused by this spatial spread variation.\label{fig:CSF}}
\end{figure}

However, it is difficult to deal with the FSF for post-processing operations because of its complex shape, notably due to the atmospheric turbulence effects and AO system corrections. We thus seek to reverse its effect in the spectral direction in which it is expected to behave more simply. For this purpose, we define the chromatic spread function (CSF), a chromatic PSF in a way. Its associated vector is determined from the FSF tensor $\tensorPhi=\left(\boldsymbol{\phi}_{i}\right)_{i\in[\ell]} \in \mathbb{R}^{\ell \times m \times n}$, constructed from the FSF matrices at each of the observed wavelengths. Whereas the set of the FSF matrices corresponds to the $\ell$ two-dimensional spatial entries of this tensor, the set of the CSF vectors corresponds to the $m \times n$ one-dimensional spectral entries of this tensor. We note $\boldsymbol{\alpha}_{xy}\in\mathbb{R}^\ell$ for these vectors at each position $(x,y)\in[m]\times[n]$.

As illustrated by Fig. \ref{fig:CSF}, the $\boldsymbol{\alpha}_{xy}$ can be understood as a set of deformation coefficients, which allows us to explicit all the expressions of the spaxels from objects' spectra. By defining the vectors $\boldsymbol{\alpha}_{xy}^{(j)}$ as above, for $j\in[u]$, from the tensors $\tensorPhi^{(j)}$ built with the set of the $\boldsymbol{\phi}_{i}^{(j)}$ matrices, the spaxels of the observation cube can be written as spectra $\boldsymbol{a}^{(j)}$ deformed by coefficients $\boldsymbol{\alpha}_{xy}^{(j)}$:
\begin{equation}
\label{o_aalpha}
\boldsymbol{o}_{xy} = \sum_{j=1}^{u} \boldsymbol{a}^{(j)} \cdot \boldsymbol{\alpha}_{xy}^{(j)},\quad\forall {(x,y)\in[m]\times[n]}\,,
\end{equation}
where $\cdot$ stands for the Hadamard (entrywise) product.

More specifically, the studied observations show two types of unresolved astronomical bodies. To illustrate this duality, the observation cube is decomposed as $\tensorO = \tensorS + \tensorP$ with $\tensorP$ a signal cube associated to $u_p\geq0$ unresolved sources (typically from protoplanets or other accreting companions like young stars) and $\tensorS$ a nuisance cube (typically from one bright star). Both take into account the spatio-spectral deformations previously described.

Finally, in addition to all the optical effects described above, the cubes are also affected by various noise sources (read-out, thermal fluctuations, and photon noises). Although they follow various probability laws, these noises are (relatively) realistically encompassed in a centered, symmetrically distributed Gaussian noise (see Appendix \ref{noise} for additional details on this assumption). We let $\tensorE$ denote the cube whose elements contain the random noise realizations that are assumed to be additive. Furthermore, for the sake of simplicity, these realizations are assumed to be spatially independent and identically distributed (although this amounts to ignoring the correlation between nearby pixels).

Then, the expression of the data, $\tensorD$, that we consider for this entire study can be expressed by the sum of three components:
\begin{equation}
\label{D_SPE}
\tensorD = \tensorS+\tensorP+\tensorE\,.
\end{equation}
The values of the noise component, $\tensorE$, being assumed centered, it can be mitigated by multiple approaches: median of various exposures, sigma-clipping, or even matched filtering (proposed in Sect. \ref{pds70bc_match}). A low-noise framework is thus considered for the theoretical problem solving described below, where $\tensorS$ and $\tensorP$ are the dominant contributions.

\subsection{Problem statement}\label{meths_pb}

The planetary signal estimate, $\tensorPest$, is determined by subtraction from the data, $\tensorD$, of an estimate, $\tensorSest$, of the stellar nuisance:
\begin{equation}
\label{Pest}
\tensorPest = \tensorD-\tensorSest\,.
\end{equation}
This is the so-called “stellar halo subtraction,” a classic method for detecting planets masked by the halo of their star in direct imaging. This technique has many variants, notably depending on how the stellar component is estimated (e.g., from temporal diversity \citep{Marois_2005} or from spectral diversity \citep{Hoeijmakers_2018}). Both solutions described in the following are based on the spectral diversity, the stellar nuisance estimate being based on the data model of Eq. \eqref{o_aalpha}. The deformation, $\boldsymbol{\alpha}_{xy}$, is assumed to be smooth (although more complex in the case of VLT/MUSE or VLT/SINFONI observations than in the case of diffraction limited observations because of the system of AO notably providing higher Strehl at the longest wavelengths), allowing for its estimation by simple models. A spaxel-by-spaxel estimation of the stellar nuisance contribution (i.e., without any spatial a priori) is enabled by this approach.

We let $\boldsymbol{s}_{xy}$ be the stellar nuisance contribution to the data spaxel at position $(x,y)$, and $\boldsymbol{\hat{s}}_{xy}$ its estimation. According to the CSF-based model of the observations given by Eq. \eqref{o_aalpha}, $\boldsymbol{s}_{xy}=\boldsymbol{s}\cdot\boldsymbol{\alpha}_{xy}$, where $\boldsymbol{s}$ stands for the stellar nuisance spectrum (distorted by the LSF). The estimation problem can thus come down to determining both $\boldsymbol{\hat{s}}$ the estimate of $\boldsymbol{s}$ for the whole field and $\boldsymbol{\hat{\alpha}}_{xy}$ the estimate of $\boldsymbol{\alpha}_{xy}$ at each position $(x,y)$:
\begin{equation}
\label{sest}
\boldsymbol{\hat{s}}_{xy} = \boldsymbol{\hat{s}}\cdot\boldsymbol{\hat{\alpha}}_{xy}\,.
\end{equation}
As it is assumed that the optical system preserves the total flux (all $\boldsymbol{\phi}_i$ sum to 1), $\boldsymbol{s}$ is nothing more than the integrated flux over the field in the absence of planet and noise. The random noise being centered, and a potential planet being faint, the integration of the flux over the field at each spectral channel is already a good estimate $\boldsymbol{\hat{s}}$ of $\boldsymbol{s}$ (the weight of the planetary contribution to this integrated flux being expected to be weak). In practice, the spaxels from the edges of the field are masked because the stellar signal is too low there to be reliable. Similarly, in practice, the spaxels around the star center are masked because more likely to be considered as outliers by the MUSE pipeline and, therefore, replaced by too rough approximations interpolating neighboring values (in particular when the Strehl ratio is high). The integrated flux thus no longer matches the total flux of the star, but this does not rule out the usefulness of Eq. \eqref{sest}. Indeed, estimating both $\boldsymbol{s}$ and $\boldsymbol{\alpha}_{xy}$ jointly can only be done up to a multiplicative indeterminacy (at the cost of losing the flux conservation). Thus, for the same reasons, this sum can even be replaced by a mean (this only changes the estimate by the factor $m\times n$) and therefore also by a median for increased robustness.

Although much remains to be done to improve the estimate $\boldsymbol{\hat{s}}$ of the stellar spectrum, $\boldsymbol{s}$, this article focuses on the improvement of the estimate $\boldsymbol{\hat{\alpha}}_{xy}$ of the deformation coefficients, $\boldsymbol{\alpha}_{xy}$.

\subsection{State-of-the-art solution}\label{meths_sgf}

\begin{figure}[t!]
\includegraphics[width=\columnwidth]{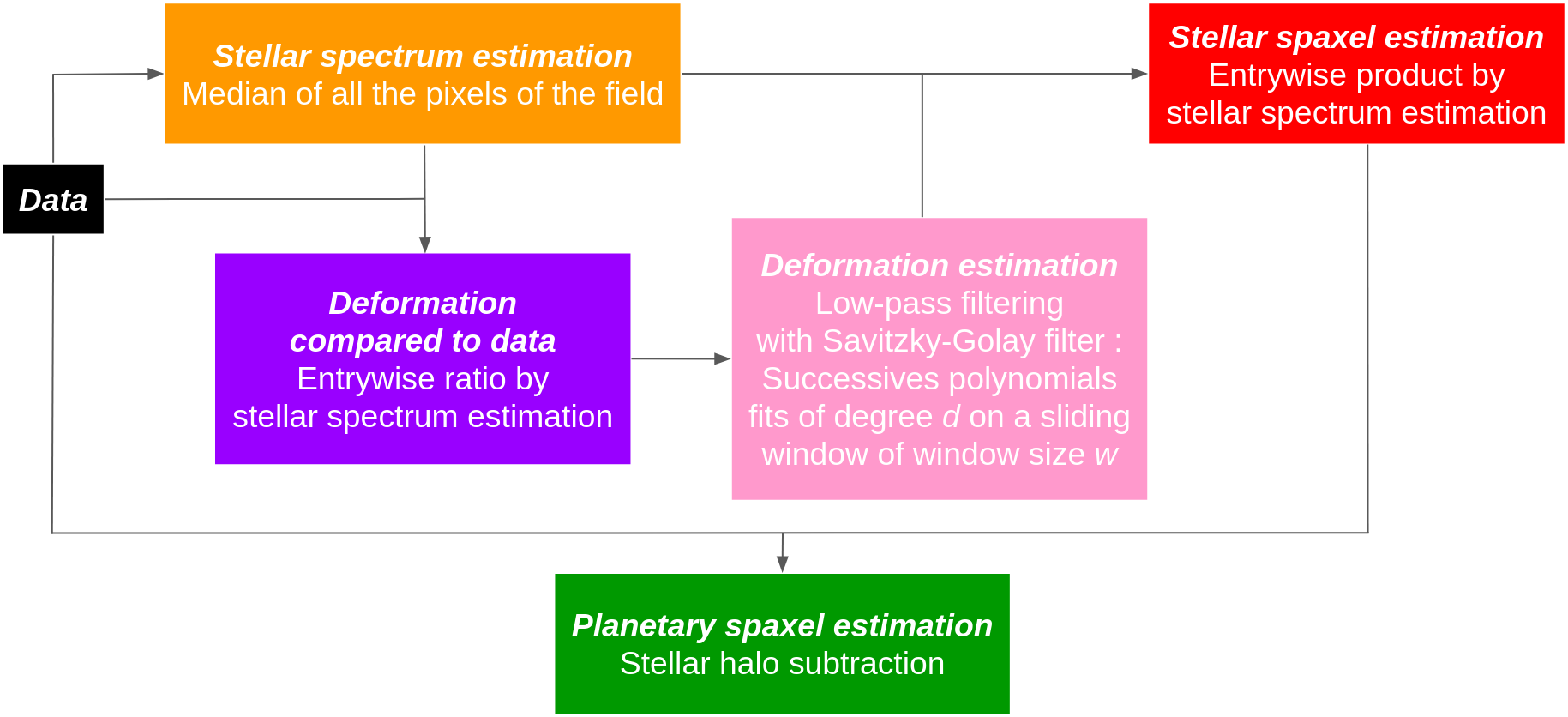}
\caption{General principle of the SGF stellar halo subtraction method presented in \cite{Haffert_2019}. Each planetary spaxel (in green) is estimated by subtraction from the data (in black) of an estimate of the stellar spaxel (in red). This latter is estimated by entrywise product of the stellar spectrum estimate (in orange) by the deformation coefficient estimate (in pink). This estimate is made by low-pass filtering of the stellar spectrum estimate to data spaxel ratio (in purple).\label{fig:SGF}}
\end{figure}

While \cite{Keppler_2018} reported the first detection of a protoplanet (a.k.a. PDS70 b) using classical detection techniques (i.e., angular/polarimetric differential imaging with VLT/SPHERE, VLT/NaCo, and Gemini/NICI instruments), \cite{Haffert_2019} used a spaxel-by-spaxel approach adapted from \cite{Hoeijmakers_2018} for the first detection of the second known protoplanet (a.k.a. PDS70 c, in the same system).
The principle of the method, which we refer to as the Savitzky-Golay filtering (SGF) method, is sketched in Fig. \ref{fig:SGF}. It can also be expressed using the quantities $\boldsymbol{\hat{\alpha}}_{xy}$, $\boldsymbol{d}_{xy}$, and $\boldsymbol{\hat{s}}$ described in the previous section. The following equation sums up the process of estimation of the geometric deformation in this case:
\begin{equation}
\label{alphaest_SGF}
\boldsymbol{\hat{\alpha}}_{xy} = \text{SG}\left(\boldsymbol{d}_{xy}./\boldsymbol{\hat{s}}\right),\,
\end{equation}
where the rationale consists in inverting Eq. \eqref{sest} using the Hadamard (entrywise) division $./$. The spaxel $\boldsymbol{d}_{xy}$ can be seen as a rough estimate of the unknown stellar component spaxel, $\boldsymbol{s}_{xy}$, at position $(x,y)$. The low-pass filtering (by a Savitzky-Golay filter of degree $\mathrm{d}$ and spectral window size $\bar{\b{W}}$) aims at removing high frequencies of $\boldsymbol{p}_{xy}$ to derive a smooth estimate $\boldsymbol{\hat{\alpha}}_{xy}$ of $\boldsymbol{\alpha}_{xy}$.

\subsection{Our proposed solution}\label{meths_lpm}

\begin{figure}[t!]
\includegraphics[width=\columnwidth]{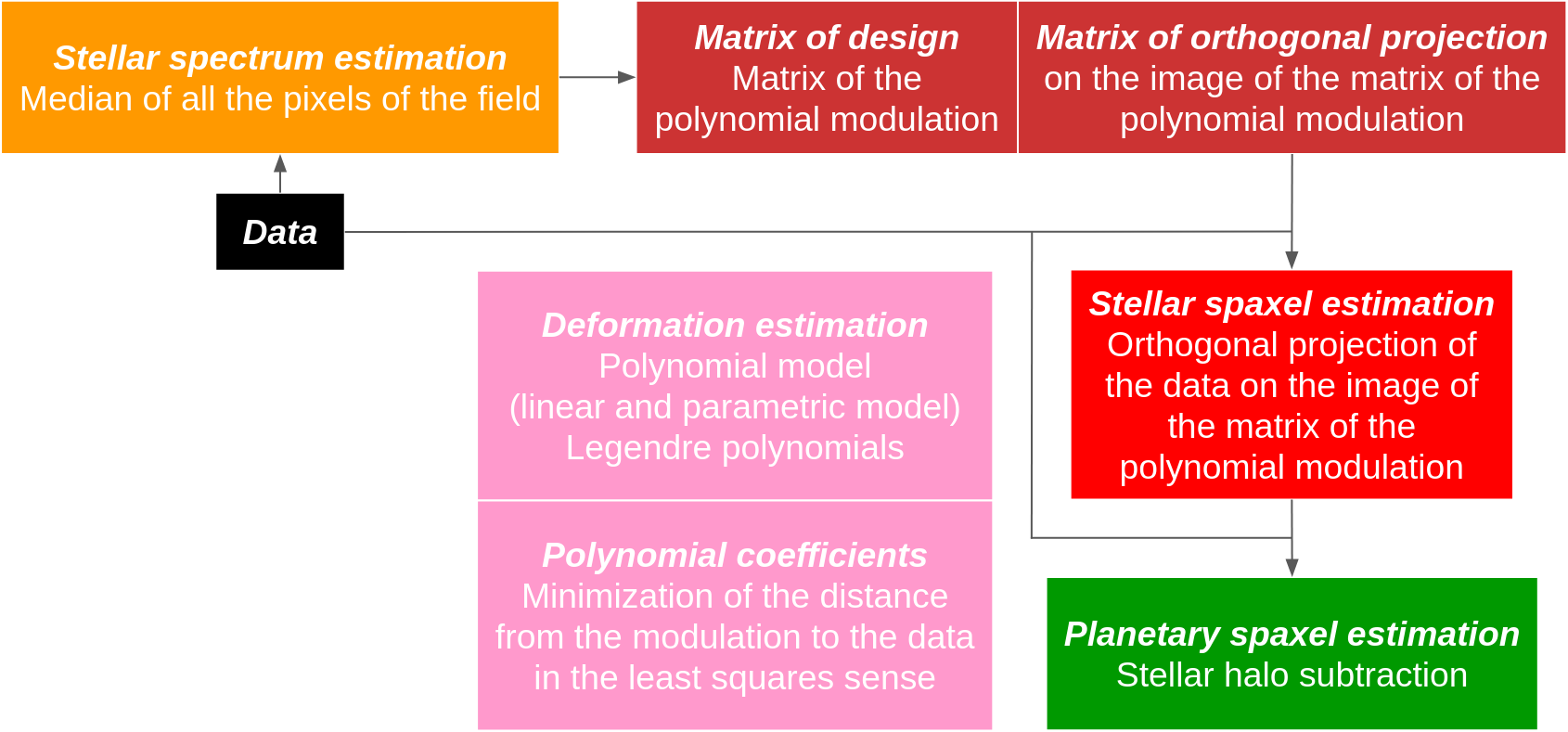}
\caption{General principle of the LPM stellar halo subtraction method we propose. Each planetary spaxel (in green) is estimated by subtraction from the data (in black) of an estimate of the stellar spaxel (in red). This latter is estimated by polynomial modulation (in pink) of the stellar spectrum estimate (in orange). In practice, this is done by multiplication of the data spaxel by the orthogonal projection matrix onto the column space of the design matrix of the polynomial modulation (in brown).\label{fig:Mod}}
\end{figure}

To tackle the problem of self-subtraction and spectral distortion, we propose to estimate $\boldsymbol{\hat{\alpha}}_{xy}$ as a smooth function of the wavelengths, $\lambda_i$, expressed as a polynomial function of degree $\partial$:
\begin{equation}
\label{alphaest_Mod}
\boldsymbol{\hat{\alpha}}_{xy} = \left(\sum_{k=0}^{\partial}\lambda_i^{k}\, \hat{\beta}_{xyk}\right)_{i\in[\ell]}\,,
\end{equation}
where $\hat{\beta}_{xyk}$ is the $k^{th}$ regression coefficient at position $(x,y)$. Equivalently (using Eq. \eqref{sest}), the stellar spaxels estimates, $\boldsymbol{\hat{s}}_{xy}$, are modeled as polynomial modulations of degree $\partial$ of the stellar spectrum estimate, $\boldsymbol{\hat{s}}$:
\begin{equation}
\label{sest_Mod}
\boldsymbol{\hat{s}}_{xy} = \left(\sum_{k=0}^{\partial}\lambda_i^{k}\hat{\beta}_{xyk}\hat{s}_i\right)_{i\in[\ell]}\,.
\end{equation}
The complexity of the learning model is driven by an only hyperparameter: the degree of the polynomial model, $\partial$. Its value must be chosen wisely for a good stellar nuisance estimate. Moreover, this linear regression model can be written in a matrix form as:
\begin{equation}
\label{sest_Mbeta}
\boldsymbol{{s}}_{xy} = \boldsymbol{M}\boldsymbol{{\beta}}_{xy}+\boldsymbol{\varepsilon}_{xy}\,,
\end{equation}
where $\boldsymbol{M} \in \mathbb{R}^{\ell \times \partial+1}$ is the design matrix defined for $k\in[\![\partial]\!]$\footnote{The set of the $\partial+1$ first integers including $0$ is denoted as $[\![\partial]\!]$} by columns $\boldsymbol{m}_k = \boldsymbol{\lambda}^{\cdot k} \cdot \boldsymbol{\hat{s}} \in \mathbb{R}^{\ell}$, $\boldsymbol{\lambda}^{\cdot k}=(\lambda_1^k,\ldots,\lambda_\ell^k)$. The column space of $\boldsymbol{M}$ is the set of polynomial modulations of the stellar spectrum estimate. The vector $\boldsymbol{{\beta}}_{xy}$ is the vector of the unknown regression coefficients, defined as $\boldsymbol{{\beta}}_{xy}=\left(\beta_{xyk}\right)_{k\in [\![\partial]\!]} \in \mathbb{R}^{\partial+1}$ and estimated by ordinary least squares:
\begin{equation}
\label{betaest}
\boldsymbol{\hat{\beta}}_{xy} = \underset{\boldsymbol{\beta}}{\operatorname{argmin}}\,\left\lVert\boldsymbol{d}_{xy}-\boldsymbol{M}\boldsymbol{\beta}\right\rVert_2^2 = (\boldsymbol{M}^\top\boldsymbol{M})^{-1}\boldsymbol{M}^\top\boldsymbol{d}_{xy}\,.
\end{equation}
By defining $\boldsymbol{\Pi_M}=\boldsymbol{M}(\boldsymbol{M}^\top\boldsymbol{M})^{-1}\boldsymbol{M}^\top$ the orthogonal projection matrix onto the column space of $\boldsymbol{M}$, the expression of the stellar nuisance estimate finally reduces to:
\begin{equation}
\label{sest_Pid}
\boldsymbol{\hat{s}}_{xy} = \boldsymbol{\Pi_M}\boldsymbol{d}_{xy}\,.
\end{equation}
The whole estimation process is summarized in Fig. \ref{fig:Mod} and named the Legendre polynomial modulation (LPM) method.

Indeed, it is well-known that the Gram matrix $\boldsymbol{M}^\top\boldsymbol{M}$ is badly conditioned when using the canonical monomial basis \citep{Seber_2012}. Thus, we proposed to expand the data spaxel, $\boldsymbol{d}_{xy}$, on an orthogonal basis, with Legendre polynomials. More details are provided in \cite{Julo_2024}. This change in the basis has the major advantage, in addition, of restoring physical interpretation of the $d+1$ coefficients of the estimation polynomials, as in this way each of them explains different FSF components at different spectral frequencies (see Fig. \ref{fig:PDS70_PSF} for a real FSF decomposition).

\section{Analytical calculations}\label{analytics}

\subsection{Low-pass filtering effect with the SGF method}\label{filter_effect}

\begin{figure}[h]
\raggedleft
\begin{tikzpicture}
\draw (1.5*0*0.74,0*0.74) -- (1.5*2*0.74,0*0.74);
\draw (1.5*2*0.74,0*0.74) -- (1.5*2*0.74,3*0.74);
\draw (1.5*2*0.74,3*0.74) -- (1.5*3*0.74,3*0.74);
\draw (1.5*3*0.74,3*0.74) -- (1.5*3*0.74,0*0.74);
\draw (1.5*3*0.74,0*0.74) -- (1.5*5*0.74,0*0.74) node [right] {$h_\mathrm{min}$};
\draw[dotted] (1.5*3*0.74,3*0.74) -- (1.5*5*0.74,3*0.74) node [right] {$h_\mathrm{max}$};
\draw[dotted] (1.5*2.5*0.74,3*0.74) -- (1.5*2.5*0.74,-0.5*0.74) node [below] {$l_0$};
\draw[dotted] (1.5*2*0.74,0*0.74) -- (1.5*2*0.74,-1*0.74) node [below] {$l_{-w_\text{peak}}$};
\draw[dotted] (1.5*3*0.74,0*0.74) -- (1.5*3*0.74,-1*0.74) node [below] {$l_{+w_\text{peak}}$};
\draw[dotted] (1.5*1*0.74,0*0.74) -- (1.5*1*0.74,-0.5*0.74) node [below] {$l_{-w_\mathrm{max}}$};
\draw[dotted] (1.5*4*0.74,0*0.74) -- (1.5*4*0.74,-0.5*0.74) node [below] {$l_{+w_\mathrm{max}}$};
\draw[->] (-1.5*0.1*0.74,-1.8*0.74) -- (1.5*5.2*0.74,-1.8*0.74) node [right] {Wavelengths in $\AA$};
\draw[->] (-1.5*0.1*0.74,-1.8*0.74) -- (-1.5*0.1*0.74,3.2*0.74) node [right, anchor=north west, align=left] {Data spaxel\\to stellar\\spectrum\\estimation\\ratio in $\mathrm{cm}^{-2}$};
\end{tikzpicture}
\caption{Simple peak model sampled in wavelengths and deformations.}\label{fig:detailed_model}
\end{figure}
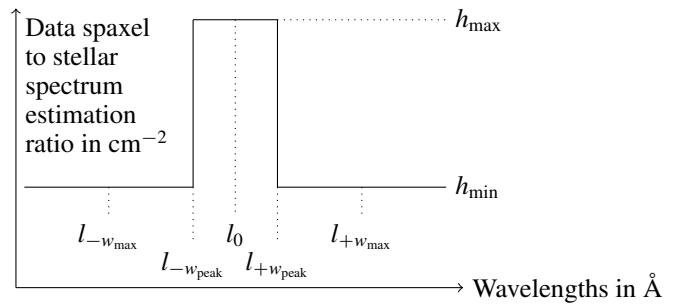

We study in this section the effect of a first-order Savitzky-Golay filtering of window width $2w_\mathrm{max}+1$ on a simple flat signal with a single peak of width $2w_\mathrm{peak}+1$, base $h_\mathrm{min}$ and height $h_\mathrm{max}-h_\mathrm{min}$. In particular, for the window centered on the peak, the linear regression is performed on the following signal, represented as two sets accounting for abscissa and ordinate, respectively:
\begin{align*}
\mathrm{L} = \left(l_w\right)_{w\in W} \text{ and } \mathrm{H} = \left(h_w\right)_{w\in W} \text{ for } W = \ldbrack -w_\mathrm{max},w_\mathrm{max} \rdbrack\,.
\end{align*}
More specifically, assuming that the abscissa axis is regularly sampled at step $l_{\mathrm{step}}$ and that the peak center is correctly sampled at abscissa $l_{\mathrm{peak}}$, we are now interested in the signal following the simple constraints below (and depicted in Fig. \ref{fig:detailed_model}):
\begin{align*}
l_w=l_{\mathrm{peak}} + wl_{\mathrm{step}}
\mbox{ and }
h_w = \left\{ \begin{array}{ll} h_\mathrm{max} \mbox{ if } w \in \ldbrack -w_\mathrm{peak},w_\mathrm{peak} \rdbrack \\ h_\mathrm{min} \mbox{ for all the other points} \end{array} \right.\,.
\end{align*}

Linear regression with the ordinary least squares method then consists in determining $(b_0,b_1)\in\mathbb{R}^2$ minimizing
\begin{align*}
\sum_{i=-w_\mathrm{max}}^{i=+w_\mathrm{max}} (h_i-g_i)^2 \mbox{ with } g_i = b_0+b_1l_i\,.
\end{align*}
It is known that the solution to this problem is
\begin{align*}
\left\{ \begin{array}{ll}
b_1 = \mathrm{Cov}(\mathrm{L},\mathrm{H})~/~\mathrm{Var}(\mathrm{L}) \\
b_0 = \mathrm{Mean}(\mathrm{H}) - b_1\mathrm{Mean}(\mathrm{L})
\end{array} \right.\,,
\end{align*}
with
\begin{align*}
\mathrm{Mean}(\mathrm{L}) = l_{\mathrm{peak}}\,,
\end{align*}
and
\begin{align*}
\mathrm{Mean}(\mathrm{H}) = \frac{(2w_\mathrm{peak}+1)h_\mathrm{max}+2(w_\mathrm{max}-w_\mathrm{peak})h_\mathrm{min}}{2w_\mathrm{max}+1}\,,
\end{align*}
as well as
\begin{align*}
\mathrm{Var}(\mathrm{L}) = \frac{1}{3}w_\mathrm{max}(w_\mathrm{max}+1)l_{\text{step}}^2\,,
\end{align*}
and
\begin{align*}
\mathrm{Cov}(\mathrm{L},\mathrm{H}) = \frac{2w_w(w_\mathrm{max}+1)(w_\mathrm{max}-w_\mathrm{peak})}{2w_\mathrm{max}+1}(h_\mathrm{max}+h_\mathrm{min})l_{\text{step}}\,,
\end{align*}
following straightforward computations and leading to
\begin{align*}
b_1 = \frac{6(w_\mathrm{max}-w_\mathrm{peak})(h_\mathrm{max}+h_\mathrm{min})}{(2w_\mathrm{max}+1)l_{\text{step}}}\,,
\end{align*}
and
\begin{align*}
b_0 &= \frac{(2w_\mathrm{peak}+1)h_\mathrm{max}+2(w_\mathrm{max}-w_\mathrm{peak})h_\mathrm{min}}{2w_\mathrm{max}+1} - b_1l_{\mathrm{peak}}\,.
\end{align*}

In particular, the new value returned by the Savitzky-Golay algorithm at the peak abscissa is given by

\begin{equation*}
g_{\mathrm{peak}} = h_\mathrm{min}+\frac{2w_\mathrm{peak}+1}{\hphantom{.}2w_\mathrm{max}+1}(h_\mathrm{max}-h_\mathrm{min})\,.
\end{equation*}
This expression can be simplified as follows:
\begin{equation*}
g_{\mathrm{peak}} = \b{H}+\mathrm{R}\bar{\b{H}}\,,
\end{equation*}

with
\begin{align*}
\b{H}&=h_\mathrm{min} \mbox{ the base of the peak,}\\
\mathrm{R}&=\frac{2w_\mathrm{peak}+1}{\hphantom{.}2w_\mathrm{max}+1} \mbox{ the ratio of the peak to window widths,}\\
\bar{\b{H}}&=h_\mathrm{max}-h_\mathrm{min} \mbox{ the height of the peak relative to its base.}\\
\end{align*}

\subsection{Residual distribution with the LPM method}\label{resids}

\subsubsection{Analytic expression of the spaxels estimates}\label{resids_expr}

The use of the orthogonal projection matrix, $\boldsymbol{\Pi_M}$, allows us to identify the contribution of each of the components of the data in the stellar spectrum estimate (using Eq. \eqref{D_SPE} and \eqref{sest}):
\begin{equation}
\label{sest_ortho}
\boldsymbol{\hat{s}}_{xy} = \boldsymbol{\Pi_M}\boldsymbol{s}_{xy}+\boldsymbol{\Pi_M}\boldsymbol{p}_{xy}+\boldsymbol{\boldsymbol{\Pi_{\boldsymbol{M}}}}\boldsymbol{e}_{xy}\,.
\end{equation}
Indeed, the components of the true stellar spaxel, $\boldsymbol{s}_{xy}$, that are orthogonal to the modulation of the stellar spectrum estimate, $\boldsymbol{\hat{s}}$, cannot contribute to the stellar spaxel estimate, $\boldsymbol{\hat{s}}_{xy}$. Conversely, the components of the planetary and noise spaxels, $\boldsymbol{p}_{xy}$ and $\boldsymbol{e}_{xy}$, that can be (over)fitted by the modulation model can contribute to the stellar spaxel estimate, $\boldsymbol{\hat{s}}_{xy}$.

In the same way, it is possible to identify the contribution of each of the components of the data in the post-subtraction residuals (using the Eq. \eqref{Pest}):
\begin{equation}
\label{pest_ortho}
\boldsymbol{\hat{p}}_{xy} = \boldsymbol{\Pi}^\perp_{\boldsymbol{M}}\boldsymbol{s}_{xy}+\boldsymbol{\Pi}^\perp_{\boldsymbol{M}}\boldsymbol{p}_{xy}+\boldsymbol{\Pi}^\perp_{\boldsymbol{M}}\boldsymbol{e}_{xy}\,,
\end{equation}
where the matrix $\boldsymbol{\Pi}^\perp_{\boldsymbol{M}}=\boldsymbol{\mathrm{I}}_{\ell}-\boldsymbol{\Pi_M}$ is the matrix of the projection onto the orthogonal complement of the column space of $\boldsymbol{M}$.

To go a step further, assuming the noise components, $e_{ixy}$, to be Gaussian, distributed with standard deviation $\sigma$,  as well as independent and identically distributed, the distribution of each of the residuals spaxels, $\boldsymbol{\hat{p}}_{xy}$, is given by:
\begin{equation}
\label{pest_gaussian}
\boldsymbol{\hat{p}}_{xy} \sim \mathcal{N}\left(\boldsymbol{\Pi}^\perp_{\boldsymbol{M}}\boldsymbol{s}_{xy} + \boldsymbol{\Pi}^\perp_{\boldsymbol{M}}\boldsymbol{p}_{xy}, \sigma^2\boldsymbol{\Pi}^\perp_{\boldsymbol{M}}\right)\,.
\end{equation}
These results are used in Sect. \ref{params_analytics} to select an optimal degree of polynomial modulation, $\partial$, on simulated datasets, expecting the generalization of the performances to the real data case.

\subsubsection{Decomposition of the Mean Square Error}\label{resids_mse}

The mean square error (MSE) defined by:
\begin{equation}
\label{MSE_PP}
\textrm{MSE} = \mathbb{E}\left(\left\lVert\boldsymbol{\hat{p}}_{xy}-\boldsymbol{p}_{xy}\right\rVert_2^2\right)\,,
\end{equation}
can be decomposed by expectation linearity using the analytical expression of the planetary component estimation (Eq. \ref{pest_ortho}):
\begin{equation}
\label{MSE_SPE}
\textrm{MSE} = \left\lVert\boldsymbol{\Pi}^\perp_{\boldsymbol{M}}\boldsymbol{s}_{xy}\right\rVert_2^2 + \left\lVert\boldsymbol{\Pi_{\boldsymbol{M}}}\boldsymbol{p}_{xy}\right\rVert_2^2 + \left(\ell-(\partial+1)\right)\sigma^2\,,
\end{equation}
as the square error (SE) is decomposable as follows:
\begin{align*}
\textrm{SE} &= \left\lVert\boldsymbol{\hat{p}}_{xy}-\boldsymbol{p}_{xy}\right\rVert_2^2
= \left\lVert\boldsymbol{\Pi}^\perp_{\boldsymbol{M}} \boldsymbol{s}_{xy} + \boldsymbol{\Pi}^\perp_{\boldsymbol{M}} \boldsymbol{p}_{xy} + \boldsymbol{\Pi}^\perp_{\boldsymbol{M}} \boldsymbol{e}_{xy} - \boldsymbol{p}_{xy}\right\rVert_2^2
\\&= \left\lVert\boldsymbol{\Pi}^\perp_{\boldsymbol{M}} \left(\boldsymbol{s}_{xy} + \boldsymbol{e}_{xy} \right) + \boldsymbol{\Pi_{\boldsymbol{M}}} \boldsymbol{p}_{xy}\right\rVert_2^2
= \left\lVert\boldsymbol{\Pi}^\perp_{\boldsymbol{M}} \left(\boldsymbol{s}_{xy} + \boldsymbol{e}_{xy} \right)\right\rVert_2^2 + \left\lVert\boldsymbol{\Pi_{\boldsymbol{M}}} \boldsymbol{p}_{xy}\right\rVert_2^2
\\&= \left\lVert\boldsymbol{\Pi}^\perp_{\boldsymbol{M}} \boldsymbol{s}_{xy}\right\rVert_2^2 + \left\lVert\boldsymbol{\Pi}^\perp_{\boldsymbol{M}} \boldsymbol{e}_{xy}\right\rVert_2^2 + 2\left(\boldsymbol{\Pi}^\perp_{\boldsymbol{M}} \boldsymbol{s}_{xy}\right)^\top\left(\boldsymbol{\Pi}^\perp_{\boldsymbol{M}} \boldsymbol{e}_{xy}\right) + \left\lVert\boldsymbol{\Pi_{\boldsymbol{M}}} \boldsymbol{p}_{xy}\right\rVert_2^2\,,
\end{align*}
\begin{equation*}
\mbox{with}
\left\{ \begin{array}{ll}
    \mathbb{E}\left(\left\lVert\boldsymbol{\Pi}^\perp_{\boldsymbol{M}} \boldsymbol{s}_{xy}\right\rVert_2^2\right) = \left\lVert\boldsymbol{\Pi}^\perp_{\boldsymbol{M}} \boldsymbol{s}_{xy}\right\rVert_2^2 \\
    \mathbb{E}\left(\left\lVert\boldsymbol{\Pi}^\perp_{\boldsymbol{M}} \boldsymbol{p}_{xy}\right\rVert_2^2\right) = \left\lVert\boldsymbol{\Pi}^\perp_{\boldsymbol{M}} \boldsymbol{p}_{xy}\right\rVert_2^2
\end{array} \right. \mbox{by determinism,}
\end{equation*}
\begin{align*}
\mbox{as well as } \mathbb{E}\left(2\left(\boldsymbol{\Pi}^\perp_{\boldsymbol{M}} \boldsymbol{s}_{xy}\right)^\top\left(\boldsymbol{\Pi}^\perp_{\boldsymbol{M}} \boldsymbol{e}_{xy}\right)\right) = 0 \mbox{ by noise centering,}
\end{align*}
\begin{align*}
\mbox {and }
\mathbb{E}\left(\left\lVert\boldsymbol{\Pi}^\perp_{\boldsymbol{M}} \boldsymbol{e}_{xy}\right\rVert_2^2\right)
&= \mathbb{E}\left(\mathrm{tr}\left(\boldsymbol{\Pi}^\perp_{\boldsymbol{M}}\boldsymbol{e}_{xy}\boldsymbol{e}_{xy}^\top\boldsymbol{\Pi}^{\perp \top}_{\boldsymbol{M}}\right)\right)
= \mathbb{E}\left(\mathrm{tr}\left(\boldsymbol{\Pi}^\perp_{\boldsymbol{M}}\boldsymbol{e}_{xy}\boldsymbol{e}_{xy}^\top\boldsymbol{\Pi}^\perp_{\boldsymbol{M}}\right)\right)
\\&= \mathrm{tr}\left(\boldsymbol{\Pi}^\perp_{\boldsymbol{M}}\mathbb{E}\left(\boldsymbol{e}_{xy}\boldsymbol{e}_{xy}^\top\right)\boldsymbol{\Pi}^\perp_{\boldsymbol{M}}\right)
= \mathrm{tr}\left(\boldsymbol{\Pi}^\perp_{\boldsymbol{M}}\sigma^2\boldsymbol{\mathrm{I}}_\ell\boldsymbol{\Pi}^\perp_{\boldsymbol{M}}\right)
\\&= \sigma^2\mathrm{tr}\left(\boldsymbol{\Pi}^\perp_{\boldsymbol{M}}\boldsymbol{\Pi}^\perp_{\boldsymbol{M}}\right)
= \sigma^2\mathrm{tr}\left(\boldsymbol{\Pi}^\perp_{\boldsymbol{M}}\right) = \sigma\left(\ell-(\partial+1)\right)\,,
\end{align*}
\begin{align*}
\mbox{as }
\left\lVert\boldsymbol{\Pi}^\perp_{\boldsymbol{M}} \boldsymbol{e}_{xy}\right\rVert_2^2 &= \left(\boldsymbol{\Pi}^\perp_{\boldsymbol{M}}\boldsymbol{e}_{xy}\right)^\top\left(\boldsymbol{\Pi}^\perp_{\boldsymbol{M}} \boldsymbol{e}_{xy}\right)
= \boldsymbol{e}_{xy}^\top\boldsymbol{\Pi}^{\perp \top}_{\boldsymbol{M}}\boldsymbol{\Pi}^\perp_{\boldsymbol{M}}\boldsymbol{e}_{xy}
\\&= \mathrm{tr}\left(\boldsymbol{e}_{xy}^\top\boldsymbol{\Pi}^{\perp \top}_{\boldsymbol{M}}\boldsymbol{\Pi}^\perp_{\boldsymbol{M}}\boldsymbol{e}_{xy}\right)
= \mathrm{tr}\left(\boldsymbol{\Pi}^\perp_{\boldsymbol{M}}\boldsymbol{e}_{xy}\boldsymbol{e}_{xy}^\top\boldsymbol{\Pi}^{\perp \top}_{\boldsymbol{M}}\right)\,.
\end{align*}

\subsection{Estimation quality with the LPM method}\label{qual}

The orthogonal decomposition of the data spaxels into planetary and stellar components being the basis of our proposed method, the angles between the associated vectors are the most important variable controlling the efficiency of the two sources separation.

Their effect can even be highlighted analytically by using the model of a data spaxel, $\boldsymbol{d}_{xy}$, only composed of a stellar spaxel, $\boldsymbol{s}_{xy}$, perfectly estimable by polynomial modulation of the stellar spectrum estimate (i.e., with $\boldsymbol{\Pi}^\perp_{\boldsymbol{M}}\boldsymbol{s}_{xy} = 0$) and of a planetary spaxel, $\boldsymbol{p}_{xy}$, orthogonally decomposable as follows:
\begin{align*}
\boldsymbol{p}_{xy} = \boldsymbol{p}_{xy}^\parallel + \boldsymbol{p}_{xy}^\perp \mbox{ with } \left\{ \begin{array}{ll} \boldsymbol{p}_{xy}^\parallel\in \mathrm{span}\left(\{\boldsymbol{s}_{xy}\}\right) \\ \boldsymbol{p}_{xy}^\perp\in \mathrm{span}\left(\{\boldsymbol{s}_{xy}\}\right)^\perp \end{array} \right.\,.
\end{align*}
Under these conditions, $\exists c\in\mathbb{R}_+$ so that $\boldsymbol{p}_{xy}^\parallel=c\boldsymbol{s}_{xy}$.\newline
The planetary spaxel estimate thus can be simplified as:
\begin{align*}
\boldsymbol{\hat{p}}_{xy} &= \boldsymbol{\Pi}^\perp_{\boldsymbol{M}}\boldsymbol{s}_{xy} + \boldsymbol{\Pi}^\perp_{\boldsymbol{M}}\boldsymbol{p}_{xy}^\parallel + \boldsymbol{\Pi}^\perp_{\boldsymbol{M}}\boldsymbol{p}_{xy}^\perp
\\&= c\boldsymbol{\Pi}^\perp_{\boldsymbol{M}}\boldsymbol{s}_{xy} + \boldsymbol{\Pi}^\perp_{\boldsymbol{M}}\boldsymbol{p}_{xy}^\perp
= \boldsymbol{\Pi}^\perp_{\boldsymbol{M}}\boldsymbol{p}_{xy}^\perp\,.
\end{align*}

\subsubsection{Relative norms}\label{qual_norms}

Noting that, consequently,
\begin{align*}
\left\lVert\boldsymbol{\hat{p}}_{xy}\right\rVert = \left\lVert\boldsymbol{\Pi}^\perp_{\boldsymbol{M}}\boldsymbol{p}_{xy}^\perp\right\rVert = \left\lVert\boldsymbol{p}_{xy}^\perp\right\rVert - \left\lVert\boldsymbol{\Pi}_{\boldsymbol{M}}\boldsymbol{p}_{xy}^\perp\right\rVert \le \left\lVert\boldsymbol{p}_{xy}^\perp\right\rVert\,,
\end{align*}
its norm squared can be majorized as
\begin{align*}
\left\lVert\boldsymbol{\hat{p}}_{xy}\right\rVert^2 &\le \left\lVert\boldsymbol{p}_{xy} - c\boldsymbol{s}_{xy}\right\rVert^2 = \left\lVert\boldsymbol{p}_{xy}\right\rVert^2 +\left\lVert c\boldsymbol{s}_{xy}\right\rVert^2 - 2~\langle\boldsymbol{p}_{xy},c\boldsymbol{s}_{xy}\rangle \\
&\le \left\lVert\boldsymbol{p}_{xy}\right\rVert^2 + \left\lVert c\boldsymbol{s}_{xy}\right\rVert^2 - 2\left(\langle c\boldsymbol{s}_{xy},c\boldsymbol{s}_{xy}\rangle+\langle\boldsymbol{p}_{xy}^\perp,c\boldsymbol{s}_{xy}\rangle\right) \\
&\le \left\lVert\boldsymbol{p}_{xy}\right\rVert^2 + \left\lVert c\boldsymbol{s}_{xy}\right\rVert^2 - 2\left\lVert c\boldsymbol{s}_{xy}\right\rVert^2 = \left\lVert\boldsymbol{p}_{xy}\right\rVert^2 - \left\lVert c\boldsymbol{s}_{xy}\right\rVert^2\,.
\end{align*}
Thus, the squared ratio of the planetary spaxel estimate norm to the planetary spaxel norm can be expressed as
\begin{align*}
\frac{\left\lVert\boldsymbol{\hat{p}}_{xy}\right\rVert^2}{\left\lVert\boldsymbol{p}_{xy}\right\rVert^2} \le \frac{\left\lVert\boldsymbol{p}_{xy}\right\rVert^2 - \left\lVert c\boldsymbol{s}_{xy}\right\rVert^2}{\left\lVert\boldsymbol{p}_{xy}\right\rVert^2} = 1 - \frac{\left\lVert c\boldsymbol{s}_{xy}\right\rVert^2}{\left\lVert\boldsymbol{p}_{xy}\right\rVert^2} = 1 - c^2\frac{\left\lVert\boldsymbol{s}_{xy}\right\rVert^2}{\left\lVert\boldsymbol{p}_{xy}\right\rVert^2}\,.
\end{align*}
To go further, noting that
\begin{align*}
\langle\boldsymbol{p}_{xy},\boldsymbol{s}_{xy}\rangle = \langle c\boldsymbol{s}_{xy},\boldsymbol{s}_{xy}\rangle + \langle\boldsymbol{p}_{xy}^\perp,\boldsymbol{s}_{xy}\rangle = c~\langle \boldsymbol{s}_{xy},\boldsymbol{s}_{xy}\rangle + 0 = c\left\lVert\boldsymbol{s}_{xy}\right\rVert^2\,,
\end{align*}
it is possible to express c as
\begin{align*}
c = \frac{\langle\boldsymbol{p}_{xy},\boldsymbol{s}_{xy}\rangle}{\left\lVert\boldsymbol{s}_{xy}\right\rVert^2} = \frac{\left\lVert\boldsymbol{p}_{xy}\right\rVert\left\lVert\boldsymbol{s}_{xy}\right\rVert\mbox{cos}(\theta_{\boldsymbol{p}_{xy}/\boldsymbol{s}_{xy}})}{\left\lVert\boldsymbol{s}_{xy}\right\rVert^2} = \frac{\left\lVert\boldsymbol{p}_{xy}\right\rVert}{\left\lVert\boldsymbol{s}_{xy}\right\rVert}\mbox{cos}(\theta_{\boldsymbol{p}_{xy}/\boldsymbol{s}_{xy}})\,,
\end{align*}
with $\theta_{\boldsymbol{p}_{xy}/\boldsymbol{s}_{xy}}$ the angle between the stellar and planetary spaxels.
\begin{align*}
\mbox{It comes }
\frac{\left\lVert\boldsymbol{\hat{p}}_{xy}\right\rVert^2}{\left\lVert\boldsymbol{p}_{xy}\right\rVert^2} \le 1 - \mbox{cos}(\theta_{\boldsymbol{p}_{xy}/\boldsymbol{s}_{xy}})^2 = \mbox{sin}(\theta_{\boldsymbol{p}_{xy}/\boldsymbol{s}_{xy}})^2\,,
\end{align*}
\begin{align*}
\mbox{or even }
\frac{\left\lVert\boldsymbol{\hat{p}}_{xy}\right\rVert}{\left\lVert\boldsymbol{p}_{xy}\right\rVert} \le \left\lvert\mbox{sin}(\theta_{\boldsymbol{p}_{xy}/\boldsymbol{s}_{xy}})\right\rvert = \mbox{sin}\left(\left\lvert\theta_{\boldsymbol{p}_{xy}/\boldsymbol{s}_{xy}}\right\rvert\right)\,,
\end{align*}
as $\theta_{\boldsymbol{p}_{xy}/\boldsymbol{s}_{xy}} \in [-\frac{\pi}{2},\frac{\pi}{2}]$ assuming that $(\boldsymbol{p}_{xy},\boldsymbol{s}_{xy})\in\mathbb{R}_+^\ell\times\mathbb{R}_+^\ell$.

\subsubsection{Relative angles}\label{qual_angles}

Similarly, noting that, consequently,
\begin{align*}
\theta_{\boldsymbol{\hat{p}}_{xy}/\boldsymbol{p}_{xy}} = \theta_{\boldsymbol{\hat{p}}_{xy}/\boldsymbol{s}_{xy}} + \theta_{\boldsymbol{s}_{xy}/\boldsymbol{p}_{xy}} = \theta_{\boldsymbol{\Pi}^\perp_{\boldsymbol{M}}\boldsymbol{p}_{xy}^\perp/\boldsymbol{s}_{xy}} - \theta_{\boldsymbol{p}_{xy}/\boldsymbol{s}_{xy}}\,,
\end{align*}
its angle relative to the planetary spaxel can be expressed by
\begin{align*}
\theta_{\boldsymbol{\hat{p}}_{xy}/\boldsymbol{p}_{xy}} = \pm\frac{\pi}{2} - \theta_{\boldsymbol{p}_{xy}/\boldsymbol{s}_{xy}} \text{ (depending on axis orientation),}
\end{align*}
since
\begin{align*}
\lvert\theta_{\boldsymbol{\Pi}^\perp_{\boldsymbol{M}}\boldsymbol{p}_{xy}^\perp/\boldsymbol{s}_{xy}}\rvert = \mathrm{Arccos}\left(\frac{\langle\boldsymbol{\Pi}^\perp_{\boldsymbol{M}}\boldsymbol{p}_{xy}^\perp,\boldsymbol{s}_{xy}\rangle}{\left\lVert\boldsymbol{\Pi}^\perp_{\boldsymbol{M}}\boldsymbol{p}_{xy}^\perp\right\rVert \left\lVert\boldsymbol{s}_{xy}\right\rVert}\right) = \frac{\pi}{2}\,,
\end{align*}
knowing that
\begin{align*}
\langle\boldsymbol{\Pi}^\perp_{\boldsymbol{M}}\boldsymbol{p}_{xy}^\perp,\boldsymbol{s}_{xy}\rangle = \left(\boldsymbol{\Pi}^\perp_{\boldsymbol{M}}\boldsymbol{p}_{xy}^\perp\right)^\top\boldsymbol{s}_{xy} = \boldsymbol{p}_{xy}^{\perp \top}\boldsymbol{\Pi}^{\perp \top}_{\boldsymbol{M}}\boldsymbol{s}_{xy} = \boldsymbol{p}_{xy}^{\perp \top}\boldsymbol{\Pi}^\perp_{\boldsymbol{M}}\boldsymbol{s}_{xy} = 0\,,
\end{align*}
using again the simplifying assumption $\boldsymbol{\Pi}^\perp_{\boldsymbol{M}}\boldsymbol{s}_{xy} = 0$.

\section{On-sky data sum up}\label{obs}

We used observations from 3 systems, each on one single epoch. A complete description of these data is available in Sect. \ref{data}.

\begin{table}[ht]
    \renewcommand{\arraystretch}{1.11}
    \begin{tabular}{ c c c c }
      \hline
      Target & Date & UT Start& Exposure time \\ 
              & [YYYY-MM-DD] & [hh:mm:ss] & [s] \\
        \hline\hline
      PDS70 & 2018-06-20 & 00:02:15* & 300 \\
      PDS70 & 2018-06-20 & 00:09:34* & 300 \\
      PDS70 & 2018-06-20 & 00:16:53° & 300 \\
      PDS70 & 2018-06-20 & 00:24:13° & 300 \\
      PDS70 & 2018-06-20 & 00:31:33° & 300 \\
      PDS70 & 2018-06-20 & 00:38:54° & 300 \\
      \hline
      HTLup & 2021-03-26 & 08:49:56° & 20 \\
      HTLup & 2021-03-26 & 08:52:49° & 20 \\
      HTLup & 2021-03-26 & 08:55:54° & 20 \\
      HTLup & 2021-03-26 & 08:59:29° & 20 \\
      HTLup & 2021-03-26 & 09:03:05° & 20 \\
      HTLup & 2021-03-26 & 09:06:41° & 20 \\
      HTLup & 2021-03-26 & 09:10:17° & 20 \\
      HTLup & 2021-03-26 & 09:13:52° & 20 \\
      HTLup & 2021-03-26 & 09:17:29° & 20 \\
      HTLup & 2021-03-26 & 09:21:07° & 20 \\
      HTLup & 2021-03-26 & 09:24:00* & 20 \\
      \hline
      YSES1 & 2023-04-27 & 01:44:26° & 300 \\
      YSES1 & 2023-04-27 & 01:51:07° & 300 \\
      YSES1 & 2023-04-27 & 01:57:48° & 300 \\
      YSES1 & 2023-04-27 & 02:04:30° & 300 \\
      YSES1 & 2023-04-27 & 02:11:12° & 300 \\
      YSES1 & 2023-04-27 & 02:20:40° & 300 \\
      YSES1 & 2023-04-27 & 02:27:21° & 300 \\
      YSES1 & 2023-04-27 & 02:34:03° & 300 \\
      YSES1 & 2023-04-27 & 02:40:45° & 300 \\
      \hline
    \end{tabular}
    \caption{Information on the IFS cubes used for the work presented in this article. Two PDS70 cubes (annotated with *) are rejected due to poor Strehl \citep[as in][]{Hashimoto_2020}. Then, one HTLup cube (still annotated with *) is rejected because of an extension of the AO band on the side of the \Halpha line. Finally, no YSES1 observation cube is rejected.
    \label{table:obs}}
\end{table}

\onecolumn

\section{Fake planets injections}\label{fake}

\begin{figure*}[ht]
\begin{center}
\vspace{0.25cm}\par
\begin{minipage}{0.475\textwidth}
\centering
\includegraphics[width=\linewidth]{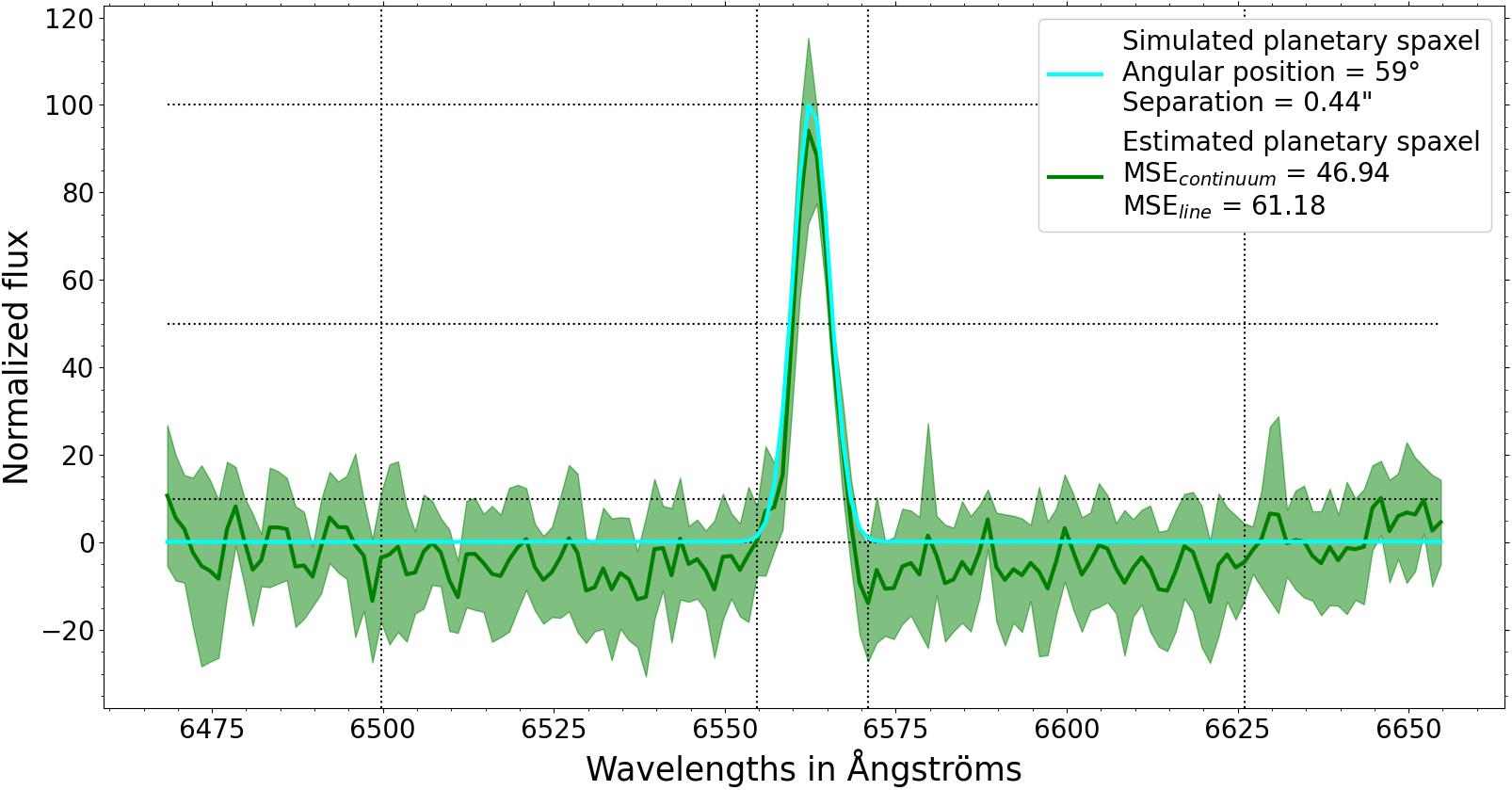}
\end{minipage}
\hspace{0.025\textwidth}
\begin{minipage}{0.475\textwidth}
\centering
\includegraphics[width=\linewidth]{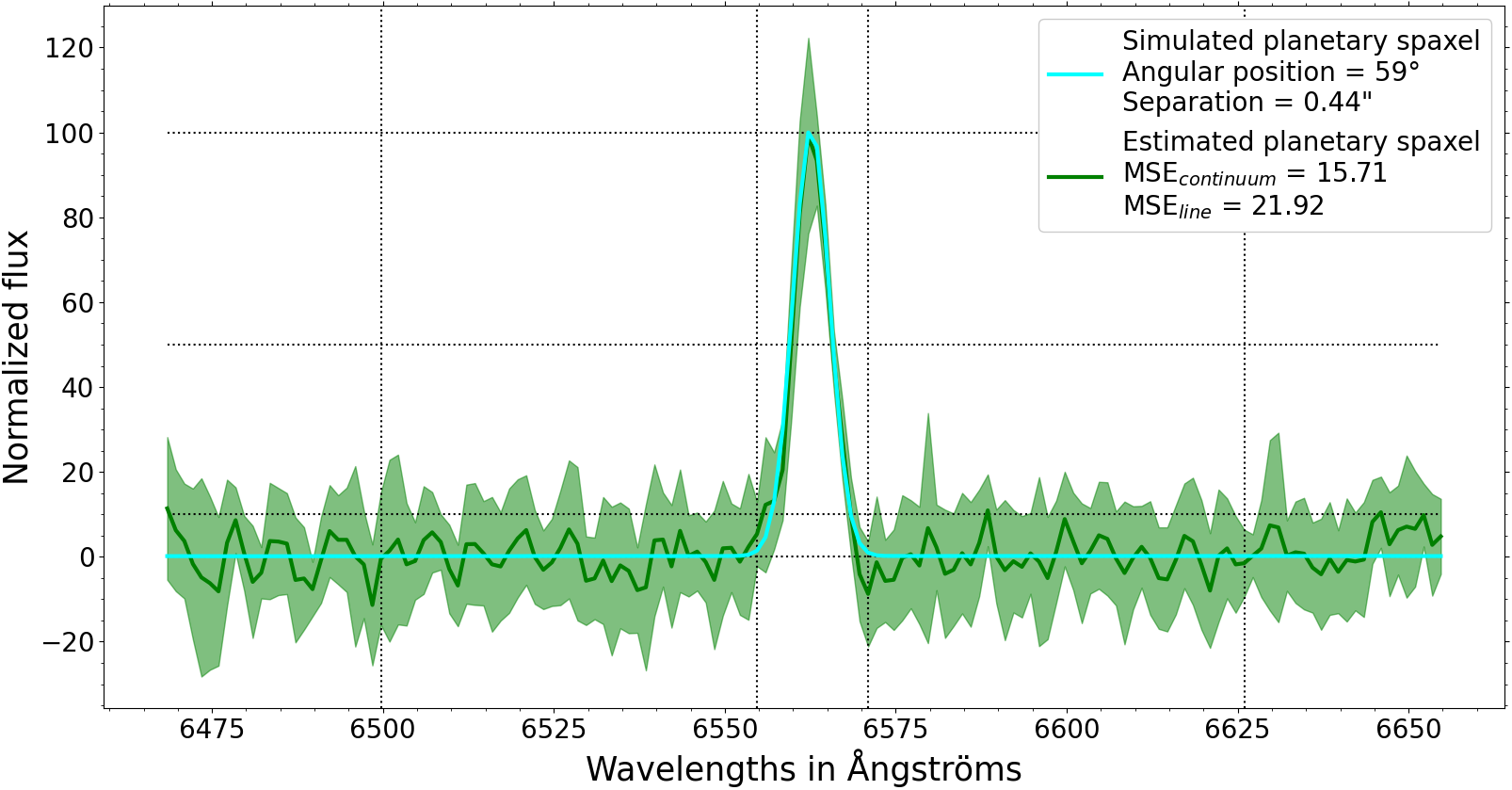}
\end{minipage}
\vspace{0.25cm}\par
\begin{minipage}{0.475\textwidth}
\centering
\includegraphics[width=\linewidth]{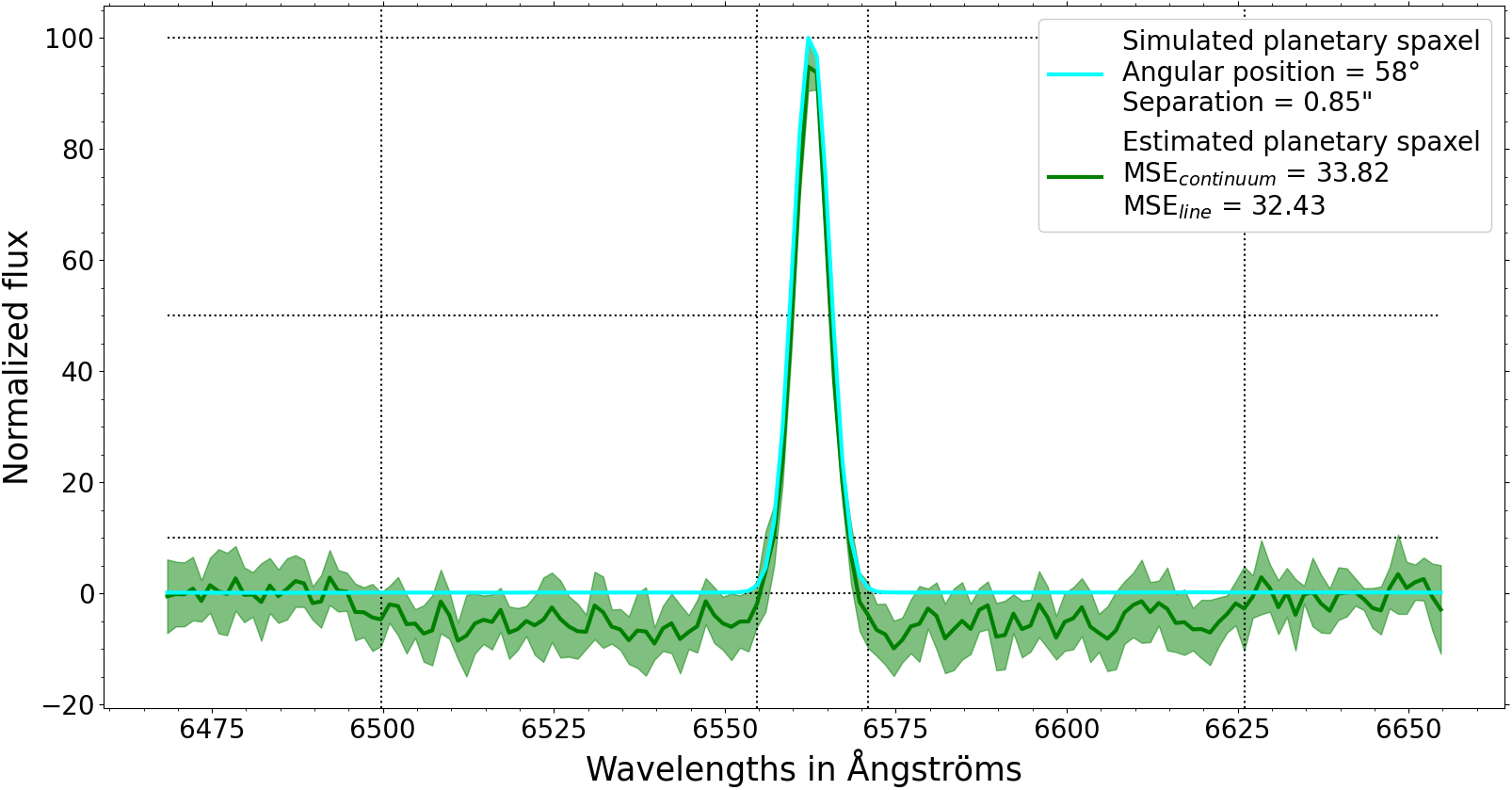}
\end{minipage}
\hspace{0.025\textwidth}
\begin{minipage}{0.475\textwidth}
\centering
\includegraphics[width=\linewidth]{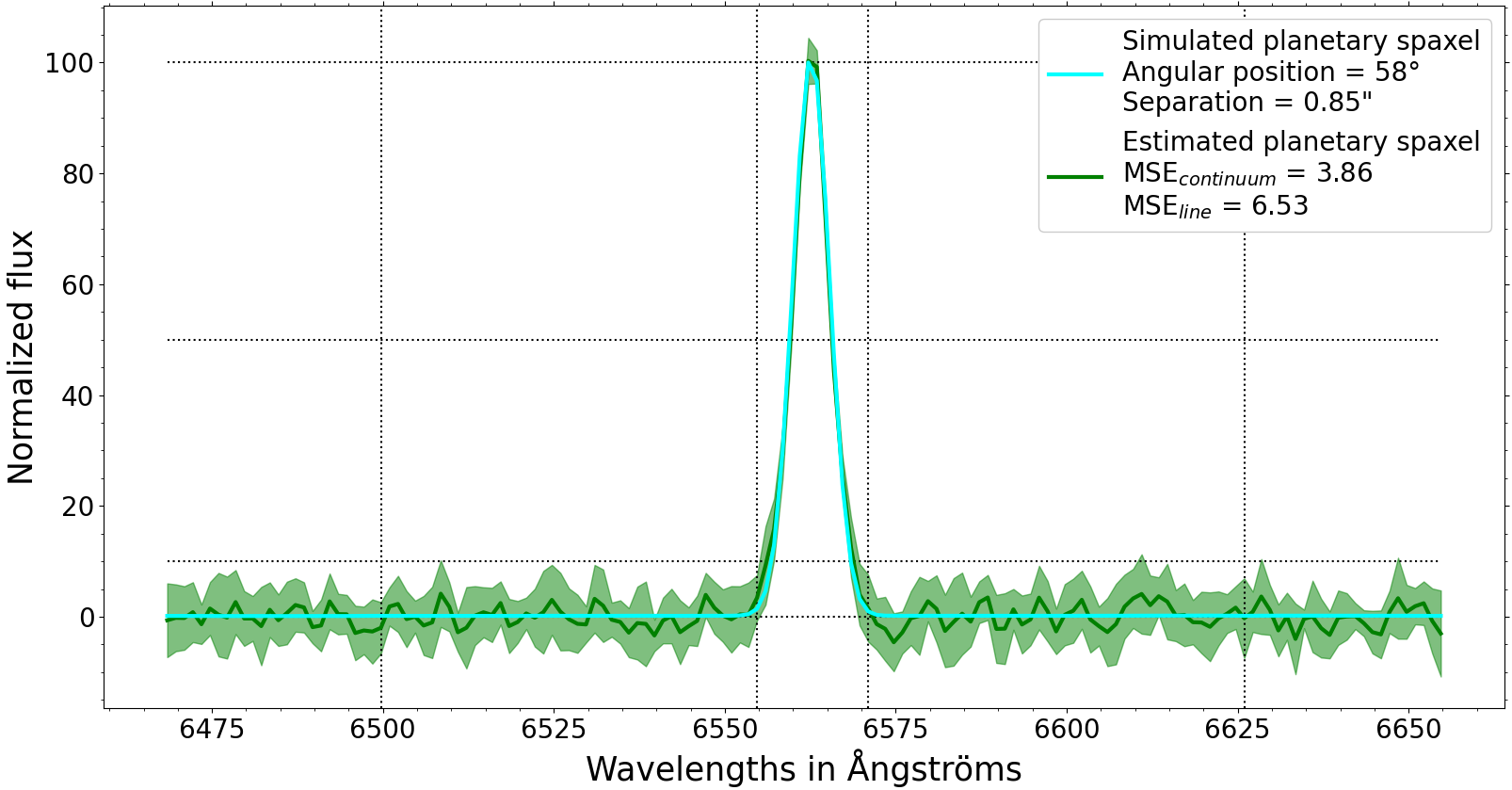}
\end{minipage}
\vspace{0.25cm}\par
\begin{minipage}{0.475\textwidth}
\centering
\includegraphics[width=\linewidth]{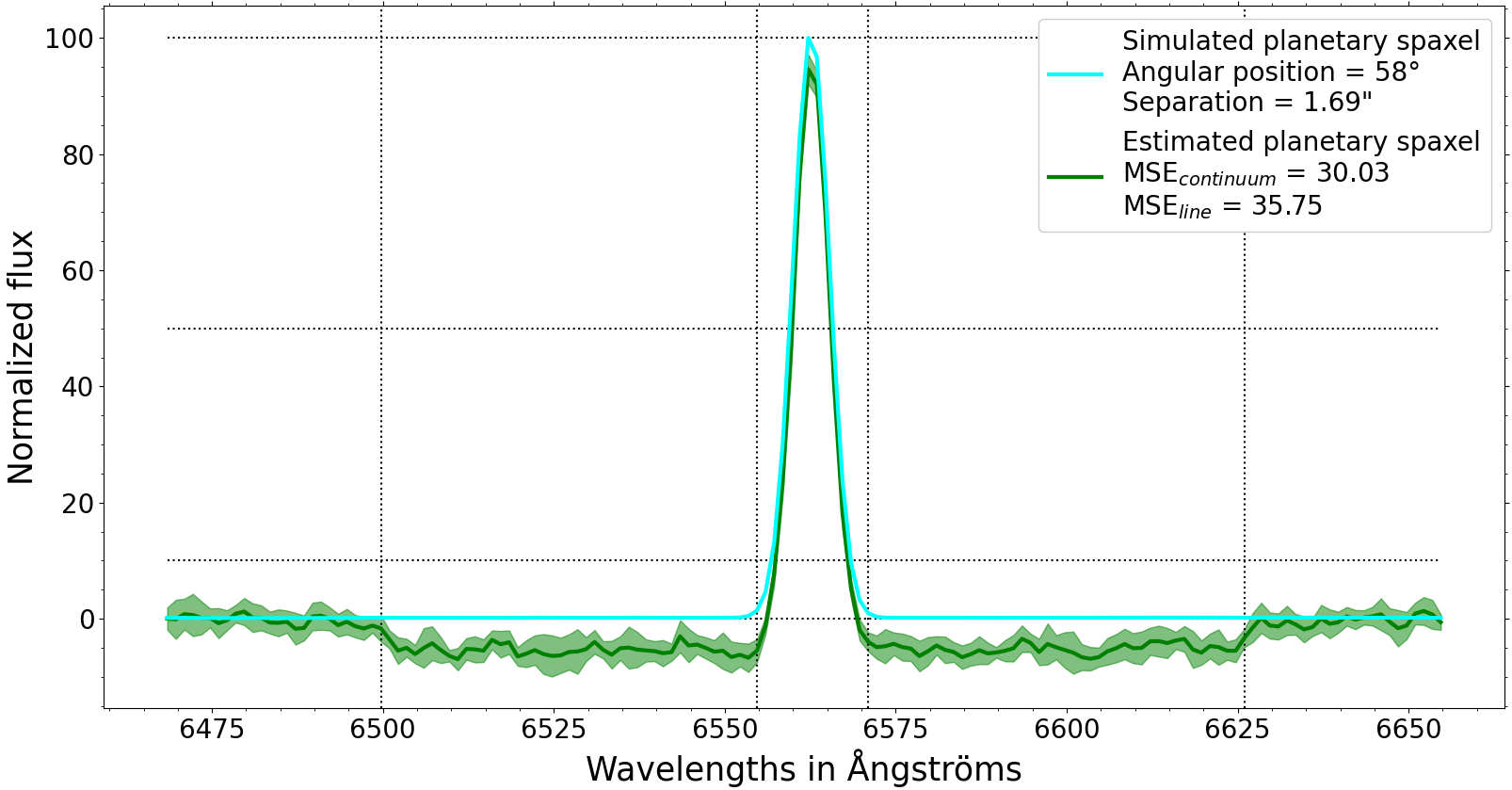}
\end{minipage}
\hspace{0.025\textwidth}
\begin{minipage}{0.475\textwidth}
\centering
\includegraphics[width=\linewidth]{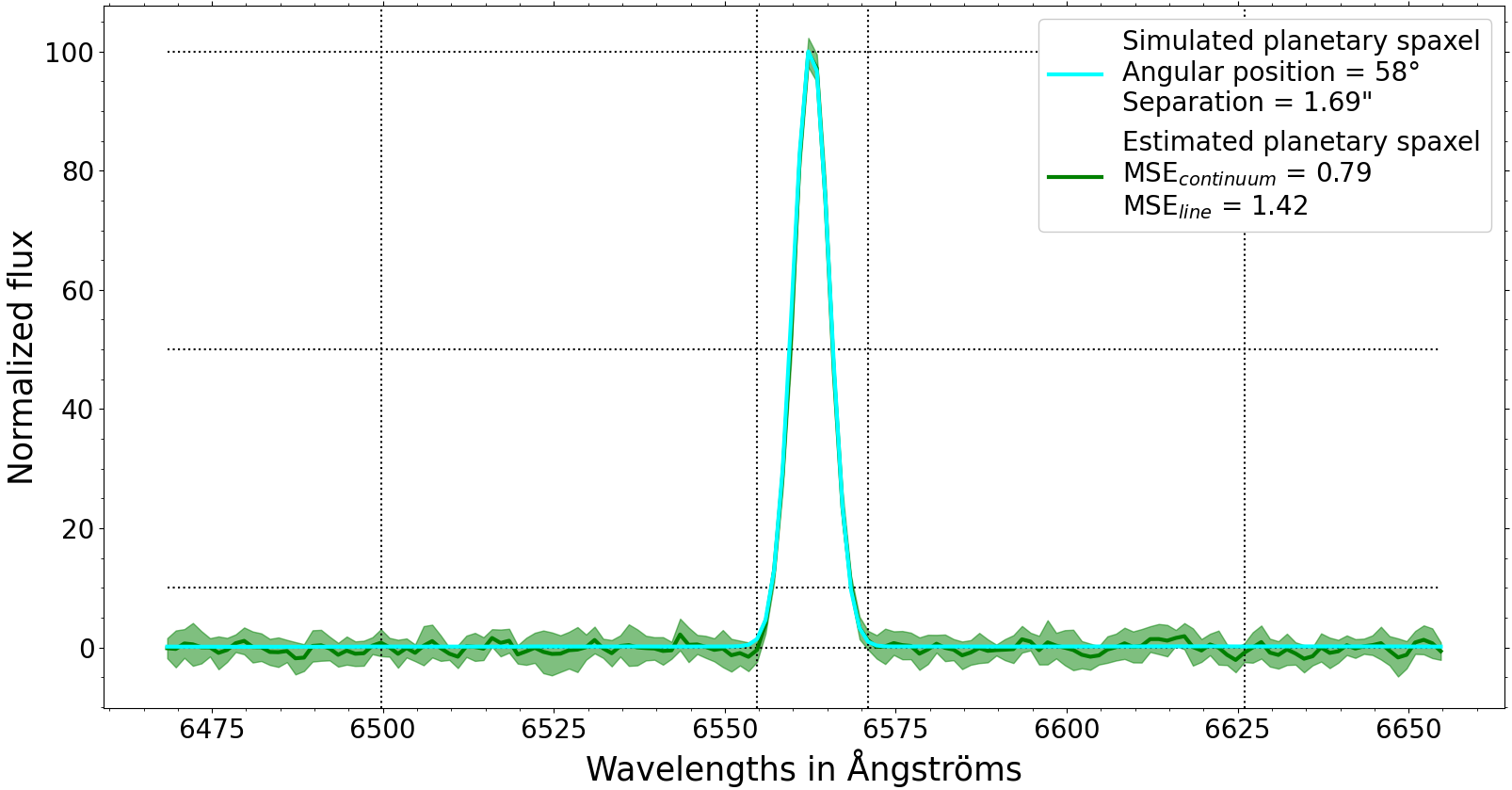}
\end{minipage}
\vspace{0.25cm}\par
\begin{minipage}{0.475\textwidth}
\centering
\includegraphics[width=\linewidth]{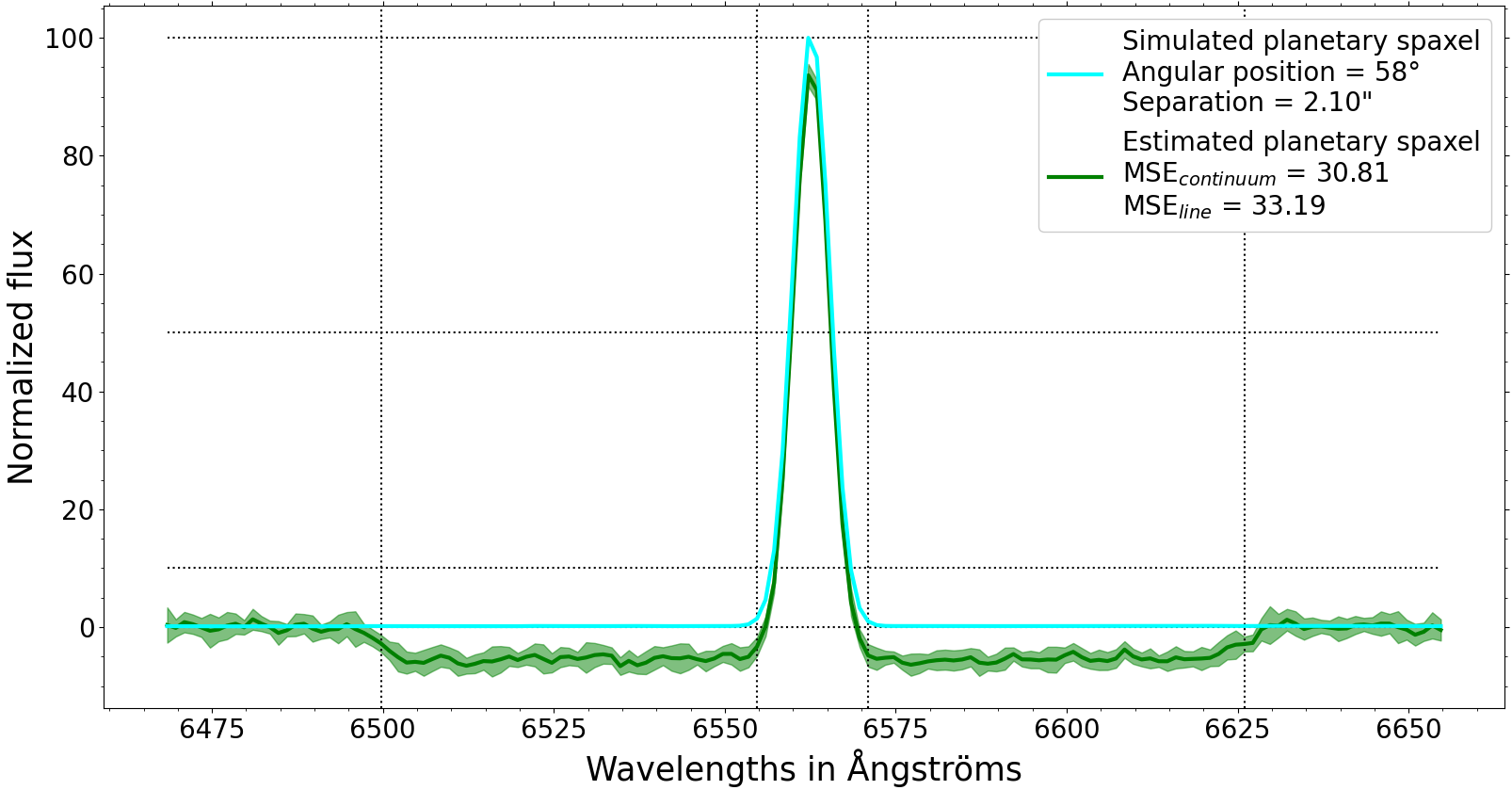}
\end{minipage}
\hspace{0.025\textwidth}
\begin{minipage}{0.475\textwidth}
\centering
\includegraphics[width=\linewidth]{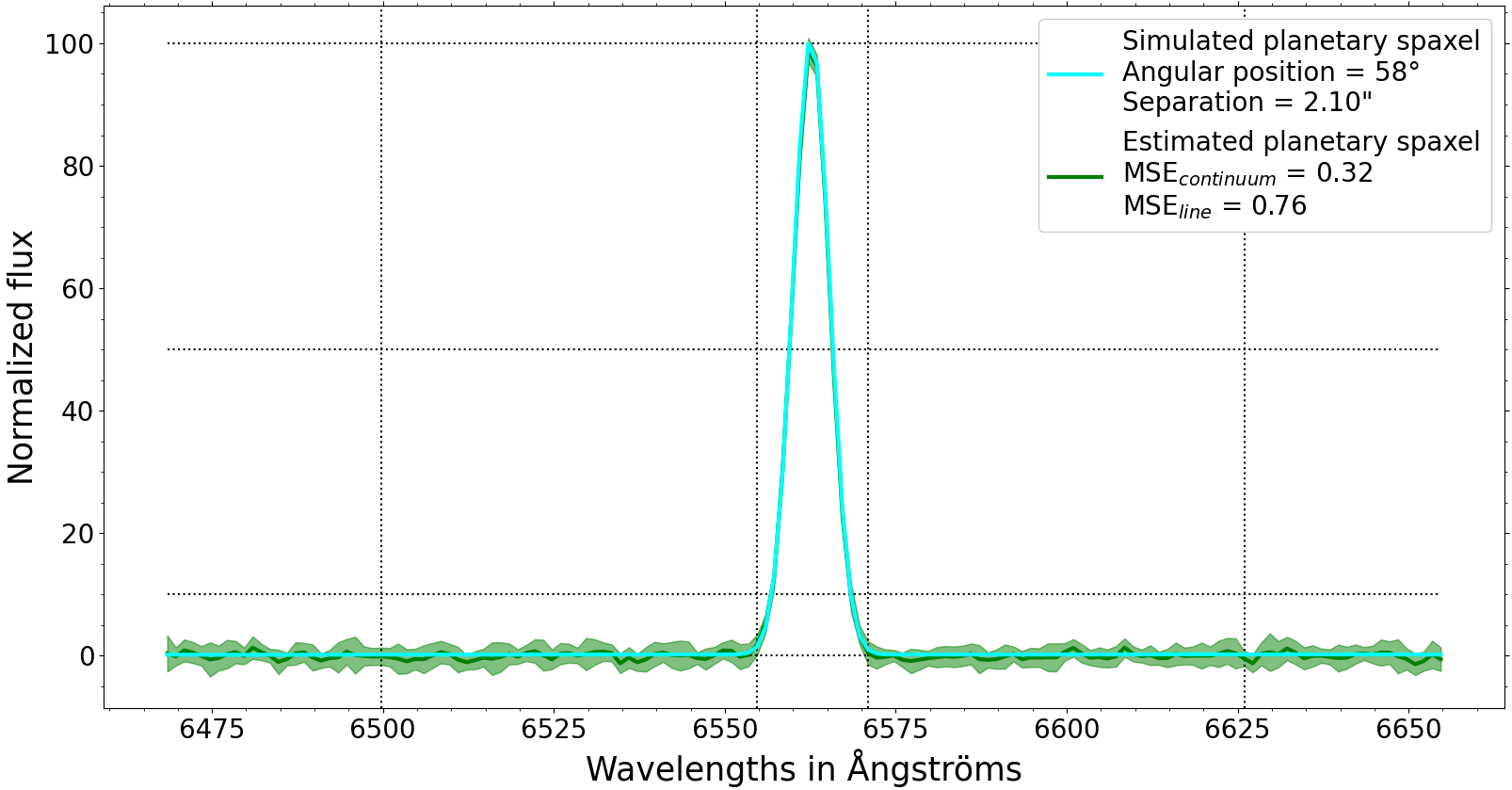}
\end{minipage}
\end{center}
\caption{Fake planet recovery by both stellar halo subtraction methods (SGF on the left and LPM on the right) after injection in YSES1 b observation cubes. The cyan spaxel is the one of the simulated planet injected in the cubes. The green spaxel is the mean of its estimations by stellar halo subtraction along the 9 cubes. The light green area covers this mean more or less the standard deviation computed along the 9 cubes at each wavelength. The fake planets were injected at separations of 0.44” (first line), 0.85” (second line), 1.69” (third line), and 2.10” (fourth line). The horizontal dots delimit the 100\%, 50\%, 10\% and 0\% of the height of the simulated line (normalized to 100). The vertical dots delimit the line wavelengths from the continuum wavelengths used for MSE calculations.}\label{fig:fake_planets_injections}
\end{figure*}

\twocolumn

\section{Simulation process}\label{simus_process}

Although simplistic compared with real data, simulations give a better understanding of processing, defects, and improvements thanks to ground-truth knowledge. In addition, it allows for an exploration of the free parameters of each of the two methods.

We chose to simulate a cube of observation of the PDS70 system with the MUSE instrument. The scene then consists of the star PDS70 and the planets PDS70 b and c. A BT-NextGen (AGSS2009)\footnote{\samepage\sloppy\url{http://svo2.cab.inta-csic.es/theory/newov2/index.php?models=bt-nextgen-agss2009}} model spectrum \citep{Allard_2011} is chosen to approximate the continuum emission of the star. Those of the planets are presently not characterized in detail. Their emission at MUSE wavelengths should be a mix of photospheric emission and veiling \citep{Zhou_2021}. However, being predicted to be faint compared to the line \citep{Zhou_2021}, it is expected to have little impact on the estimation problem. A solar metallicity BT-Settl\footnote{\samepage\sloppy\url{http://svo2.cab.inta-csic.es/theory/newov2/index.php?models=bt-settl}} model spectrum \citep{Allard_2012} is finally chosen to approximate it. The surface gravity is set each time at 3.5 dex, while the effective temperatures are set at 4000 K for PDS70 and 1500 K for PDS70 b and c. The atmospheric parameters of the two protoplanets should be considered as rough guesses (e.g., with a lower observed magnitude at infrared wavelengths, c might be cooler or more extincted than b). The real continua of the planets are expected to be a mix of atmospheric emission, emission at the accretion shock, and extinction by foreground material \citep{Wang_2021, Cugno_2021, Zhou_2021}. Main hint of accretion phenomena, an \Halpha line is simulated by a Gaussian of Full Width at Half Maximum (FWHM) $\sim$ $3.75\text{\AA}$ (for Doppler spreading, informative of accretion dynamics) and added to the stellar and planetary spectra. These spectra are then multiplied by a telluric absorption spectrum generated with the SKYCALC tool\footnote{\samepage\sloppy\url{https://www.eso.org/observing/etc/bin/gen/form?INS.MODE=swspectr+INS.NAME=SKYCALC}} \citep{Moehler_2014}.

Finally, the spectra are degraded to the resolution of MUSE by convolution with an LSF modeled as a Gaussian of total flux 1 and FWHM $\sim$ $3.75\text{\AA}$, discretized by interpolation into $\ell=3681$ spectral channels from $\sim4750\text{\AA}$ to $\sim9350\text{\AA}$ in steps of $1.25\text{\AA}$ (as set by the data handling pipeline for wavelength increment) and set to Not a Number (NaN) on the band $\sim5578\text{\AA}-6050\text{\AA}$ (because of the pollution by the sodium line of the AO laser).

A simple Moffat function, with simple hand-built parameter evolution models (FWHM$(\lambda)\sim 1/\lambda$ notably) is used for the FSF simulation. It models both the heart and the wings of the FSF.  

Then, even if data cubes originally have more than $300$ rows and $300$ columns, we only use a smaller zoomed cube centered on the host star with $m=41$ and $n=41$ ($\sim1"$ in the sky) to speed up processing. Considering this small sub-cube also allows for the use of the small-field hypothesis. No planet is expected to be revealed by stellar halo subtraction beyond this, since the adaptive optics only corrects turbulence over a radius of $\sim0.5"$. With this configuration, the PDS70 b and c planets are respectively in positions $(17,14)$ and $(29,22)$ in the $(x,y)$ plane, i.e., around $0.17$” and $0.23$” separations with $153$° and $-77.5$° angular positions (knowing that 1 pixel corresponds to 0.025”).

Finally, an additive noise following a Gaussian distribution is generated. The signal-to-noise ratio (S/N) is defined as being the ratio between the planetary flux at \Halpha line wavelength and the noise standard deviation, $\sigma$. It is set to 20, as reported in the literature, but also at 200 or 2000 depending on the application.

\begin{figure}[t!]
\includegraphics[width=\columnwidth]{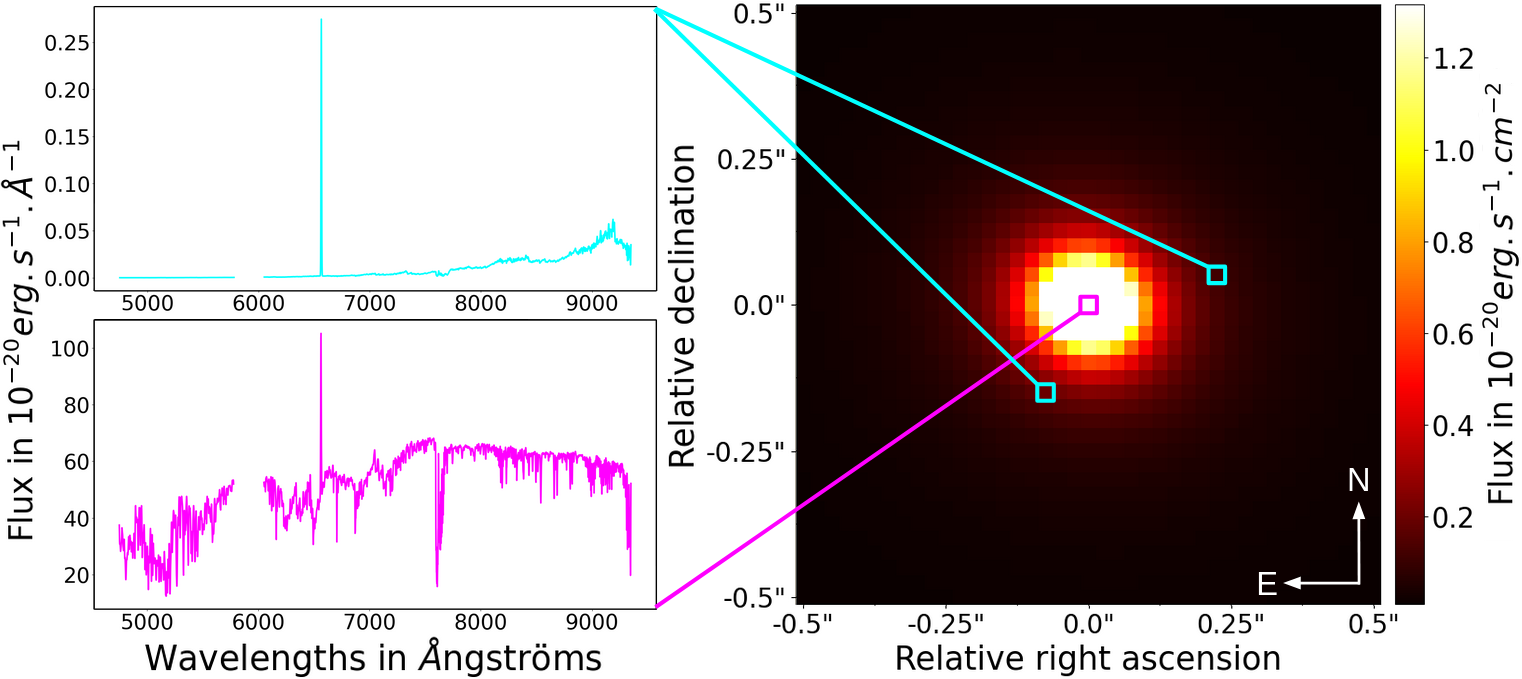}
\caption{Simulated scene of the PDS70 system, with the PDS70 star and the PDS70 b and c protoplanets. Top-left:\ Spectrum of the protoplanets. Bottom-left:\ Spectrum of the star. They are both degraded by the LSF of the instrument as well as multiplied by a telluric absorption spectrum modeling the effect of the Earth atmosphere. Right: Spectrally integrated image of all the monochromatic images. Stellar and planetary fluxes are spread by the FSF of the instrument resulting in the masking of the planets because of their very low contrast compared to the star. The random noise is negligible compared with the stellar nuisance.\label{fig:Sim}}
\end{figure}

Fig. \ref{fig:Sim} sums up the simulated scene, with the planetary and stellar spectra (including LSF spreads), as well as the spectrally integrated image of the three objects (including FSF spreads).

\section{ROC curve construction process}\label{roc_process}

\begin{figure}[h!]
\includegraphics[width=\columnwidth]{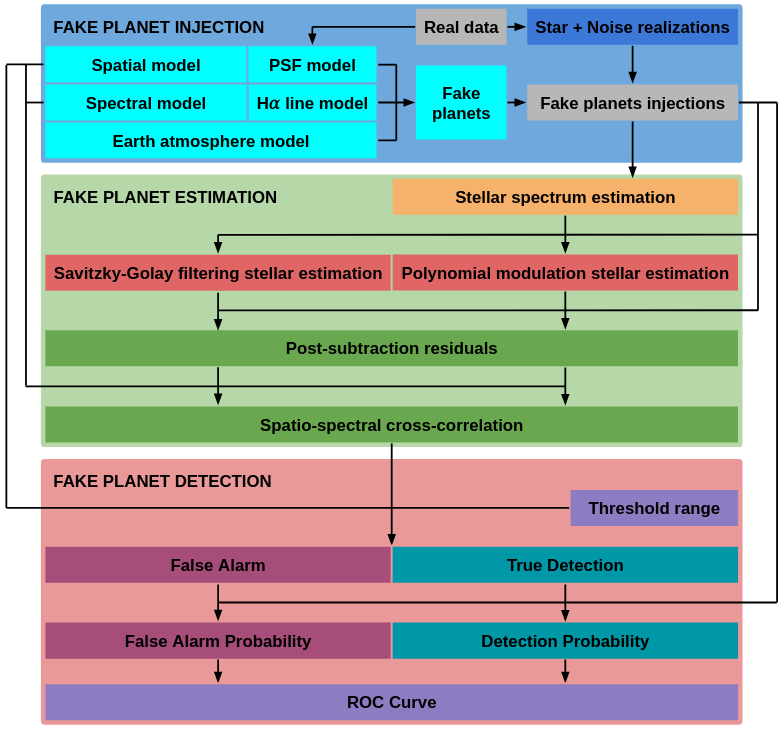}
\caption{General ROC curve construction diagram. The first step is the injection of fake planets into real data cubes, at different separations and contrasts. The second step is the recovery of the planetary information from these data cubes using the two stellar halo subtraction methods. The third step is the determination of the probabilities of false alarm and detection following the variation of the planet detection validation threshold. The ROC curve depends directly on these last two variables.\label{fig:ROC_process}}
\end{figure}

\onecolumn

\section{Noise statistics}\label{noise}

\begin{figure*}[ht]
\includegraphics[width=\linewidth]{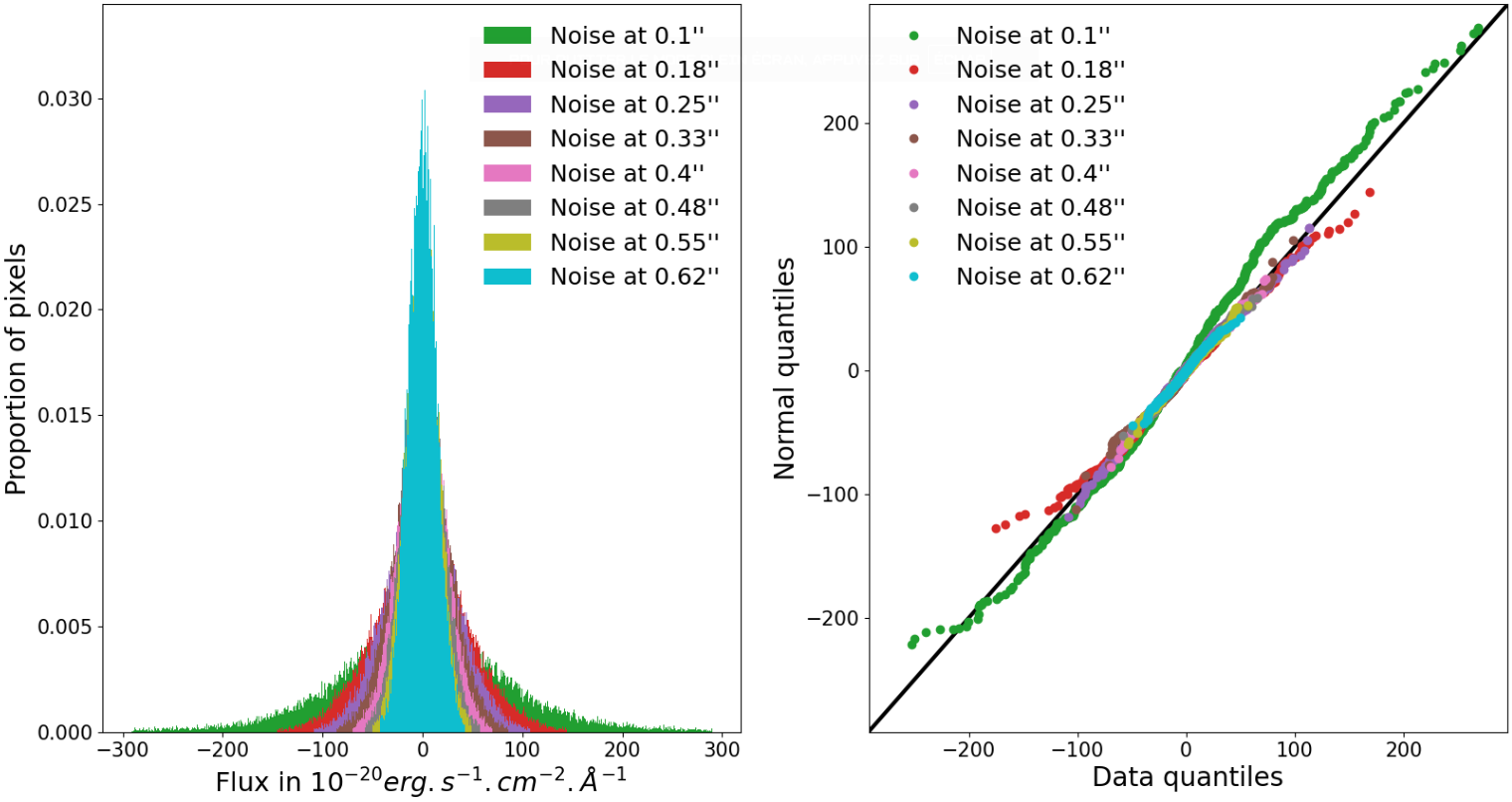}
\caption{Normalized histograms and quantile-quantile plots of post-subtraction residual noise (from the 821 spectral channels of the median cube of the post-subtraction residuals of PDS70 images indicated in App. \ref{obs}, with our proposed method following the parameters indicated in Tab. \ref{table:param}). Each monochromatic image is separated into concentric rings \citep{Andres_1994} grouped by 3 to distinguish distributions by separation at the center. Due to stellar residuals and possibly photon noise, the noise standard deviation is greater in the center, as shown in the figure on the left, and its distribution deviates from a Gaussian distribution, as shown in the figure on the right. On the contrary, as the separation increases, the standard deviation decreases and the distribution approaches a Gaussian distribution.\label{fig:PDS70_noise}}
\end{figure*}

\label{LastPage}

\end{appendix}

\end{document}